\documentclass[pra,twocolumn,superscriptaddress]{revtex4-1}

\usepackage{graphicx}
\usepackage{bm}
\usepackage{amsmath}                          
\usepackage{amssymb}                          
\usepackage{amsthm} 
\usepackage{mathtools}
\usepackage{longtable}
\usepackage{hhline}

\newcommand{\bvec}[1]{{\mathbf #1}}

\newcommand{\ket}[1]{\left| #1 \right>}

\newcommand{\beq}{\begin{eqnarray}}
\newcommand{\eeq}{\end{eqnarray}}

\newcommand{\vast}{\bBigg@{4}}
\newcommand{\Vast}{\bBigg@{5}}

\begin{document}

\title{Charge glass in an extended dimer Hubbard model}

\author{Meldon B. Deglint}
\affiliation{Department of Geoscience, University of Calgary, Calgary, Alberta, T2N 1N4, Canada}

\author{Krishant Akella}
\affiliation{Department of Physics, Simon Fraser University, Burnaby, British Columbia, V5A 1S6, Canada}

\author{Malcolm P. Kennett}
\affiliation{Department of Physics, Simon Fraser University, Burnaby, British Columbia, V5A 1S6, Canada}

\date{\today}
\begin{abstract}
The charge degrees of freedom in several different organic charge transfer salts display slow or glassy dynamics.
In order to gain insight into this behaviour, we obtain the low energy theory for an extended dimer Hubbard model, taking into account 
the occupations of sites on neighbouring dimers. We take a classical limit of the resulting effective model of coupled spins
and dimers and study it using classical Monte Carlo simulations.  We find that frustration induced by intra- and inter-dimer interactions 
leads to glassiness in the charge degress of freedom in the absence of ordering of the spin degrees of freedom.
Our results may have relevance to experimental observations of relaxor ferroelectric behaviour in the dynamics 
of organic charge transfer salts.
\end{abstract}

\maketitle

\section{Introduction}

Strong interactions between charge and spin degrees of freedom are responsible for numerous 
novel phases in strongly correlated electron materials.  One particularly attractive class
of materials for investigating such effects are organic charge transfer salts \cite{Powell2011,Hotta2012,Dressel2020}.
These materials display phenomena such as unconventional superconductivity and spin liquid behaviour \cite{Shimizu2003,Powell2006}.

In addition to low temperature phenomena, at temperatures on the orders of tens of kelvins, slow and 
glassy charge dynamics have been observed, particularly in the dielectric relaxation of the $\kappa$-(BEDT-TTF)$_2$X family of organic charge transfer 
salts \cite{Lunkenheimer2015a,Tomic2015}.   In the organic charge transfer salt  $\kappa$-(BEDT-TTF)$_2$Cu$_2$(CN)$_3$, which shows spin liquid
behaviour at low temperatures, there is broad in-plane dielectric relaxation below $\sim 60$ K, and
out of plane relaxor-like dielectric response below about 60 K \cite{Abdel-Jawad2010}.
Glassy response has also been observed in the dielectric function of  $\kappa$-(BEDT-TTF)$_2$Ag$_2$(CN)$_3$ \cite{Pinteric2018}
and glassy freezing of electrons at low temperatures has been suggested for  $\kappa$-(BEDT-TTF)$_2$Hg(SCN)$_2$Br \cite{Hemmida2018}.

The origin of the electric dipoles that give rise to the relaxor ferroelectric behaviour in  $\kappa$-(BEDT-TTF)$_2$X salts 
is still an active area of investigation.
There is evidence for charge disproportionation in dimers in $\kappa$-(BEDT-TTF)$_2$Hg(SCN)$_2$Cl \cite{Drichko2014}
but Sedlmeier {\it et al.} \cite{Sedlmeier2012} did not find evidence for charge disproportionation in
$\kappa$-(BEDT-TTF)$_2$Cu$_2$(CN)$_3$, $\kappa$-(BEDT-TTF)$_2$Cu[N(CN)$_2$]Cl
or  $\kappa$-(BEDT-TTF)$_2$Cu[N(CN)$_2$]Br.  Pinteri\'{c} {\it et al.} \cite{Pinteric2014} have argued that optical 
measurements of $\kappa$-(BEDT-TTF)$_2$Cu$_2$(CN)$_3$
are not consistent with local dipoles. Suggested mechanisms for glassy dynamics 
that do not involve dipoles in dimers include domain walls between dimer Mott and charge ordered phases \cite{Fukuyama2017} 
and a ``dielectric catastrophe'' \cite{Pustogow2021}.
At higher temperatures disorder in the conformational orientation of ethylene groups in BEDT-TTF \cite{Guterding2015}
has been implicated in glassy dynamics in $\kappa$-(BEDT-TTF)$_2$X materials \cite{Hartmann2015,Muller2015}.

Relaxor ferroelectric or glassy behaviour has also been seen in several 
other families of organic charge transfer salts, such as $\beta^\prime$-Pd(dmit)$_2$ salts \cite{Abdel-Jawad2013,Fujiyama2018},
$\beta^\prime$-(BEDT-TTF)$_2$ICl$_2$ \cite{Iguchi2013,Muller2020}, $\theta$-(BEDT-TTF)$_2$RbZn(SCN)$_4$ \cite{Kagawa2013} 
and $\theta$-(BEDT-TTF)$_2$CsZn(SCN)$_4$ \cite{Sato2014,Sato2016}, usually in the tens of kelvin temperature range. 
A dielectric peak similar to those seen in relaxor ferroelectrics is also seen in $\alpha$-(BEDT-TTF)$_2$I$_3$ \cite{Lunkenheimer2015b,Ivek2017}.

Many of the $\kappa$-(BEDT-TTF)$_2$X salts such as
$\kappa$-(BEDT-TTF)$_2$Cu$_2$(CN)$_3$ can be described by a triangular lattice of dimers forming 
a quarter-filled extended two dimensional Hubbard model with both intra- and inter-dimer interactions \cite{Hotta2010,Naka2010,Gomi2013}.
Quarter filled extended Hubbard models have been identified as enhancing 
geometric frustration of charge degrees of freedom \cite{Merino2005} and
geometric frustration in charge ordering has been emphasized as a factor in leading to a 
charge cluster glass with no long range order in $\theta$-(BEDT-TTF)$_2$RbZn(SCN)$_4$ \cite{Kagawa2013}.

In this paper we study a quarter-filled extended dimer Hubbard model with the same form as that proposed to 
describe $\kappa$-(BEDT-TTF)$_2$Cu$_2$(CN)$_3$.  Following Ref.~\cite{Hotta2010} we obtain the low energy theory in 
the one electron per dimer limit, which can be written in terms of spin and dipole degrees of freedom.
Going beyond Ref.~\cite{Hotta2010}, we take into account the occupation of 
next-nearest neighbour sites on the couplings in the effective Hamiltonian as was 
done for an extended Hubbard model on a square lattice \cite{Farrell2013,Farrell2014}.
This leads to a distribution of couplings between spin and dipole degrees of freedom in the low-energy theory. 
Rather than simulate the resulting model directly, which would be prohibitively computationally expensive, 
we make a classical approximation, which gives a model of vector spins coupled to Ising dipoles and study this model with classical Monte Carlo
simulations.  Our main result is that in this approximate model we find that glassiness in the charge degrees of freedom 
occurs over a range of intra-dimer and inter-dimer nearest neighbour interaction strengths.

This paper is structured as follows: in Sec.~\ref{sec:model} we introduce the model we study and describe how we obtain
the low energy theory.  In Sec.~\ref{sec:MC} we discuss our Monte Carlo simulations of the simplified model and the 
results we obtain from our simulations.  Finally, in Sec.~\ref{sec:disc} we discuss our results and conclude.

\section{Model}
\label{sec:model}
\begin{figure}[ht]
\includegraphics[width=8cm]{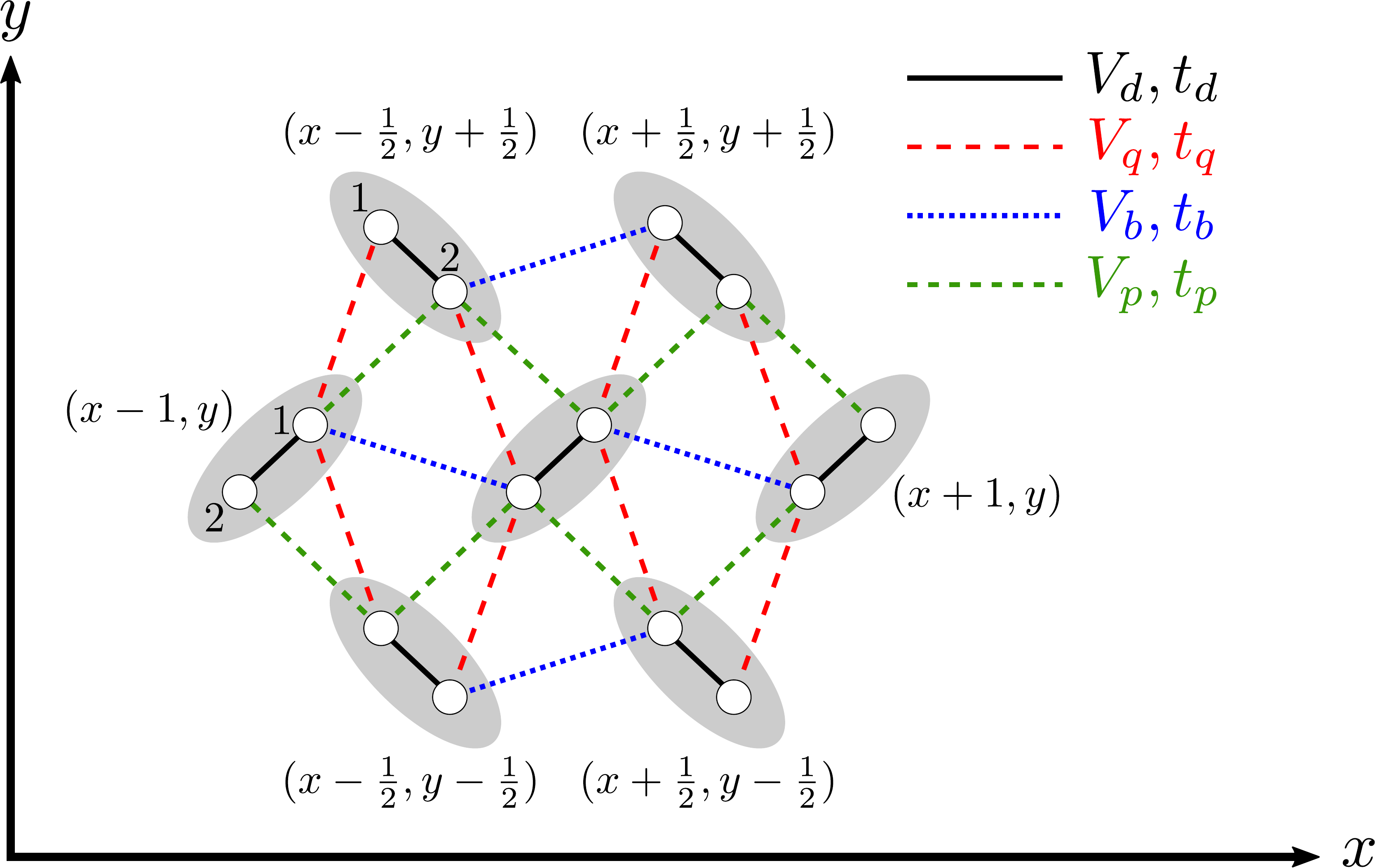}
\caption{Triangular lattice of dimers used for the extended Hubbard model. The position of the central dimer is $(x,y)$. The $t_i$ denote different types of 
	hopping and the $V_i$ denote nearest neighbour interactions.}
\label{fig:model}
\end{figure}

We consider a two dimensional extended Hubbard model of dimers on a triangular lattice, introduced by Hotta \cite{Hotta2010} for 
$\kappa$-(BEDT-TTF)$_2$Cu$_2$(CN)$_3$.
We allow for both intra-dimer and inter-dimer hopping, on-site interactions and both intra-dimer and inter-dimer
nearest neighbour interactions.  All hops and interactions are illustrated in Fig.~\ref{fig:model}.
We write the Hamiltonian as 
\begin{equation}
	H = H_T + H_U + H_V,
\label{eq:Ham}
\end{equation}
where $H_T$ is the hopping part of the Hamiltonian, $H_U$ is the on-site Hubbard interaction and $H_V$ is the 
nearest neighbour interaction term.  We now consider each term in detail. 
Let $(x,y)$ label a dimer on the triangular lattice. The hopping part of the Hamiltonian is then 
\begin{eqnarray}
	H_T &=\sum_{\alpha}t_\alpha \sum_{(x,y), i}\sum_{(x',y'),j}^\prime \sum_{\sigma}c^{\dagger}_{(x',y'), j, \sigma} c^{\vphantom{\dagger}}_{(x,y), i, \sigma}, \nonumber \\
	& 
\label{eq:hopping}
\end{eqnarray}
where $i = 1,2$ labels the lattice sites on dimer $(x,y)$, $j = 1,2$ labels the lattice sites on dimer $(x',y')$, spin is
labelled by $\sigma = \uparrow, \downarrow$ and $\alpha = d, b, p ,q$ specifies the type of hopping, following the notation of 
Ref.~\cite{Hotta2010}. 
The restriction that $(x',y'),j$ are limited to nearest neighbour sites and dimers is denoted by the prime on the sum. 
The operator $c^{\dagger}_{(x',y'), j, \sigma}$ creates an electron of spin $\sigma$ on lattice site $j$ of dimer $(x',y')$. 
Similarly $c_{(x,y), i, \sigma}$ destroys an electron of spin $\sigma$ on lattice site $i$ of dimer $(x,y)$.

The on-site interaction term is given by
\begin{equation}\label{eq:Hubbard_U}
H_U =U\sum_{(x,y)}\sum_{i}n_{(x,y), i, \uparrow} n_{(x,y), i, \downarrow},
\end{equation}
with the number operator on site $i$ of dimer $(x, y)$ for an electron of spin $\sigma$ given by $n_{(x,y), i, \sigma} = c^{\dagger}_{(x,y), i, \sigma}c_{(x,y), i, \sigma} $ and
the nearest neighbour interaction Hamiltonian is
\begin{equation}\label{eq:Hubbard_V}
H_V =\sum_{\alpha}V_\alpha \sum_{(x,y),i}\hspace{2mm}\sum_{(x',y'),j}{\vphantom{\sum}}'\sum_{\sigma, \sigma'}n_{(x',y'), j, \sigma} n_{(x,y), i, \sigma'}.
\end{equation}
Density functional theory calculations \cite{Kandpal2009,Nakamura2009,Nakamura2012} 
have predicted that the number of free electrons in $\kappa$-(BEDT-TTF)$_2$X salts is equal to the number of 
dimers. Measurements of the out-of-plane optical conductivity 
have also been performed for various $\kappa\text{-(BEDT-TTF)}_2$X salts and the the locations 
of vibrational modes were found to be consistent with one electron per dimer \cite{Sedlmeier2012}.

In the limit that $U$ and $V_d$ are much larger than other microscopic energy scales, these 
terms will strongly penalize double occupancy of dimers, and we can expect the low energy physics 
to be dominated by singly occupied dimers for
\begin{equation}\label{Eq:Low-Energy}
\frac{t_{\alpha}}{U}, \frac{t_{\alpha}}{V_\alpha}, \frac{V_{\alpha}}{U}  \ll 1.
\end{equation}
In the low energy limit, with single occupancy of dimers there are four possible states per 
dimer, which we may write as  
$\ket{\uparrow, 0}, \ket{\downarrow, 0},\ket{0,\uparrow}, \ket{0, \downarrow}$ 
where we list the occupation of site 1 before that of site 2.  We can view the dipole moment 
of the dimer as a pseudospin with site 1 corresponding to $P^z = 1/2$ and site 2 corresponding 
to $P^z = -1/2$ and switch to the $\ket{P^z, S^z}$ basis to represent the state of each 
dimer \cite{Hotta2010}.

We use a strong coupling expansion for the extended Hubbard model, Eq.~(\ref{eq:Ham}), to derive a low-energy effective model.
We project to the one electron per dimer limit and make use of the method set out in Ref.~\cite{MacDonald1988}. The effective model we derive follows the approach of Hotta 
\cite{Hotta2010} to write the Hamiltonian in the dipole-spin basis. However, we go beyond 
the expansion considered by Hotta by including all nearest neighbour interactions, leading to modified couplings 
in the low energy effective theory.

\subsection{Low energy theory}
In the standard strong-coupling expansion of the Hubbard model, e.g. \cite{MacDonald1988},
one projects on to the single electron per site limit.  The situation we consider is slightly more 
complicated in that we consider the one-electron per dimer limit.  An additional complication is the
presence of nearest neighbour interactions, since the occupation of sites on adjacent dimers will 
affect the allowed terms in the expansion.  To include these terms, we follow closely the approach used
in Refs.~\cite{Farrell2013,Farrell2014} for the extended Hubbard model on a square lattice, modifying 
their approach for a triangular lattice of dimers.

We write the Hamiltonian in the form
\begin{equation}
H = H_0 + H_T,
\end{equation}
where $H_0 = H_U + H_V$.
$H_0$ does not change the number of electrons per dimer, whereas $H_T$ 
includes hops which may change the dimer occupation (the intra-dimer hopping does not change the dimer occupation, 
but since it can change the nearest neighbour interaction energy, we do not include it in $H_0$).  We introduce
a canonical unitary transformation, $S$, so that $H_0$ remains a constant of motion to a desired order in 
$1/U$.  Let $H^\prime$ be the transformed Hamiltonian, then
\begin{equation}
	H^\prime = H_0 + H_T^\prime,
\end{equation}
where
\begin{equation}
	H_T^\prime = e^{iS}H_Te^{-iS},
\label{eq:htprime}
\end{equation}
and to ensure that $H_0$ remains a constant of motion to order $1/U$, we must have
\begin{equation}
	\left[H_0,H_T^\prime\right] = 0,
\label{eq:EoMconstraint}
\end{equation}
and expanding $S$ in powers of $1/U$, we get
\begin{equation} \label{eq:s-series}
S = -i\sum_{n=1}^{\infty}\frac{S_n}{U^n}.
\end{equation}
Since we consider the expansion to second order in perturbation theory, we must
determine both $S_1$ and $S_2$.  Expanding the hopping part of the Hamiltonian in a power
series in $1/U$ also, 
\begin{equation} \label{eq:t-series}
H_T^\prime = \sum_{n=1}^{\infty}\frac{H_{T,n}^\prime}{U^{n-1}},
\end{equation}
and using Eqs.~(\ref{eq:htprime}), (\ref{eq:s-series}) and (\ref{eq:t-series}), we obtain the following 
expressions for the first and second order corrections:
\begin{equation}\label{eq:1st-order}
H_{T,1}^\prime = H_T + [S_1,\tilde{H}_0],
\end{equation}
and
\begin{equation}\label{eq:2nd-order}
H_{T,2}^\prime = [S_1,H_T]+\frac{1}{2}\big[S_1,[S_1,\tilde{H}_0]\big]+[S_2,\tilde{H}_0],
\end{equation}
where $\tilde{H}_0 = H_0/U$.
These equations do not give a clear path for obtaining $S_1$ or $S_2$, but applying the 
equation of motion requirement Eq.~(\ref{eq:EoMconstraint}), and demanding that it apply 
at each order sequentially gives
\begin{equation} \label{eq:s1}
\big[H_0,[S_1,\tilde{H}_0]+H_T\big] = 0,
\end{equation}
and 
\begin{equation} \label{eq:s2}
\big[H_0,[S_1,[S_1,H_T]+\frac{1}{2}\big[S_1,[S_1,\tilde{H}_0]\big]+[S_2,\tilde{H}_0]\big] = 0.
\end{equation}
To determine the solutions of Eqs.~(\ref{eq:s1}) and (\ref{eq:s2}) it is helpful to decompose
the hopping term in the Hamiltonian into channels that are differentiated by whether they change the occupation of
a dimer, similar to the procedure used at the site level in Ref.~\cite{MacDonald1988}.
Define the hole occupancy $h_{(x,y),i,\sigma} = 1 - n_{(x,y),i,\sigma}$ which is 0 if site $(x,y),i$ is occupied by an electron with spin $\sigma$ and 1 otherwise.  Using the identity $ h_{(x,y),i,\sigma} + n_{(x,y),i,\sigma} = 1$, and acting from both the right and the left, we may write $H_T = \sum_\alpha (T_\alpha^1 + T_\alpha^0 + T_\alpha^{-1})$, where (with $\bar{\sigma}$ indicating the opposite spin to $\sigma$)
\begin{widetext}
\begin{eqnarray}
	T_\alpha^1 & =  & t_\alpha \sum_{(x,y),i}\hspace{2mm}\sum_{(x',y'),j}\hspace*{-2mm}{\vphantom{\sum}}^\prime \hspace{2mm}\sum_{\sigma}n^{\phantom{\dagger}}_{(x',y'),j,\bar{\sigma}}c^{\dagger}_{(x',y'), j, \sigma} c^{\vphantom{\dagger}}_{(x,y), i, \sigma} h^{\vphantom{\dagger}}_{(x,y),i,\bar{\sigma}},
\\
	T_\alpha^0 & = & t_\alpha \sum_{(x,y),i}\hspace{2mm}\sum_{(x',y'),j}\hspace*{-2mm}{\vphantom{\sum}}^\prime \hspace{2mm}\sum_{\sigma}\left\{n^{\phantom{\dagger}}_{(x',y'),j,\bar{\sigma}}c^{\dagger}_{(x',y'), j, \sigma} c^{\vphantom{\dagger}}_{(x,y), i, \sigma} n^{\vphantom{\dagger}}_{(x,y),i,\bar{\sigma}}
	+h^{\phantom{\dagger}}_{(x',y'),j,\bar{\sigma}}c^{\dagger}_{(x',y'), j, \sigma} c^{\vphantom{\dagger}}_{(x,y), i, \sigma} h^{\vphantom{\dagger}}_{(x,y),i,\bar{\sigma}}\right\}, 
\\
	T_\alpha^{-1} & = & t_\alpha \sum_{(x,y),i}\hspace{2mm}\sum_{(x',y'),j}\hspace*{-2mm}{\vphantom{\sum}}^\prime \hspace{2mm}\sum_{\sigma}h^{\phantom{\dagger}}_{(x',y'),j,\bar{\sigma}}c^{\dagger}_{(x',y'), j, \sigma} c^{\phantom{\dagger}}_{(x,y), i, \sigma} n^{\phantom{\dagger}}_{(x,y),i,\bar{\sigma}}.
\end{eqnarray}
	Writing the summand as $(T_\alpha^m)_{(x,y;\, x',y'),i,j,\sigma}$, with $m \in \{1,0,-1\}$, the hopping term may be written as 
\begin{equation} \label{eq:hopping_hub_decomp}
H_T =\sum_{\alpha} \sum_{(x,y),i}\hspace{2mm}\sum_{(x',y'),j}\hspace*{-2mm}{\vphantom{\sum}}^\prime \hspace{2mm}
	\sum_{\sigma,m}(T_\alpha^m)_{(x,y;\, x',y'),i,j,\sigma}.
\end{equation}
Note that each $T_\alpha^m$ channel changes the number of doubly occupied sites by $m$ and so
the interaction energy interaction energy in $H_0$ changes by an amount 
$mU$.  However, nearest neighbour interactions mean that hops that change site occupation also 
change nearest neighbour interaction energies.  In order to deal with this, we follow Ref.~\cite{Farrell2014}
and introduce a nearest neighbour projection operator, which projects out all states except those which have an electronic configuration $\tilde n^\beta$ neighbouring site $(x,y),i$. Formally it is defined as
\begin{eqnarray}
	O^\beta_{(x,y),i}[\tilde n^\beta] & = & \prod_{(\delta_{\beta x}, \delta_{\beta y},\delta_{\beta}), \sigma}
\{\tilde{n}_{(\delta_{\beta x}, \delta_{\beta y}), \delta_{\beta},\sigma}n_{(x+\delta_{\beta x}, y+\delta_{\beta y}), i+\delta_{\beta},\sigma} 
	+ (1-\tilde{n}_{(\delta_{\beta x}, \delta_{\beta y}), \delta_{\beta},\sigma})h_{(x+\delta_{\beta x}, y+\delta_{\beta y}), i+\delta_{\beta},\sigma}\}, 
\end{eqnarray}
where the $(\delta_{\beta x}, \delta_{\beta y},\delta_{\beta})$ connect site $(x,y),i$ to the neighbouring sites via $\beta = d, b, p, q$ and  
	$\tilde{n}_{(\delta_{\beta x}, \delta_{\beta y}), \delta_{\beta},\sigma}$ is 1 when the site at
	$(x+\delta_{\beta x}, y+\delta_{\beta y}), i+\delta_{\beta}$ is occupied and zero otherwise.
 Inserting forms of the identity $$1 = \prod_{(\delta_{\beta x}, \delta_{\beta y},\delta_{\beta}), \sigma}\left(n_{(x+\delta_{\beta x}, y+\delta_{\beta y}), i+\delta_{\beta},\sigma}+h_{(x+\delta_{\beta x}, y+\delta_{\beta y}), i+\delta_{\beta},\sigma}\right),$$ on either side of Eq.~(\ref{eq:hopping_hub_decomp}) one can find 
(similarly to Ref.~\cite{Farrell2014})
\begin{equation}
H_T = \sum_{\alpha, m,  \{M_1\},\{M_2\}} T_{\alpha}^{m, \{M_2\}, \{M_1\}},
	\label{eq:hop_chan}
\end{equation}
	where $\{M_1\} = \{M_1^d, M_1^b, M_1^p, M_1^q\}$ indicates the number of occupied neighbouring $d$, $b$, $p$, and $q$ sites before the hop
	and $\{M_2\}$ indicates the neighbours after the hop.  Hence
\begin{eqnarray}
	T_{\alpha}^{m, \{M_2\}, \{M_1\}} & = & t_\alpha  \sum_{(x,y),i}\hspace{2mm}\sum_{(x',y'),j}\hspace*{-2mm}{\vphantom{\sum}}^\prime \hspace{2mm}\sum_{\sigma}\left\{\prod_\beta \sum_{S[n_2^\beta] = M_2^\beta}\sum_{S[n_1^\beta]  = M_1^\beta}\right\} \nonumber \\
	& &\times \left\{\prod_{\gamma} O_{(x',y'),j}^\gamma[n_2^\gamma]\right\}(T_\alpha^m)_{(x,y;\, x',y'),i,j,\sigma}\left\{\prod_{\eta} O_{(x,y),i}^\eta[n_1^\eta]\right\},
\end{eqnarray}
and $S[n^\beta] = \sum_{(\delta_{\beta x}, \delta_{\beta y},\delta_{\beta}),\sigma} {n}_{(\delta_{\beta x}, \delta_{\beta y}), \delta_{\beta},\sigma}$ is the total number of occupied 
neighbouring sites of type $\beta$.
	The different hopping channel operators have the commutators (see Appendix C for a derivation)
\begin{equation}\label{eq:decomp_com}
\left[H_U + H_V, T_{\alpha}^{m, \{M_2\}, \{M_1\}}\right] = \left[mU + \sum_{\beta}V_\beta\left(M_2^\beta - M_1^\beta\right)\right] T_{\alpha}^{m, \{M_2\}, \{M_1\}},
\end{equation}
	and using the decomposition in Eq.~(\ref{eq:hop_chan}) with the commutator Eq.~(\ref{eq:decomp_com}) allows one to obtain $S_1$ and $S_2$ as outlined
	in Appendix \ref{app:strong_coup}.
The first order correction is 
\begin{equation}
H_{T,1}^\prime = \sum_{\alpha}\sum_{\{M\}}T_{\alpha}^{0, \{M\}, \{M\}},
\end{equation}
and the second order correction is
\begin{eqnarray}\label{eq.prime.example}
	H_{T,2}^\prime & = &  U \sum_{\alpha,\nu}\hspace{2mm}\tilde{\sum_{\{M_{2}\}, \{M_{1}\}, \{N\},m}}
      \frac{   T_{\alpha}^{m, \{M_{2}\}, \{M_1\}}  T_{\nu}^{-m, \{N\}, \{M_{2}\} - \{M_{1}\}
	+ \{N\}}}{mU+\sum_{\gamma}V_\gamma(M_2^\gamma - M_1^\gamma)} ,
\end{eqnarray}
	where the tilde on the sum over occupation numbers indicates that the sum excludes values of $m$, $\{M_1\}$ and $\{M_2\}$ for which the denominator vanishes.

The first order correction corresponds to a process with a single hop, which can only maintain the one
electron per dimer limit if it is an intra-dimer hop, i.e. $\alpha = d$, no other hops are allowed.  It
also requires that the number of nearest neighbour electrons is the same before and after the hop.
In summary, the full second order effective Hamiltonian is 
\begin{eqnarray}\label{eq:exp-ham-1}
	H^\prime & = & H_U + H_V + \sum_{\{M\}}T_{d}^{0, \{M\}, \{M\}} 
	+ \sum_{\alpha,\nu}\hspace{2mm}\tilde{\sum_{\{M_{2}\}, \{M_{1}\}, \{N\},m}} \frac{   T_{\alpha}^{m, \{M_{2}\}, \{M_1\}}  T_{\nu}^{-m, \{N\}, \{M_{2}\} - \{M_{1}\} + \{N\}}}{mU+\sum_{\gamma}V_\gamma(M_2^\gamma - M_1^\gamma)} .  
\end{eqnarray}
We now proceed to write the low energy model in terms of dipole and spin operators, following Hotta \cite{Hotta2010}.
The full form of the effective low energy model in the dipole-spin basis is
\begin{eqnarray}\label{eq:Ham_C_ij}
	H^\prime & =& V_q\sum_{i}\sum_{j}{\vphantom{\sum}}' P_i^z P_j^z - V_p\sum_{i}\sum_{j}{\vphantom{\sum}}' P_i^z P_j^z - V_b \sum_{i}\sum_{j}{\vphantom{\sum}}' P_i^z P_j^z  + t_d \sum_{i}(P_i^+ + P_i^-) \nonumber \\
	& & +\sum_{i}\sum_{j}{\vphantom{\sum}}'\bigg\{ C_{i,j}^0 P_i^z + C_{i,j}^1 P_j^z + C_{i,j}^2 P_i^zP_j^z + C_{i,j}^3 \vec{S_i}\cdot\vec{S_j} 
 + C_{i,j}^4 P_i^z\vec{S_i}\cdot\vec{S_j} +  C_{i,j}^5 P_j^z\vec{S_i}\cdot\vec{S_j} \nonumber \\
	& & \hspace*{1cm} + C_{i,j}^6 P_i^zP_j^z\vec{S_i}\cdot\vec{S_j} + C_{i,j}^7 \left( P_i^+ +  P_i^-\right) + C_{i,j}^8 \left( P_i^+P_j^z + P_i^-P_j^z\right) + C_{i,j}^{9} \left(P_i^+\vec{S_i}\cdot\vec{S_j} + P_i^-\vec{S_i}\cdot\vec{S_j}\right) \nonumber \\
	& & \hspace*{1cm} + C_{i,j}^{10} \left( P_i^+P_j^z\vec{S_i}\cdot\vec{S_j}  + P_i^-P_j^z\vec{S_i}\cdot\vec{S_j} \right) + C_{i,j}^{11} \left( P_j^+ + 
	P_j^- \right) + C_{i,j}^{12} \left( P_j^+P_i^z +  P_j^-P_i^z \right) \nonumber \\
	& & \hspace*{1cm} + C_{i,j}^{13} \left( P_j^+\vec{S_i}\cdot\vec{S_j} + P_j^-\vec{S_i}\cdot\vec{S_j}\right) + C_{i,j}^{14} \left( P_j^+P_i^z\vec{S_i}\cdot\vec{S_j} 
	+  P_j^-P_i^z\vec{S_i}\cdot\vec{S_j}\right)\bigg\},
\end{eqnarray}
where the various $C_{i,j}^n$ are listed in Appendix \ref{App.CoefTable}. 
When viewed as a spin model for the dipole pseudospins, the model has ferromagnetic interactions ($V_p$ and $V_b$), antiferromagnetic 
interactions ($V_q$) and interactions between dipoles that depend on the states of the physical spins of the dimers (the $C^n_{ij}$). There are also 
random field-like terms for the dipoles that depend on the occupations of nearest neighbours. Given that the model
sits on a triangular lattice, this suggests that there may be frustrating interactions between the dipoles that could possibly lead to 
slow dynamics and/or glassy behaviour.  Performing a dynamical simulation of this model for more than a small number of spins and dipoles
is not practical due to the quantum terms involving dipole raising and lowering operators $P^+$ and $P^-$.  Hence, in an effort to learn 
whether this low energy model contains glassy physics, we drop the ``quantum'' terms, and focus on a simplified classical spin model 
of dipoles coupled to spins on a triangular lattice  
\begin{eqnarray}
	H_{\rm classical} & =& V_q\sum_{i}\sum_{j}{\vphantom{\sum}}' P_i^z P_j^z - V_p\sum_{i}\sum_{j}{\vphantom{\sum}}' P_i^z P_j^z - V_b \sum_{i}\sum_{j}{\vphantom{\sum}}' P_i^z P_j^z  
	   \label{eq:eff_model}
	\\
 & & +\sum_{i}\sum_{j}{\vphantom{\sum}}'\bigg( C_{i,j}^0 P_i^z + C_{i,j}^1 P_j^z + C_{i,j}^2 P_i^zP_j^z + C_{i,j}^3 \vec{S_i}\cdot\vec{S_j} 
	+ C_{i,j}^4 P_i^z\vec{S_i}\cdot\vec{S_j} +  C_{i,j}^5 P_j^z\vec{S_i}\cdot\vec{S_j} + C_{i,j}^6 P_i^zP_j^z\vec{S_i}\cdot\vec{S_j}\bigg) . \nonumber 
\end{eqnarray}
We treat the dipoles $P_i^z$ as Ising variables and the spins $\bvec{S}_i$ as classical vectors.  This allows the model to be 
simulated using classical Monte Carlo techniques.

\end{widetext}

\section{Monte Carlo simulations}
\label{sec:MC}

In this section we present and discuss results from equilibrium Monte Carlo simulations of the model defined in Eq.~(\ref{eq:eff_model}). 
We calculate the polarization, magnetization, electric and magnetic susceptibility and Edwards-Anderson order parameters 
for both spin and charge degrees of freedom.  The couplings in $H_{\rm classical}$ depend not only on the parameters of the extended Hubbard 
model but also on the occupation of the neighbouring sites in the lattice.  This means that the couplings in a Monte Carlo simulation depend 
on the state of the system and will evolve as dipoles are flipped.  In order to obtain a more computationally tractable problem, we instead 
use the following procedure to specify the model: i) calculate the distribution of all possible values of couplings for each term in the 
Hamiltonian, ii) fit the distribution to a simplified form, iii) draw couplings between dipoles and spins randomly from the distributions 
calculated in ii), iv) average over multiple sets of couplings.

\begin{figure}[h]
\includegraphics[width=8cm]{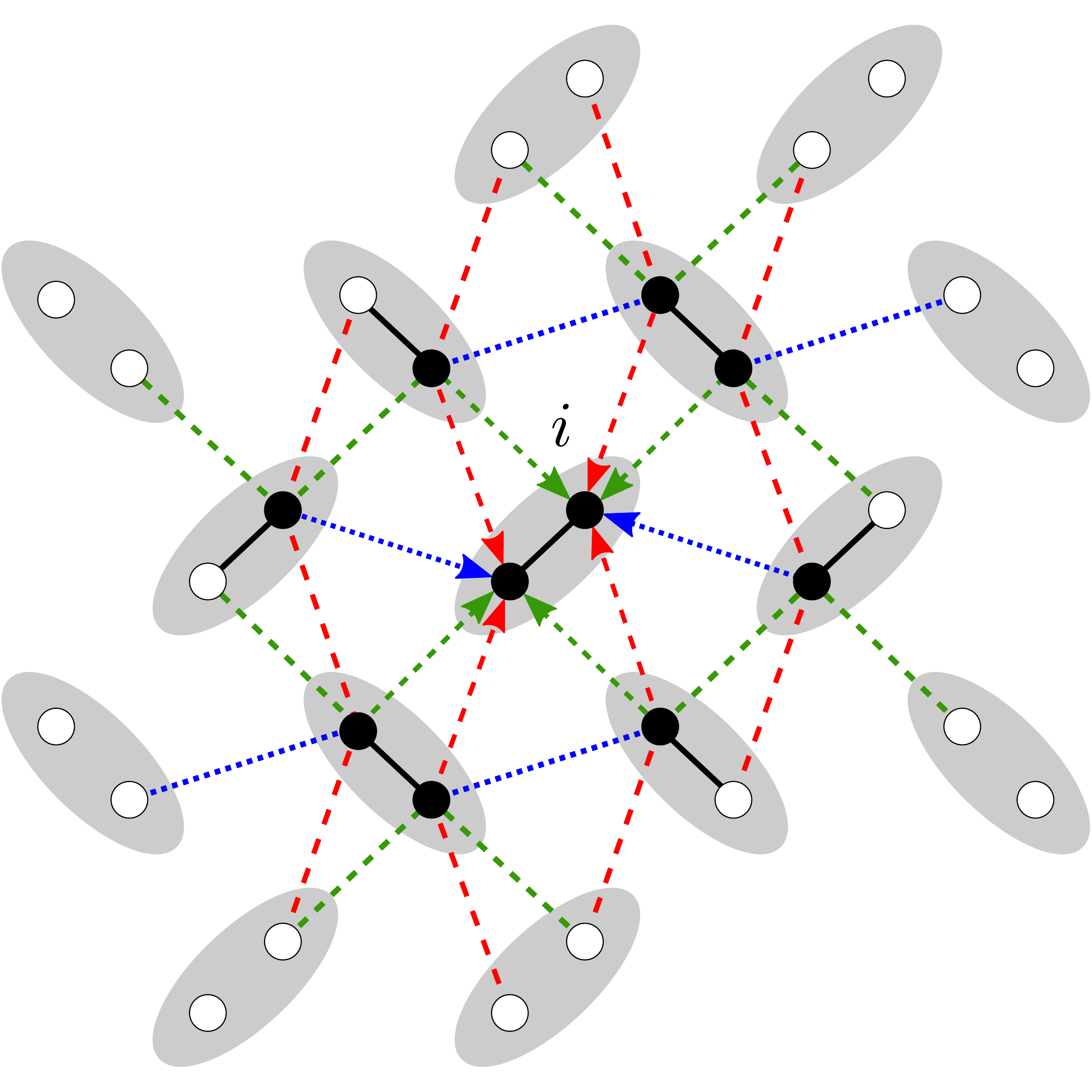}
\caption{Dimers that must be included to calculate couplings for the dimer $i$. Each black dot represents a lattice site for which the nearest neighbour occupancy
         needs to be known to properly calculate each $C_n$, meaning that a minimum of 15 dimers are needed to calculate the full set of couplings.}
\label{fig:15_con}
\end{figure}

Each of the coefficients, $C^n_{ij}$, in Eq.~(\ref{eq:eff_model}) is determined by the electronic 
configurations of the neighbouring dimers (see Tables I-\ref{tab:four} in Appendix \ref{app:couplings}). Therefore for each distinct electronic 
configuration of nearest neighbours about some central dimer $i$ there will be a different set of $C_{i,j}^n$ for that dimer. 
For large lattices, 
calculating the couplings for each dimer is a very computationally expensive task. To run Monte Carlo simulations one must first calculate 
the set of couplings $C_{i,j}^n$ from the set of $C_n$ as set out in Appendix \ref{app:couplings}. In order to simplify the calculations we find the distributions of the various 
$C_{i,j}^n$ and then sample from these distributions to set the couplings for each simulation.
In this way we aim to capture the effect of the distribution of couplings that arise from the different occupations without recalculating all of 
the couplings at each step of the calculation as dipoles flip and the charge distribution evolves with time.

We now discuss the calculation of the couplings. The denominator of each $C_n$ relating to the hopping between dimers $i$ and $j$ depends on the occupancy 
of the nearest neighbours of a lattice site on dimer $j$ as well as the occupancy of the nearest neighbours of a lattice site on the dimer $i$. 
As can be seen in Fig.~\ref{fig:15_con} this implies that to calculate a set of $C_n$ the electronic configuration of a dimer $i$ as well as its 
14 nearest and next-nearest neighbors must be known.

We calculated the distributions of the $C_{i,j}^n$ as follows: first, for a single 15 dimer electronic configuration we calculated the set of $C_n$. 
Second, we used the set of $C_n$ to calculate the corresponding set of $C_{i,j}^n$ (Appendix \ref{App.CoefTable}). We repeated this process until the electronic 
configuration space was properly sampled and binned the various $C_{i,j}^n$ to create histograms. 
There are $2^{15} = 32768$ possible electronic configurations. 

\begin{widetext}

        \begin{figure}[ht]
\centering
        \includegraphics[scale=0.31]{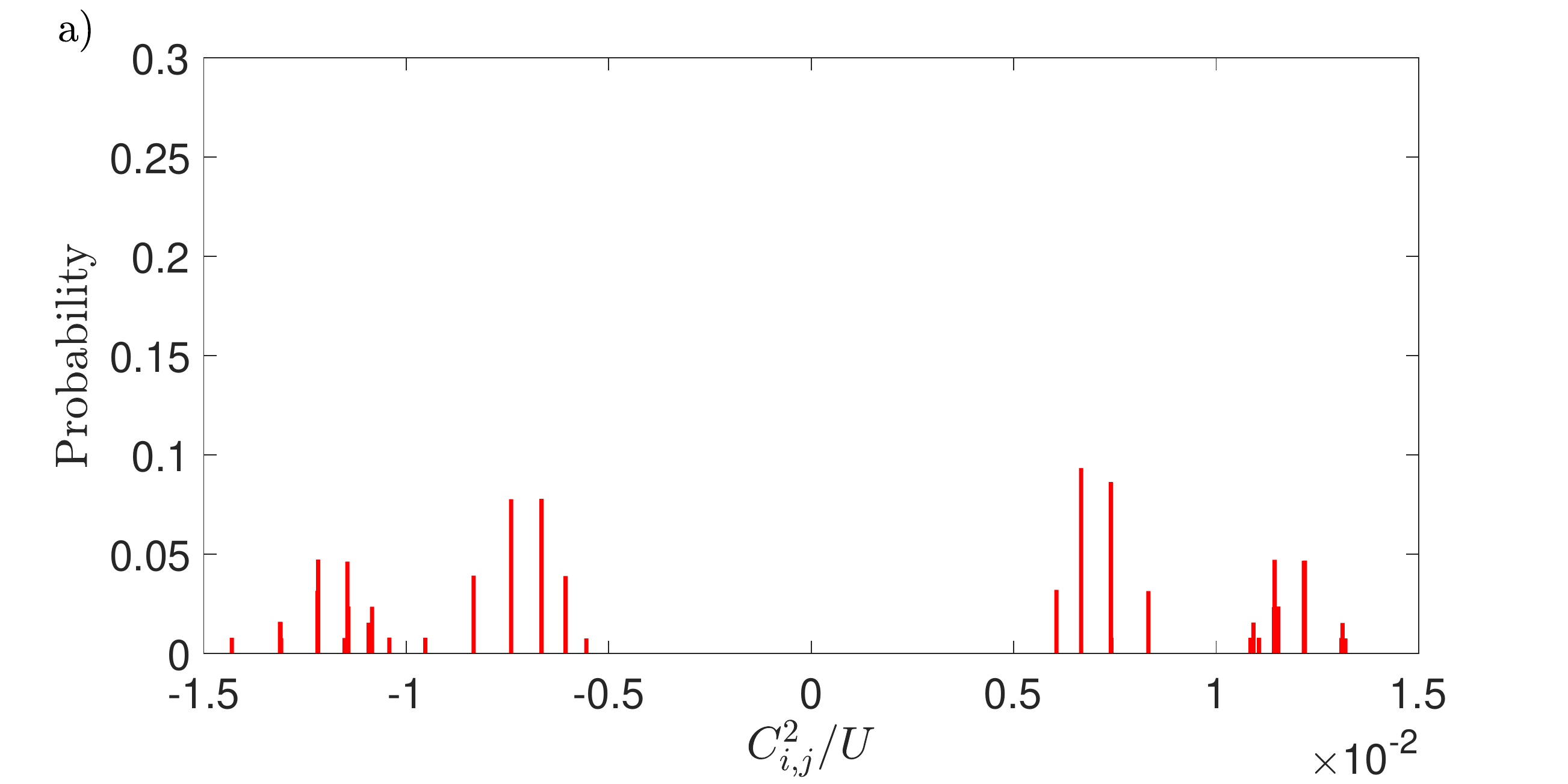}
        \includegraphics[scale=0.31]{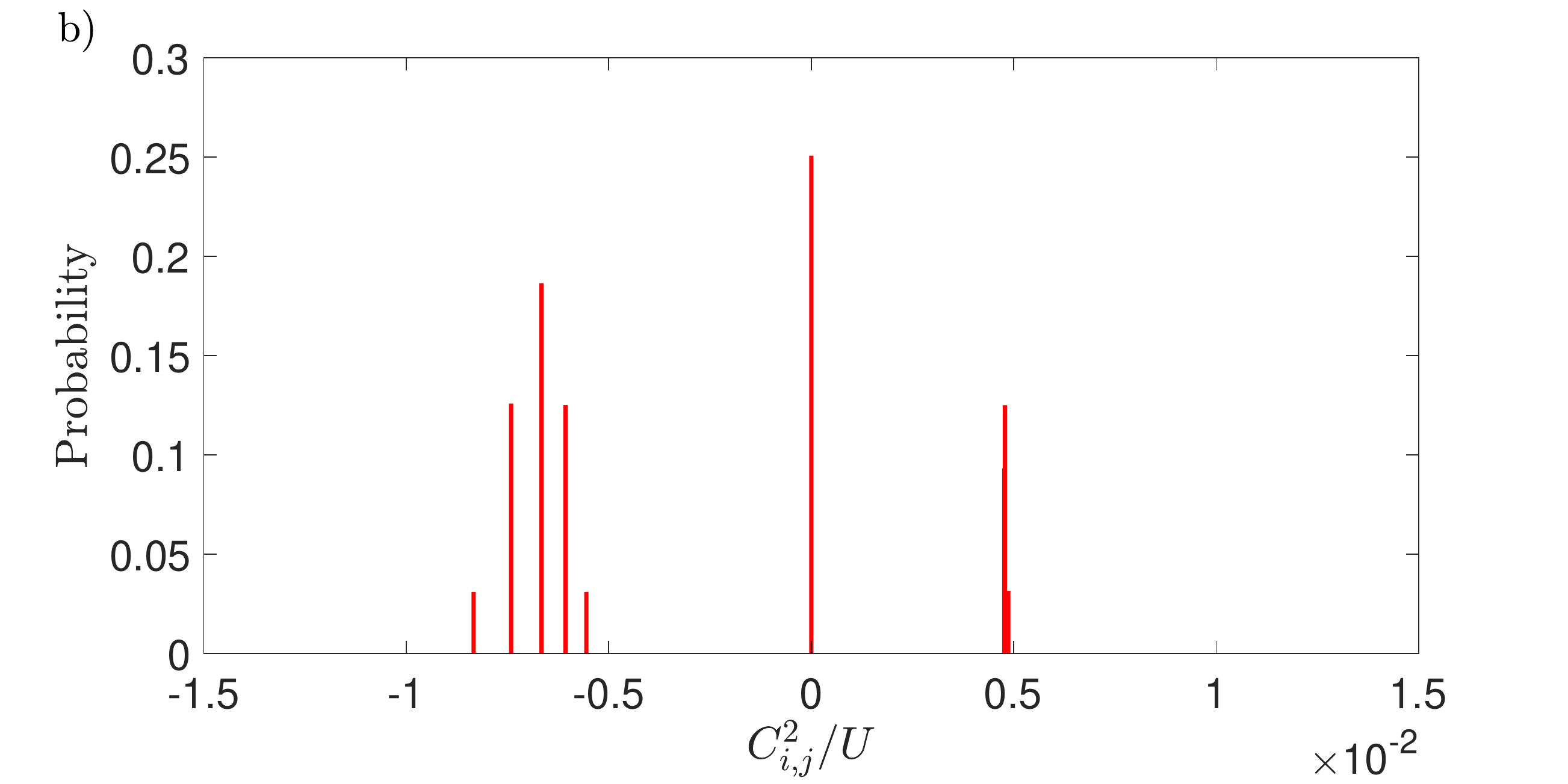}
		\caption{Normalized probability distributions for $C_{i,j}^2$ couplings to the two types of dimers neighbouring a central dimer located at $i=(x,y)$. a)  
		($x \pm \frac{1}{2}, y \pm \frac{1}{2}$) neighbours; b) ($x\pm 1$, $y$) neighbours. Each bar has a width of $1.4 \times 10^{-5}$. Parameters used are $U/t = 15.000113$, 
		$V_d/t = 10.000219$, $V_b/t = 1.000547$, $V_p/t = 1.000623$, and $V_q/t = 1.000412$.}
\label{fig:non_disorder_individual}
\end{figure}

	\begin{figure}[ht]
        \includegraphics[width=8cm]{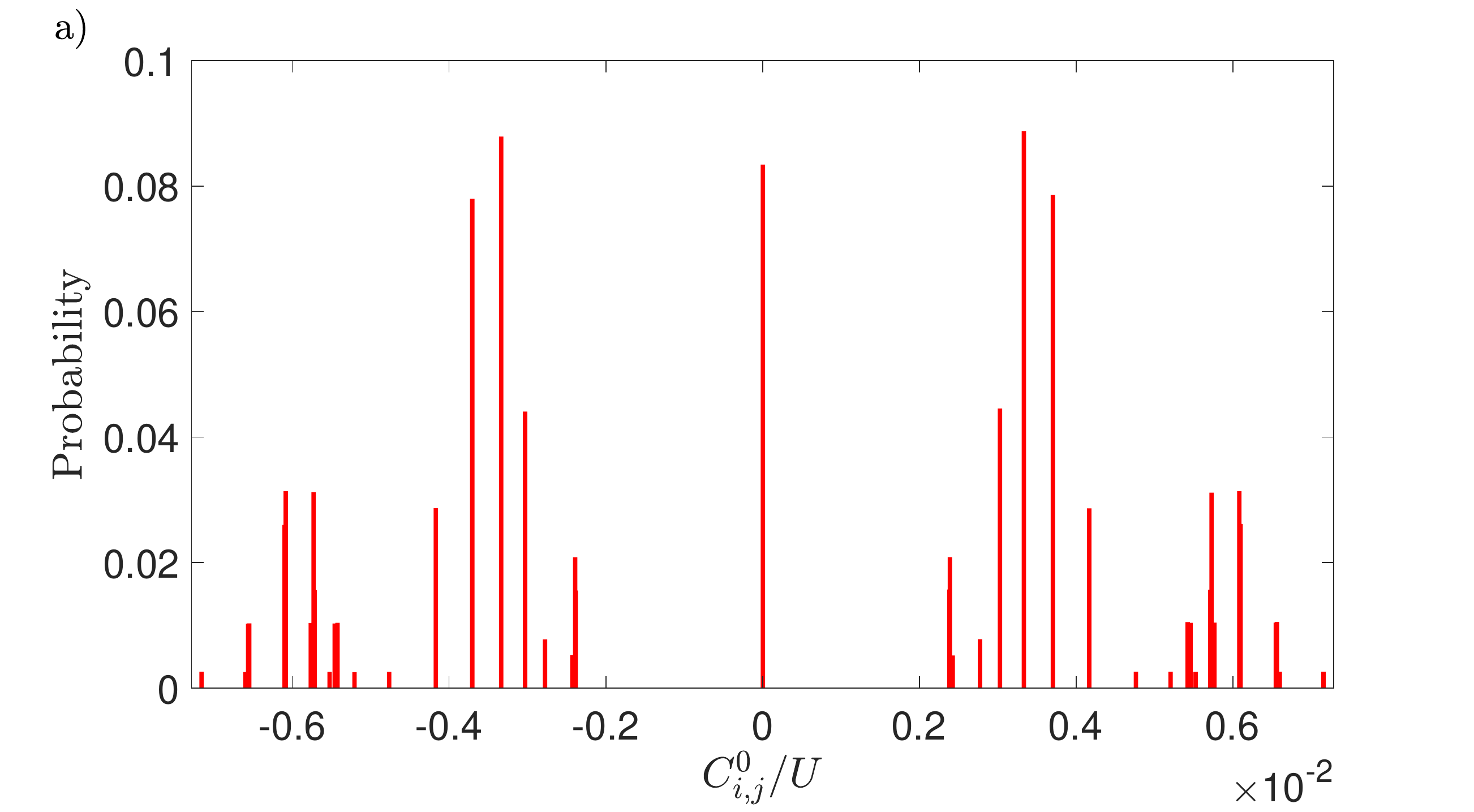}
        \includegraphics[width=8cm]{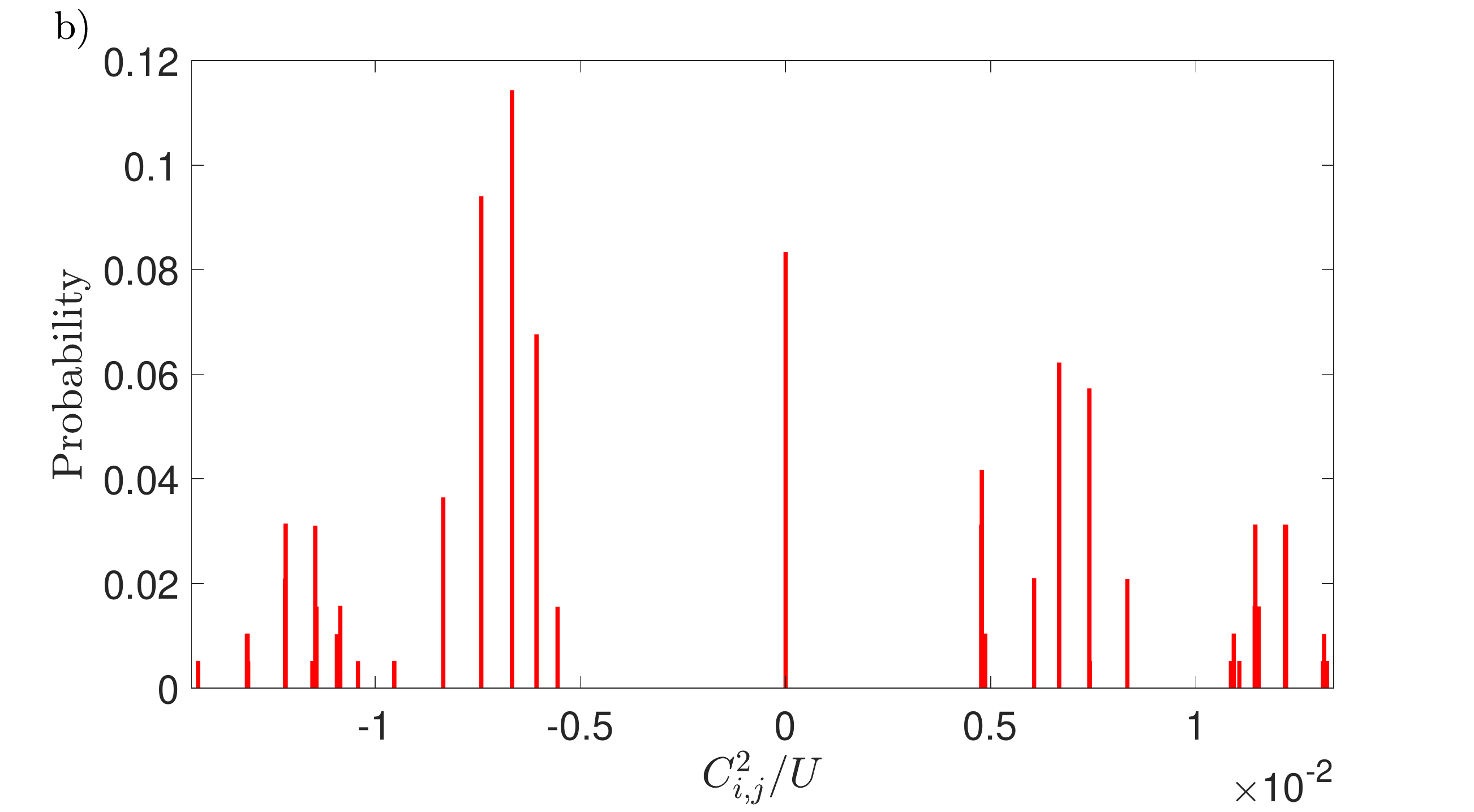}
        
        \includegraphics[width=8cm]{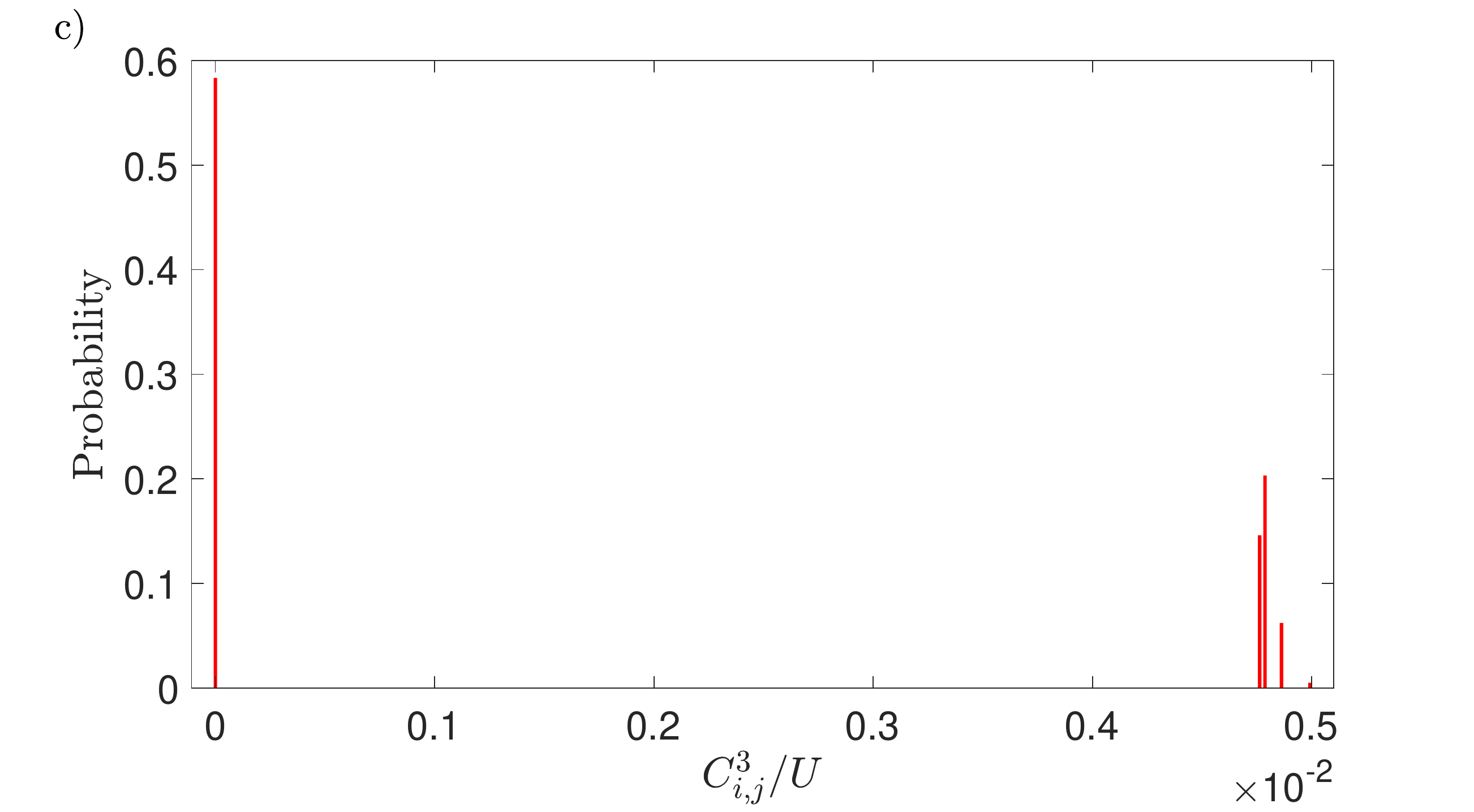}
        \includegraphics[width=8cm]{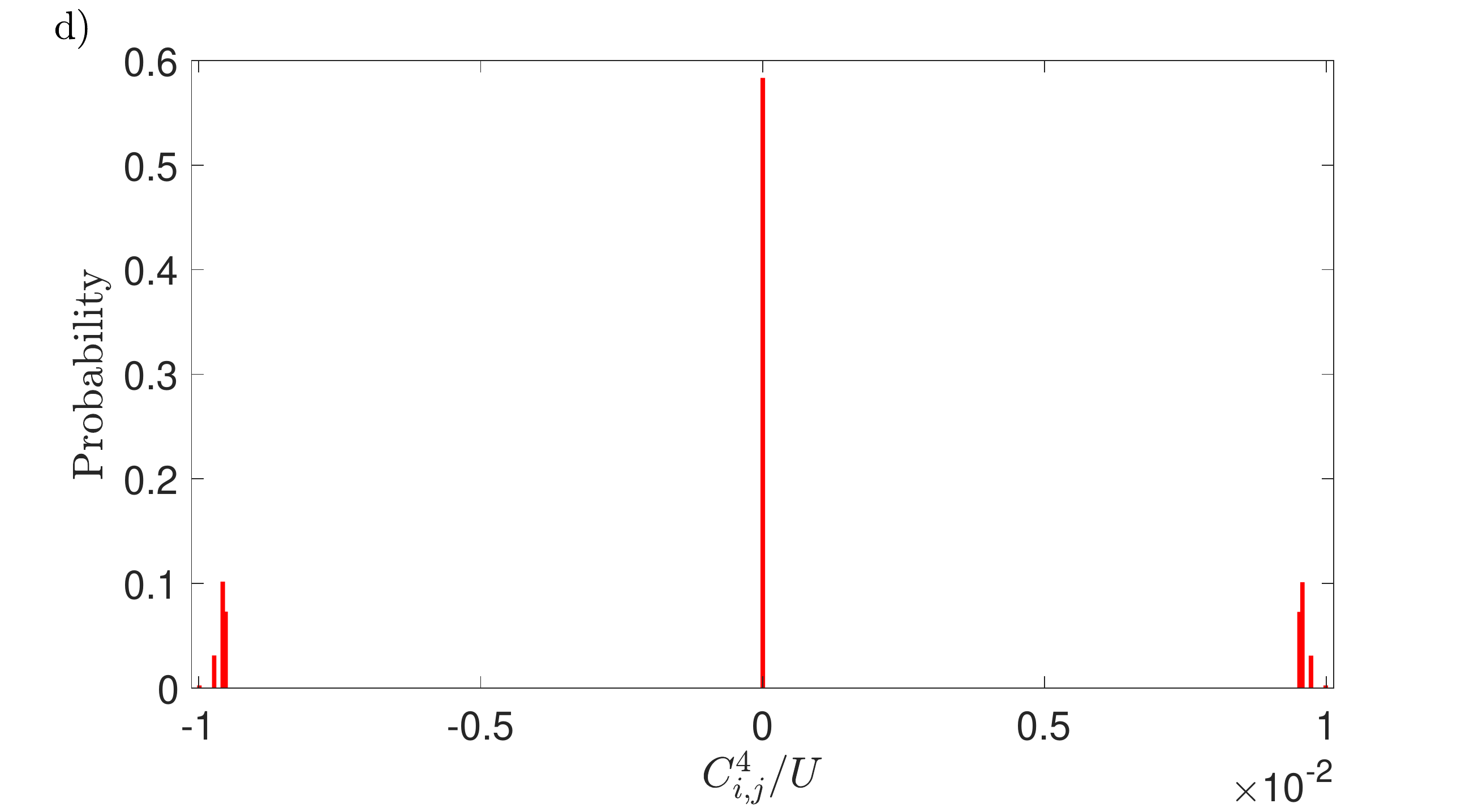}
        
\caption{Histograms combining couplings for all neighbouring dimers for various $C_{i,j}^n$ couplings: a) $C_{i,j}^0$, b) $C_{i,j}^2$, c) $C_{i,j}^3$, d) $C_{i,j}^4$.
		Parameters used are the same as Fig.~\ref{fig:non_disorder_individual}.}
\label{fig:non_disorder_mult_com}
\end{figure}

\end{widetext}

While motivated by the $\kappa$-(BEDT-TTF)$_2$X family of salts, we do not choose parameter values for any specific material, but are somewhat 
guided by the parameter choices in Ref.~\cite{Hotta2010}.  In all calculations, we take all of the hopping parameters, $t_d = t_b = t_p = t_q = t$, 
and measuring other parameters in terms of $t$, we take $U = 15 t$, with $V_d$ taking values between $4t$ and $12 t$ and $V_b, V_p$ and $V_q$ 
taking values between $t$ and $3t$.  For reference, Pinteri\'{c} {\it et al.} estimate $U/t = 7.3$ in $\kappa$-(BEDT-TTF)$_2$Cu$_2$(CN)$_3$ \cite{Pinteric2014}.
The choice of all hopping parameters being equal leads to some coupling coefficients being equal that would otherwise take on different values.

        There is a subtlety involved when calculating the denominators of the set of $C_n$. As discussed in Appendix \ref{app:couplings}, the strong coupling
expansion excludes all terms which have a zero denominator. The denominator takes the form  $mU + \sum_\gamma
V_\gamma (M_2^\gamma - M_1^\gamma)$ where $m$ and $(M_2^\gamma - M_1^\gamma)$ are integers, implying that if the $U$ and
the $V_\gamma$ are factors of each other it is possible to have a vanishing denominator.
According to the strong coupling expansion if a process leads to a vanishing denominator it does not contribute to the second order Hamiltonian.
To avoid such terms we selected parameters $U$ and $V_\gamma$ such that the denominator does not vanish.

\begin{figure}[h]
\centering
\includegraphics[width=8cm]{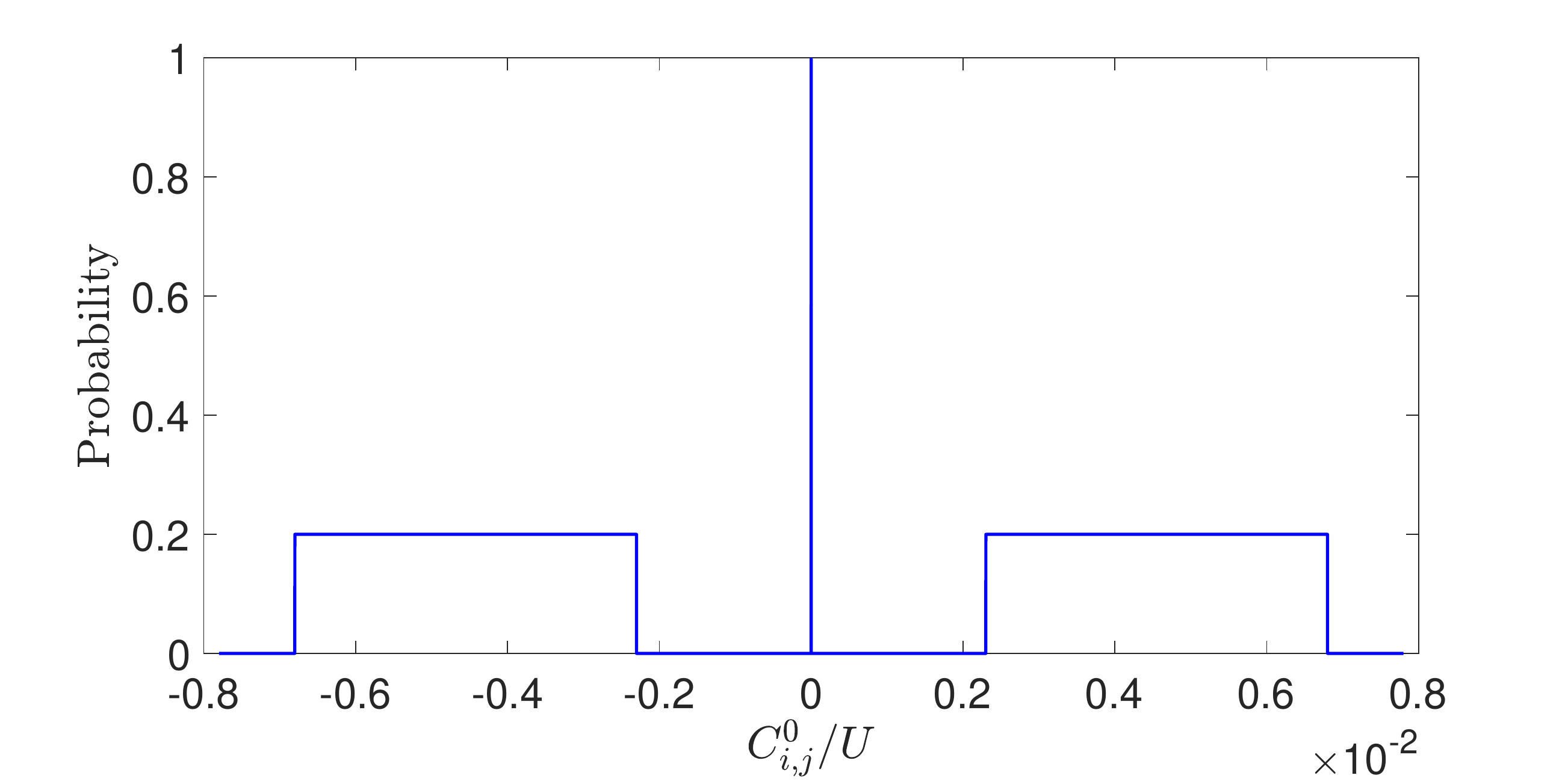}
        \caption{An example of the approximate probability density function shown in Eq. (\ref{eq:prob}) for $C^0_{ij}$:
        here $w_1 = -0.0068$, $w_2 =  -0.0023$, $w_3 = 0.0023$, $w_4 = 0.0068$, and $b_1 = b_2 = 0.2$.}
\label{fig:hist_fit}
\end{figure}

We determined histograms for all of the couplings $C^n_{ij}$ that enter into the effective model.  Examples of these histograms for 
$C_{i,j}^1$ (which give the strengths of the $P_i^z P_j^z$ interaction) are shown in Fig.~\ref{fig:non_disorder_individual}. 
The couplings to dimers $\left(x \pm \frac{1}{2}, y \pm \frac{1}{2}\right)$ are identical due to symmetry for the chosen hopping parameters as are those
to dimers located at $\left(x \pm 1, y\right)$, so there are only two types of $C^1_{ij}$ couplings for our choice of parameters. The combined
distributions of all $C^0_{ij}$, $C^2_{ij}$, $C^3_{ij}$, $C^4_{ij}$ couplings are shown in Fig.~\ref{fig:non_disorder_mult_com}.

As a further simplification we find approximate analytical forms for the distributions and draw the couplings from these approximate distributions. 
Many parameter choices lead to histograms which can be modeled adequately by a probability density function of the form

\begin{widetext}
\begin{eqnarray} \label{eq:prob}
P(x) = 
\begin{cases}
0,	& x<w_1, \\
b_1, & w_1 \leq x < w_2, \\
0, & w_2 \leq x < 0, \\
\big(1-b_2(w_4 - w_3) - b_1(w_2-w_1)\big)\delta(x), & x=0, \\
0, & 0 < x \leq w_3, \\
b_2, & w_3 < x \leq w_4, \\
0, &w_4 < x, 
\end{cases}
\end{eqnarray}
with  $w_4 > w_3 > w_2 > w_1$ and $b_1 \geq 0$, $b_2 \geq 0$. A general graphical representation of this distribution is shown in Fig. \ref{fig:hist_fit} using $C^0_{ij}$ as an example.
\end{widetext}

\begin{figure}[hbt]
\includegraphics[width=8cm]{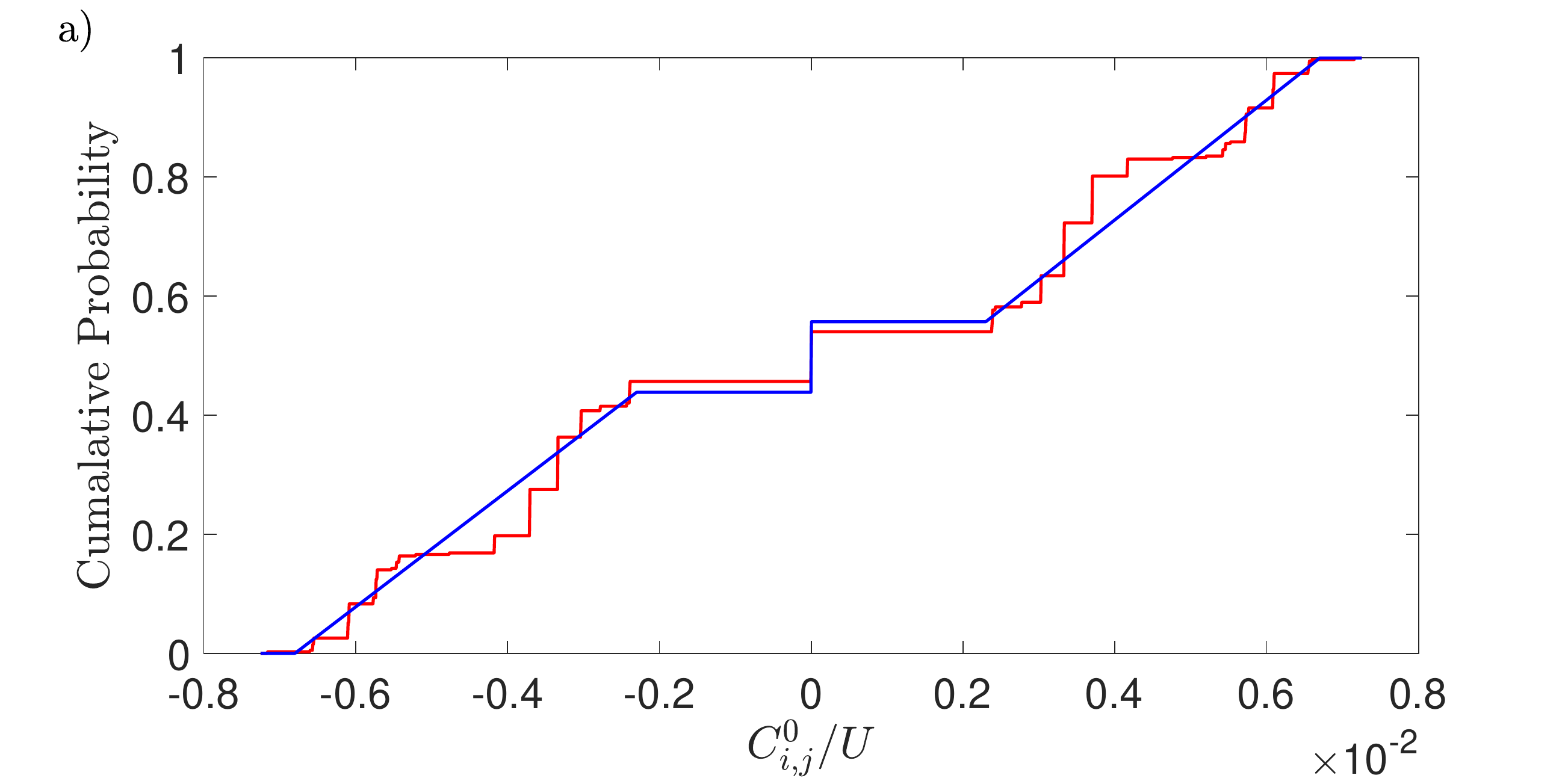}
\includegraphics[width=8cm]{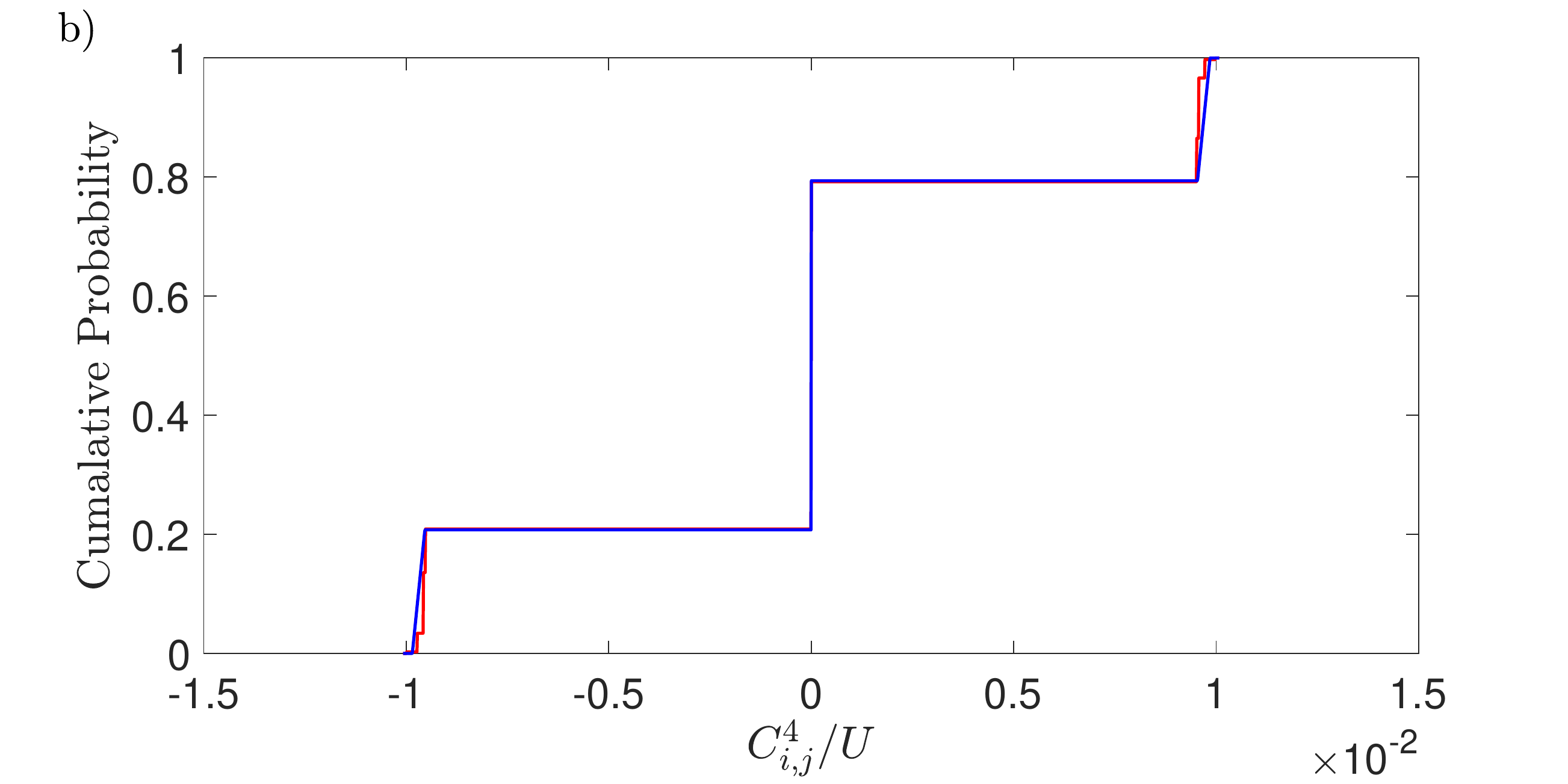}
\caption{Cumulative probability functions for the combined histograms of a) $C^{0}_{i,j}$ and b) $C^{4}_{i,j}$. 
        The parameters used are $U/t=15.0113$, $V_{d/t}=8.0219$, $V_{b}/t=1.0547$, $V_{p}/t=1.0623$, and $V_{q}/t=1.0412$.}
\label{fig:coef_cum}
\end{figure}

We select the  $w_n$ and $b_n$ parameters so that the 
corresponding cumulative probability function obtained from Eq. (\ref{eq:prob}) matches the cumulative probability function of the various histograms. 
This ensures that the behaviour of the distribution is properly captured. 
We first choose the parameters $w_n$ based on the cumulative probability function of the various histograms and then perform a least squares analysis to calculate the $b_n$ fitting parameters. This process ensures that all the desired peaks are captured and is performed for every fit. Some results of this fitting procedure are shown in Fig.~\ref{fig:coef_cum} for $C^0_{ij}$ and $C^4_{ij}$.

In the Monte Carlo simulations we create a triangular lattice of $N$ dimers and then assign couplings randomly from the calculated distributions. 
As there are many more coupling configurations than $N$, we average over couplings to obtain a representative sample of the configuration space. 
In one Monte Carlo step, we attempt to flip $N$ spins and $N$ dipoles, choosing spins and dipoles at random. A spin flip corresponds to assigning
a new direction in space for the vector spin and flipping a dipole corresponds to changing the sign of the Ising degree of freedom.
We use the Metropolis algorithm to determine the probability of whether a move that leads to an energy change $\Delta E$ is accepted:
\begin{equation}
P(\Delta E) = 
\begin{cases}
1, & \Delta E \leq 0,\\
e^{-\beta \Delta E}, & \Delta E > 0,\\
\end{cases}
\end{equation}
\noindent where $\beta = 1/(k_{B}T)$ with 
$k_{B}=1.38 \times 10^{-23}$m$^{2}$ JK$^{-1}$.  

We consider lattices of up to $N = 14^2 = 196$ dimers and establish equilibration by considering two replicas of the system, prepared with the 
same set of couplings but different thermal histories (i.e. different random numbers in the Metropolis algorithm).  The system is taken to be 
equilibrated when chosen thermodynamic variables (the electric and magnetic susceptibilities) calculated in the two replicas agree to within a 
specific tolerance.  After the system has equilibrated we sample to obtain averages of the quantities that we discuss below.

First, we calculate the average polarization per dipole at time step $t_{j}$ for a lattice size of $N$ dimers, given by 
\begin{equation}
P(t_{j}) = \dfrac{1}{N}\left| \sum_{i=1}^{N} P_{i}(t_{j}) \right| ,
\end{equation}
with $P_{i}(t_{j})$  the polarization of dimer $i$ at time step $t_{j}$. Second, we calculate the average magnetization per spin at time step $t_{j}$
\begin{equation}
M(t_{j}) = \dfrac{1}{N}\left| \sum_{i=1}^{N} \vec{S}_{i}(t_{j}) \right|, 
\end{equation}
where $\vec{S}_{i}(t_{j})$ is the spin of dimer $i$ at time step $t_{j}$.
We then calculate the average polarization $P = \left<P(t_j)\right>$ and average magnetization $M = \left<M(t_j)\right>$ where
the notation $<\ldots>$  indicates a time average over $N_{t}$ Monte Carlo time steps. We
also calculate the electric susceptibility and magnetic susceptibility, defined respectively as
\begin{equation}
	\chi _{P} = \beta \lim_{t_j \to \infty} \left( \big\langle P(t_{j})^{2} \big\rangle - \big\langle P(t_{j}) \big\rangle ^{2} \right),
\end{equation}
and
\begin{equation}
\chi _{M} = \beta  \lim_{t_j \to \infty} \left( \big\langle M(t_{j})^{2} \big\rangle - \big\langle M(t_{j}) \big\rangle ^{2} \right) .
\end{equation}
 Our equilibration criterion is that 
the electric and magnetic susceptibilities of the two replicas agree to $2\%$ tolerance or completion of 
$2 \times 10^{7}$ Monte Carlo time steps, whichever comes first.  We were unable to reach equilibrium at temperatures lower than $0.0065\, U/$k$_{B}$.

\begin{widetext}

%%%%%%%%%%%%POLARIZATION%%%%%%%%%%%%%%%%%%%
	\begin{figure}[hbt]
   \includegraphics[width=8cm]{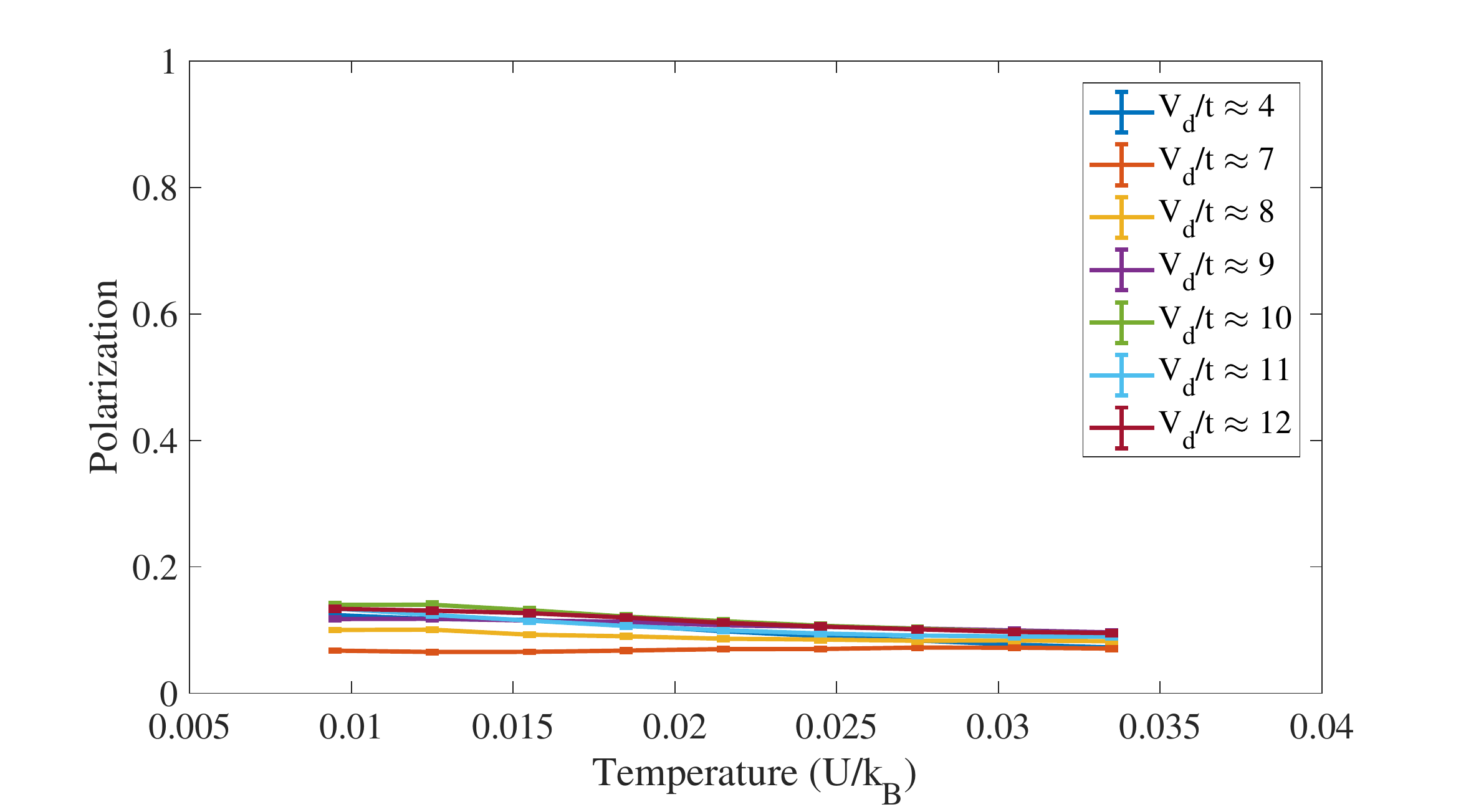}
   \includegraphics[width=8cm]{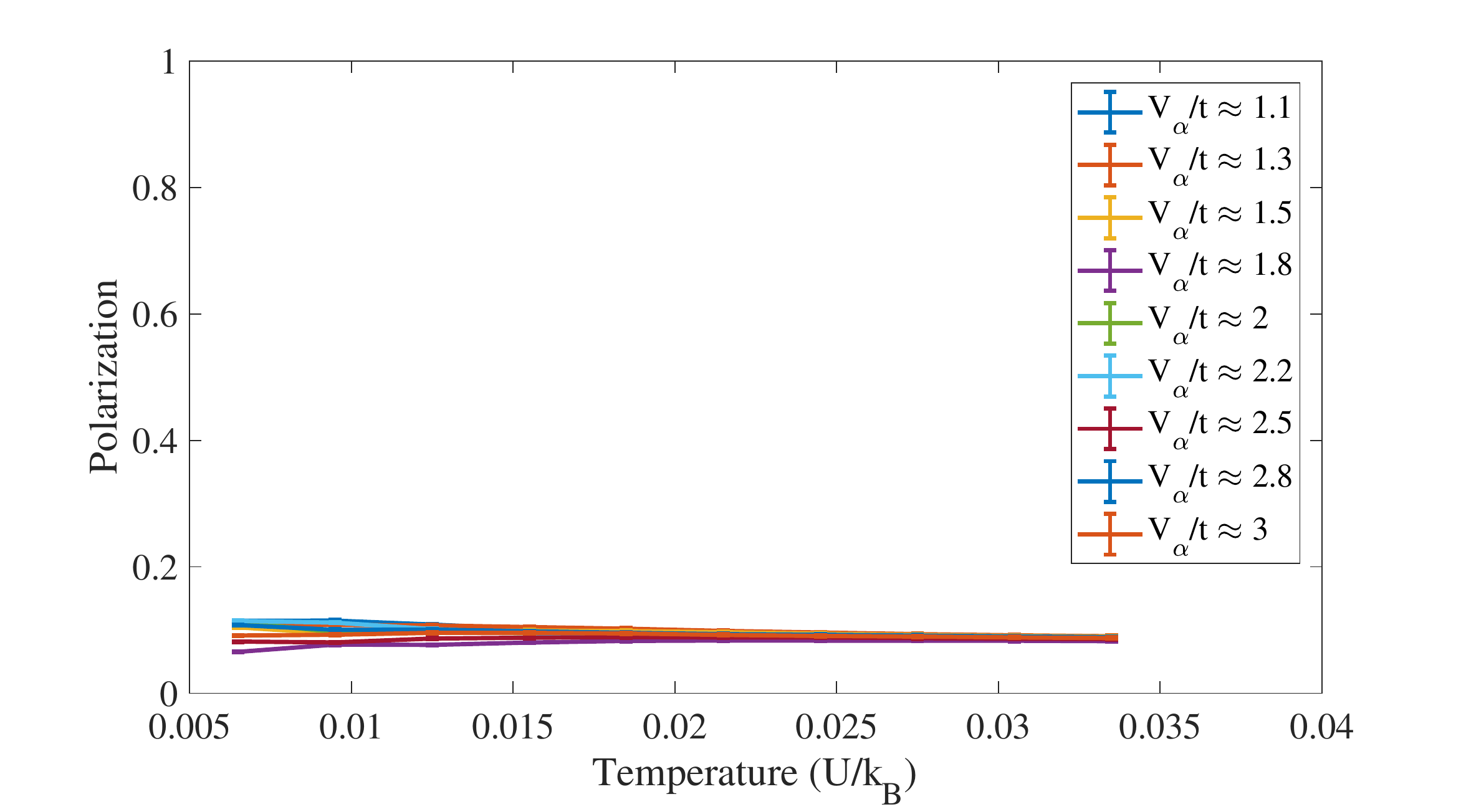}
\caption{Polarization averaged over 18 bond configurations as a function of temperature for a lattice size of $10^2 = 100$. 
		a) Fixed inter-dimer interactions:  $U/t=15.0113$, $V_{b}/t=1.0547$, $V_{p}/t=1.0623$, and $V_{q}/t=1.0412$ with varying $V_d/t$; b)
		Fixed intra-dimer coupling:  $U/t=15.0113$ and $V_{d}/t=10.0219$ with varying $V_\alpha/t$ with  $\alpha = b$, $p$ and $q$.}
\label{fig:Polarization}
\end{figure}

%%%%%%%%%%%%%MAGNETIZATION%%%%%%%%%%%%%%
\begin{figure}
   \includegraphics[width=8cm]{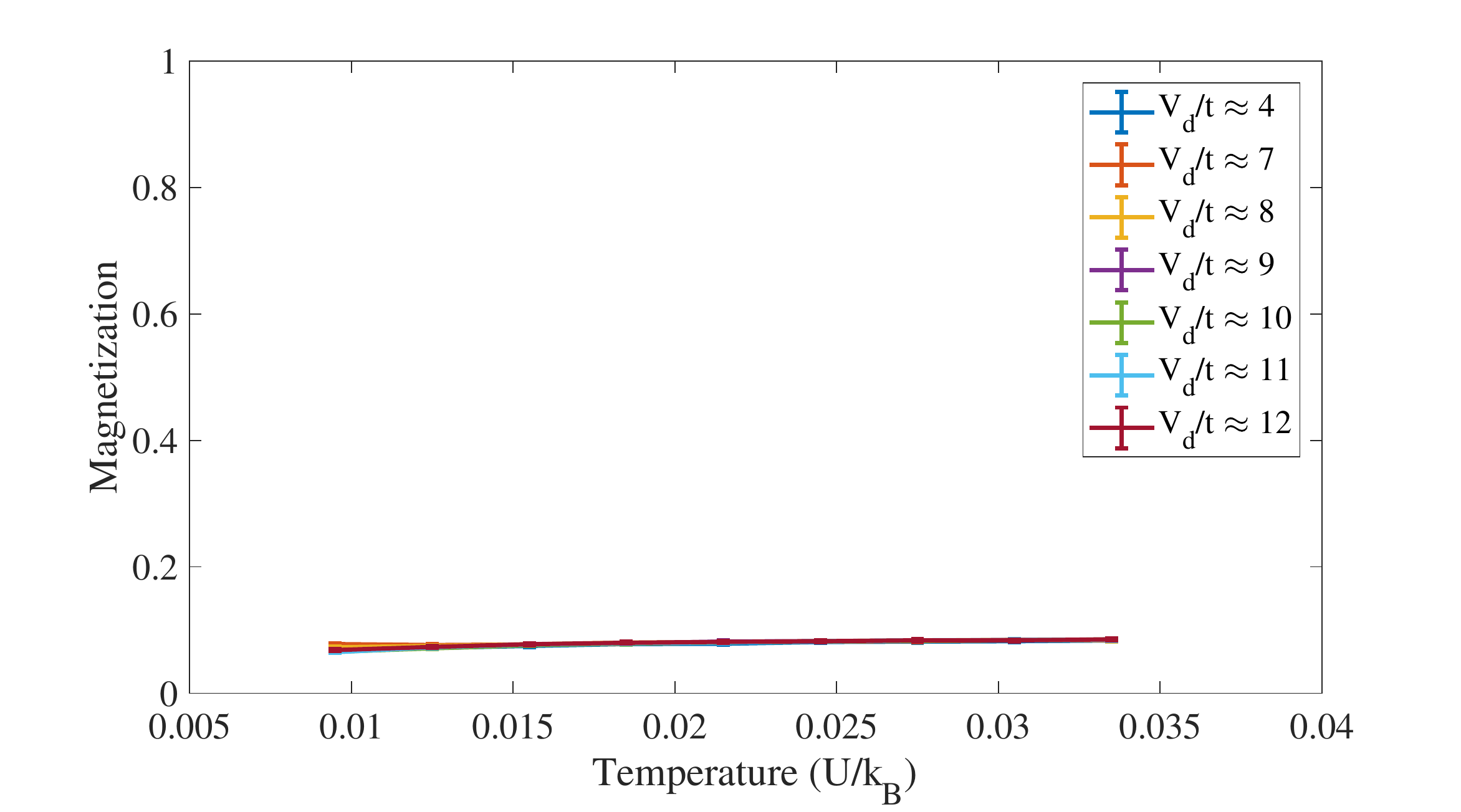}
   \includegraphics[width=8cm]{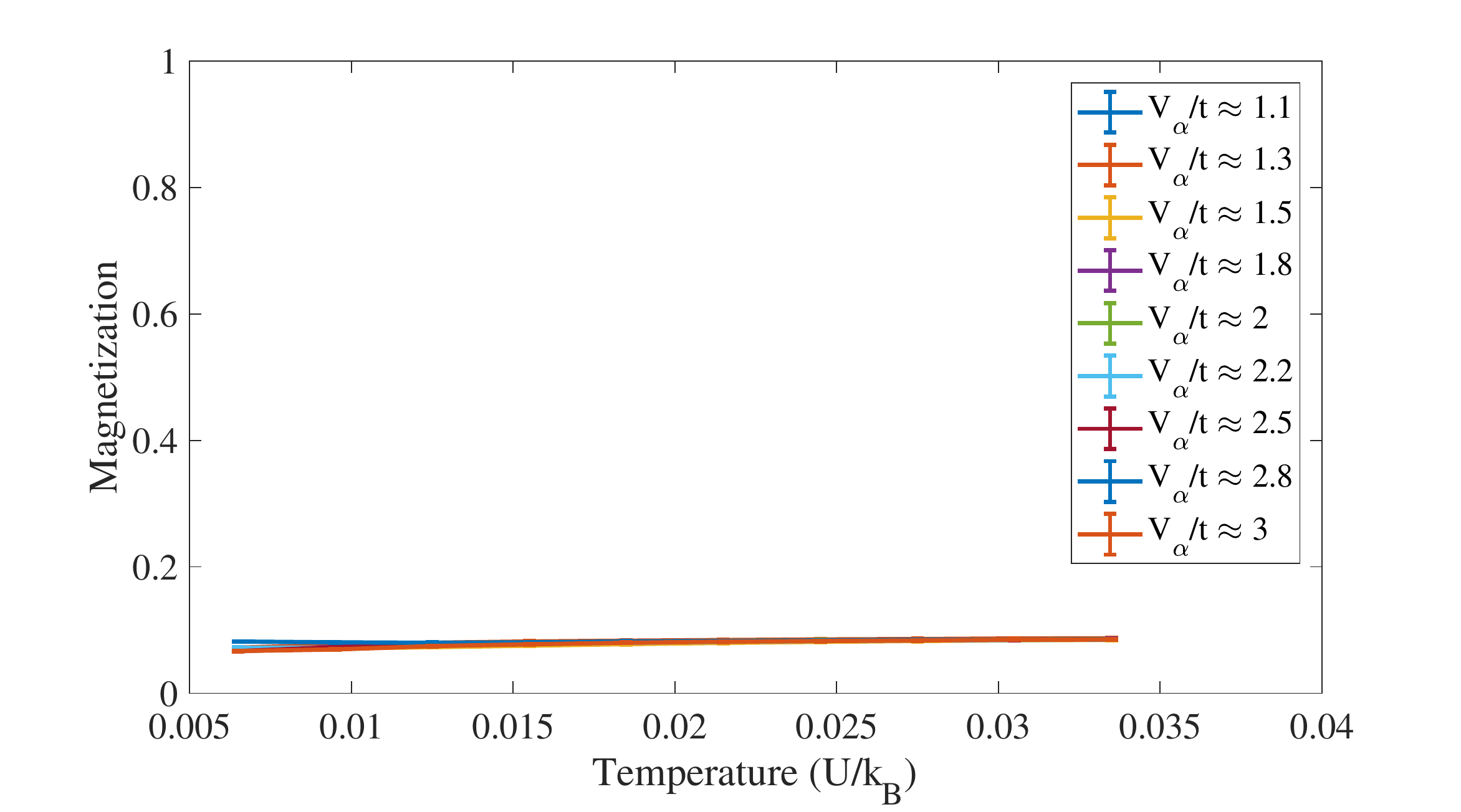}
\caption{Magnetization averaged over 18 bond configurations as a function of temperature for a lattice size of $10^2 = 100$. 
 a) Fixed inter-dimer interactions:  $U/t=15.0113$, $V_{b}/t=1.0547$, $V_{p}/t=1.0623$, and $V_{q}/t=1.0412$ with varying $V_d/t$; b)
                Fixed intra-dimer coupling:  $U/t=15.0113$ and $V_{d}/t=10.0219$ with varying $V_\alpha/t$ with  $\alpha = b$, $p$ and $q$.}
\label{fig:Magnetization}
\end{figure}

%%%%%%%%%%%%El_Sus%%%%%%%%%%%%%%%%%

\begin{figure}
   \includegraphics[width=8cm]{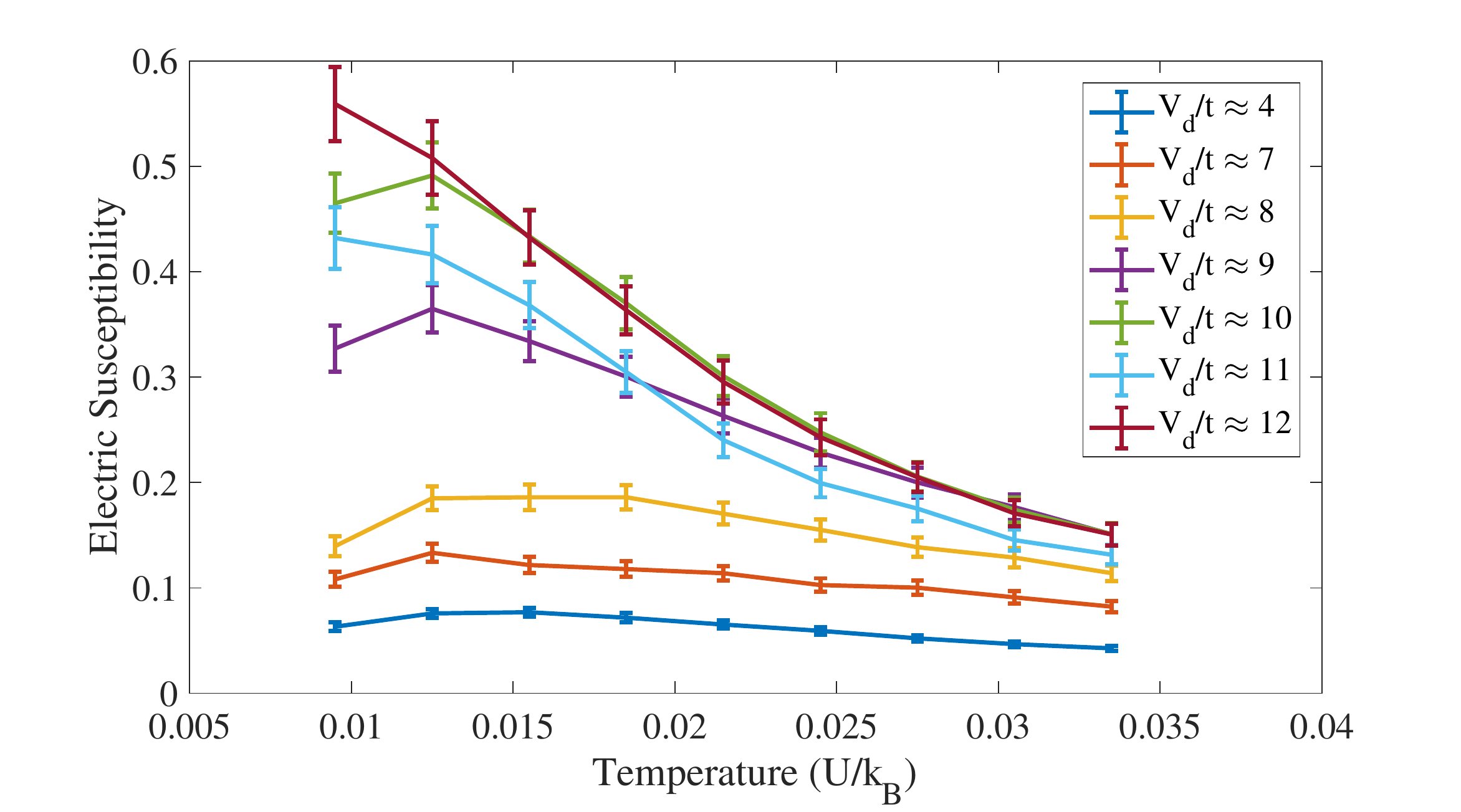}
   \includegraphics[width=8cm]{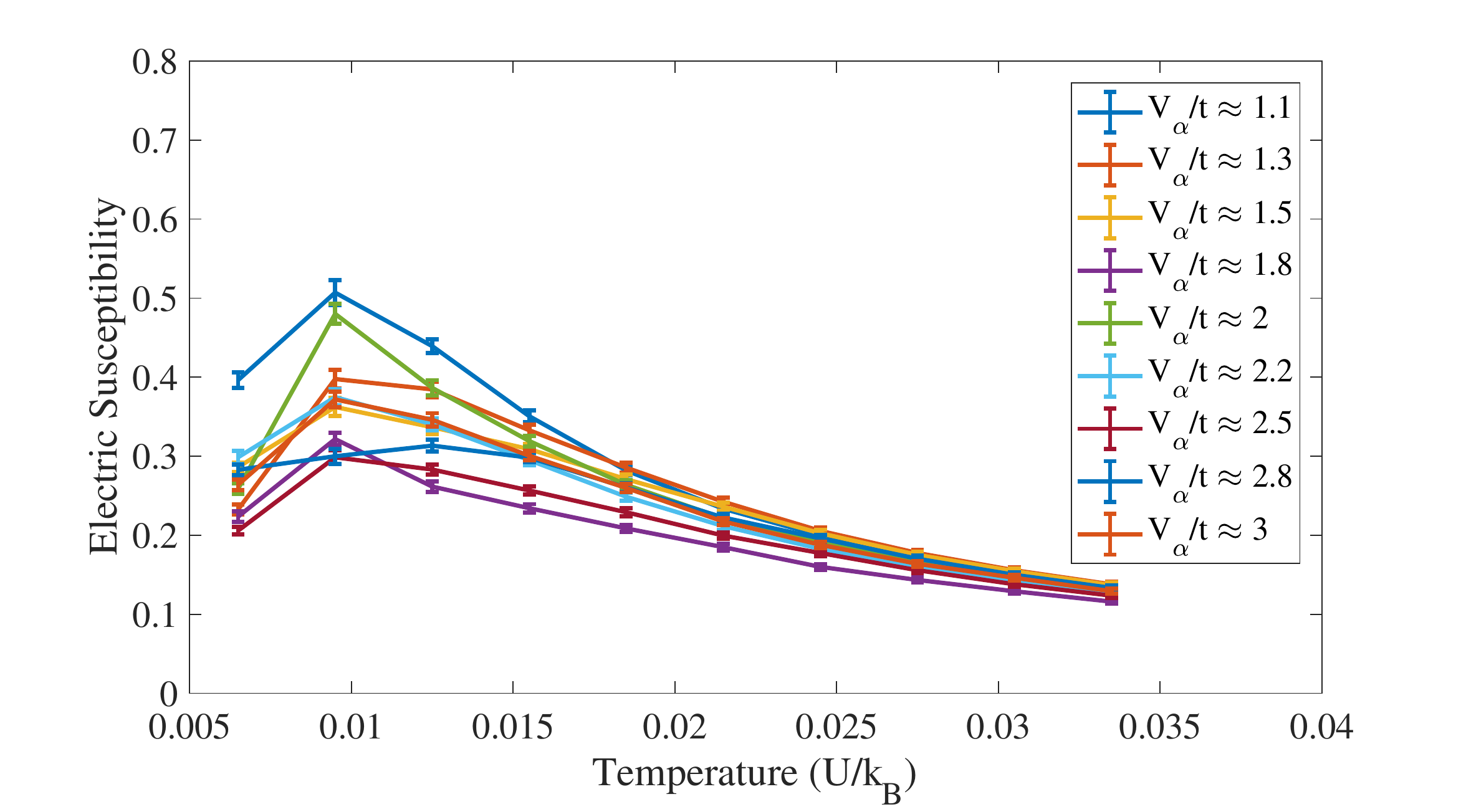}
\caption{Electric susceptibility averaged over 18 bond configurations as a function of temperature for a lattice size of $10^2 = 100$. 
	 a) Fixed inter-dimer interactions:  $U/t=15.0113$, $V_{b}/t=1.0547$, $V_{p}/t=1.0623$, and $V_{q}/t=1.0412$ with varying $V_d/t$; b)
                Fixed intra-dimer coupling:  $U/t=15.0113$ and $V_{d}/t=10.0219$ with varying $V_\alpha/t$ with  $\alpha = b$, $p$ and $q$.}
\label{fig:El_Sus}
\end{figure}
%%%%%%%%%%%%%%%Mag_Sus%%%%%%%%%%%%%%%%%%
\begin{figure}
   \includegraphics[width=8cm]{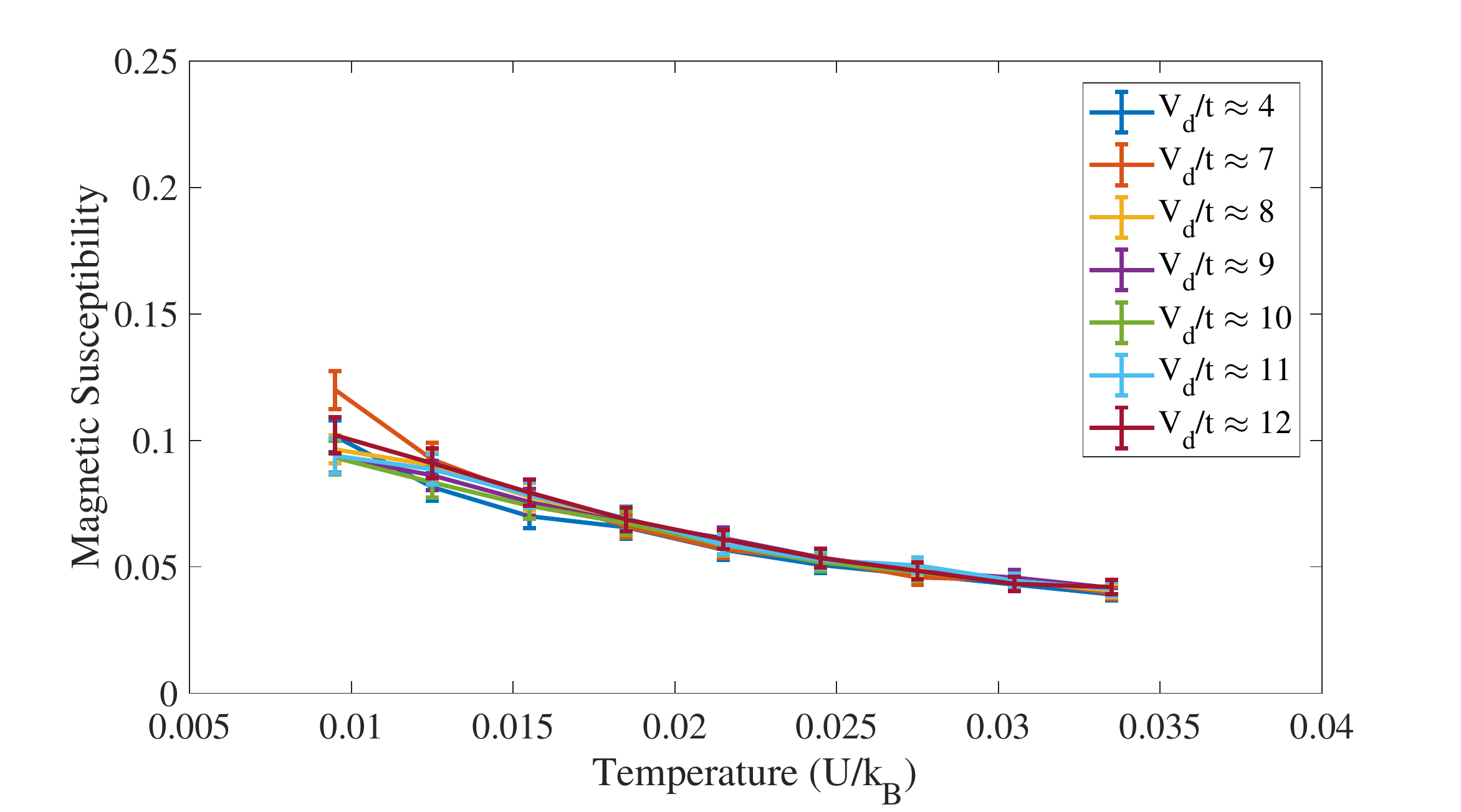}
   \includegraphics[width=8cm]{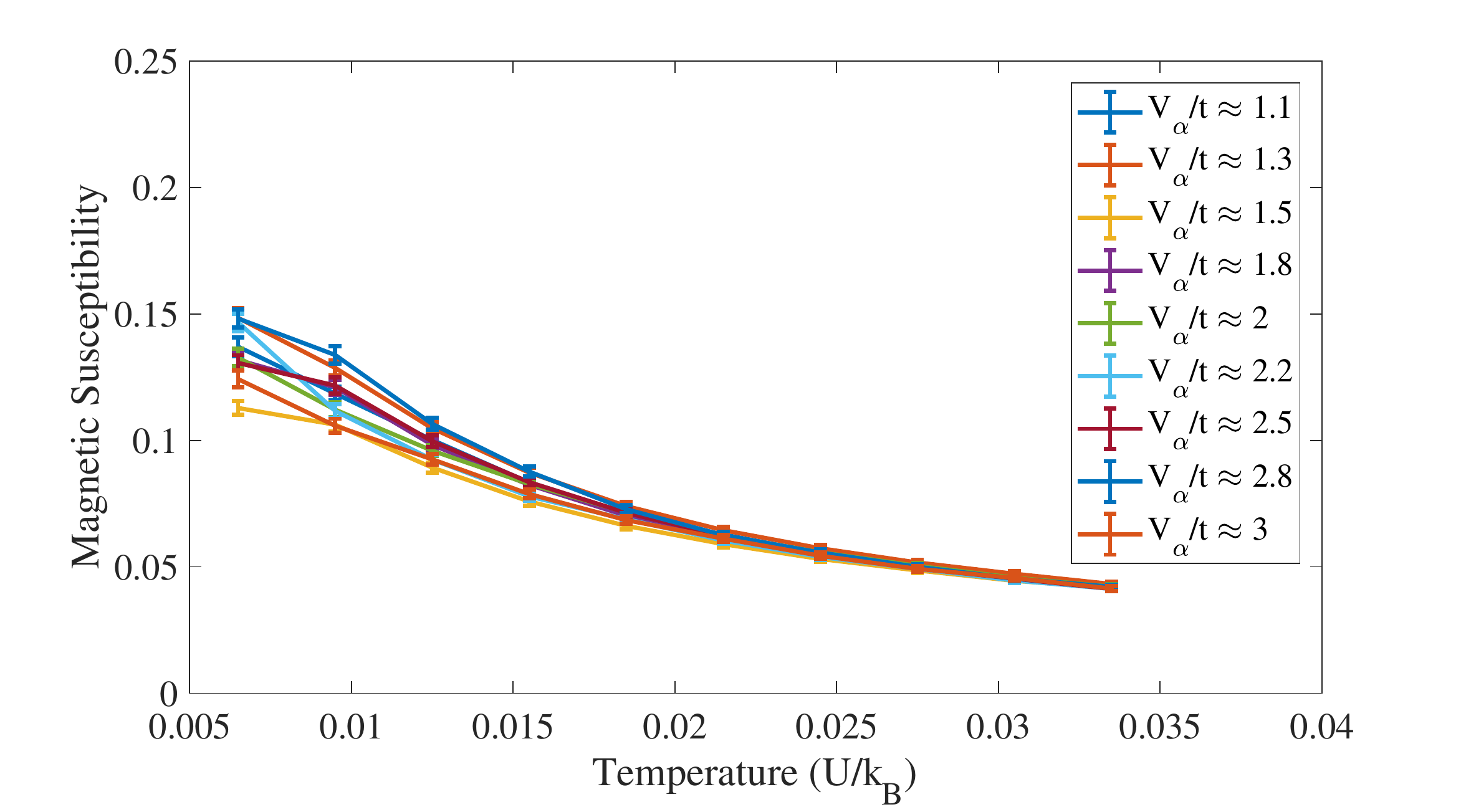}
\caption{Magnetic susceptibility averaged over 18 bond configurations as a function of temperature for a lattice size of $10^2 = 100$. 
	 a) Fixed inter-dimer interactions:  $U/t=15.0113$, $V_{b}/t=1.0547$, $V_{p}/t=1.0623$, and $V_{q}/t=1.0412$ with varying $V_d/t$; b)
                Fixed intra-dimer coupling:  $U/t=15.0113$ and $V_{d}/t=10.0219$ with varying $V_\alpha/t$ with  $\alpha = b$, $p$ and $q$.}
\label{fig:Mag_Sus}
\end{figure}

To check for glassy behaviour we calculate the Edwards-Anderson order parameters for both charge and spin degrees of freedom. 
For the Ising like dipole degrees of freedom, the Edwards-Anderson order parameter is
\begin{equation}
Q_{\rm EA}^{\rm Pol} = \lim_{t_{j}\to\infty} \left| \overline{\left(\left\langle\dfrac{1}{N}\sum_{i=1}^{N}P^{z,A}_{i}(t_{j})P^{z,B}_{i}(t_{j})\right\rangle - \left\langle \sum_{i=1}^{N}P^{z,A}_{i}(t_{j})\right\rangle \left\langle \sum_{i=1}^{N}P^{z,B}_{i}(t_{j})\right\rangle \right)}\right|.
\label{Pol_EA}
\end{equation}
Here $P^{z,A}_{i}(t_{j})$ and $P^{z,B}_{i}(t_{j})$ are the polarization on dimer $i$ at time step $t_{j}$ for replicas $A$ and $B$ respectively. The over-bar indicates an average over the bond configurations. For a lattice of $N$ dimers, the Edwards-Anderson order parameter for the spin degrees of freedom takes the form
\begin{equation}
Q_{\rm EA}^{\rm Mag} = \lim_{t_{j}\to\infty} \left|\overline{\left(\left\langle\dfrac{1}{N}\sum_{i=1}^{N}\vec{S}^{A}_{i}(t_{j})\cdot \vec{S}^{B}_{i}(t_{j})\right\rangle - \left\langle \sum_{i=1}^{N}\vec{S}^{A}_{i}(t_{j})\right\rangle \cdot \left\langle \sum_{i=1}^{N}\vec{S}^{B}_{i}(t_{j})\right\rangle \right)} \right| ,
\label{Mag_EA}
\end{equation}
where $\vec{S}^{A}_{i}(t_{j})$ and $\vec{S}^{B}_{i}(t_{j})$ are the spins at dimer $i$ at time step $t_{j}$ for replicas $A$ and $B$ respectively.

\end{widetext}

In order to establish whether there is a phase transition, and if so, to 
estimate of the critical temperature, $T_{c}$, we use the Binder cumulant statistic.  For the Edwards-Anderson order parameter, $Q$ the statistic is given by
\begin{equation}
	U_{L}(Q) = \dfrac{3}{2}\left[1 - \left(\dfrac{\langle Q^{4} \big\rangle _{L}}{3\langle Q^{2} \big\rangle ^{2} _{L}} \right) \right],
\end{equation}
where $L$ is the size of the system. We plot $U_{L}$ as a function of temperature for different system sizes.  If there is a transition to an ordered
phase at a temperature $T_c$ then in  the large $L$ limit, $U_{L}$ tends to a finite value for $T<T_{c}$ and 0 for $T> T_{c} $.
Up to small finite size corrections, the Binder cumulant curves should intersect at the critical temperature, where in the $L \to \infty$ limit, the 
Binder cumulant is independent of system size.

\begin{figure}[thb]
   \includegraphics[width=1\linewidth]{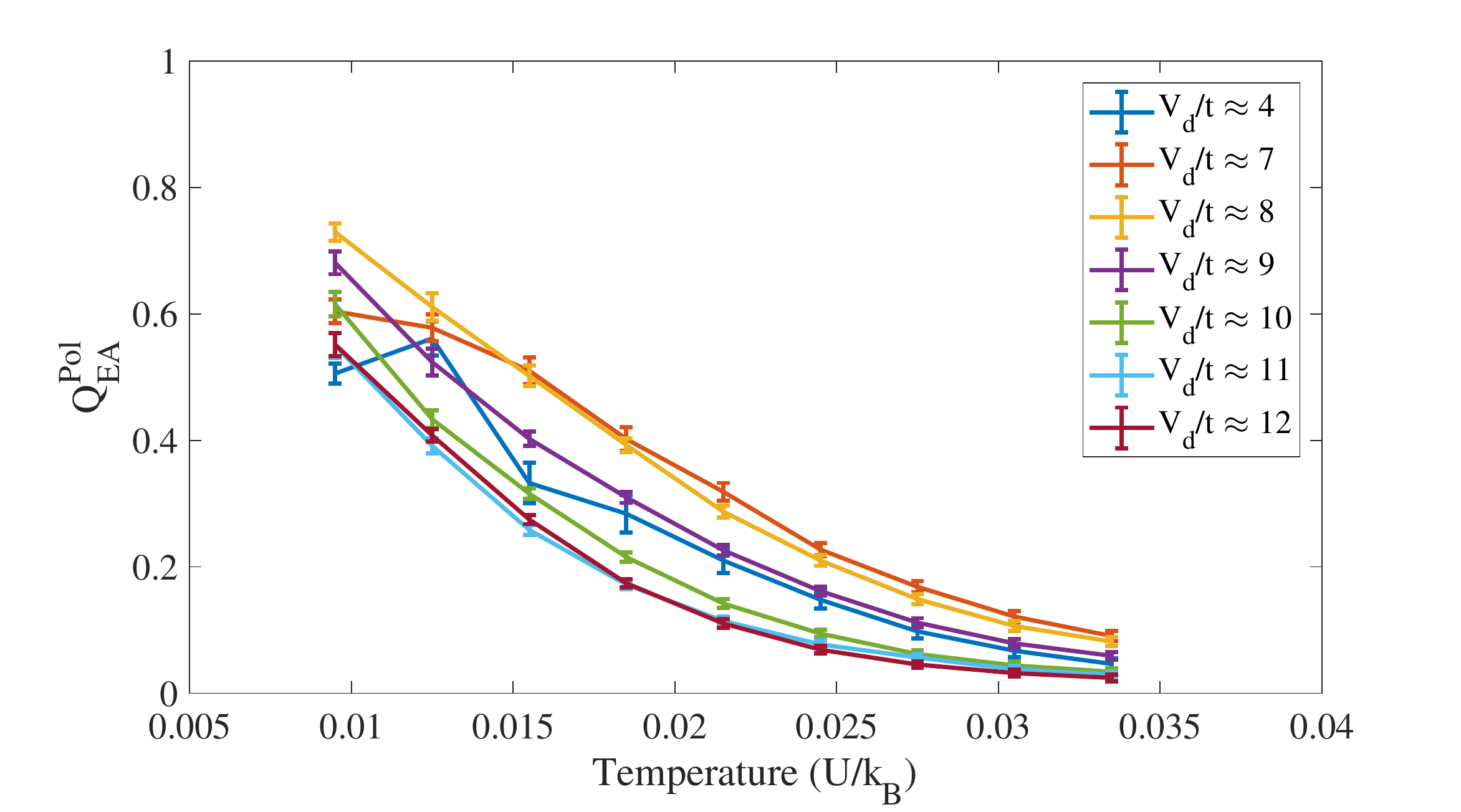}
   \includegraphics[width=1\linewidth]{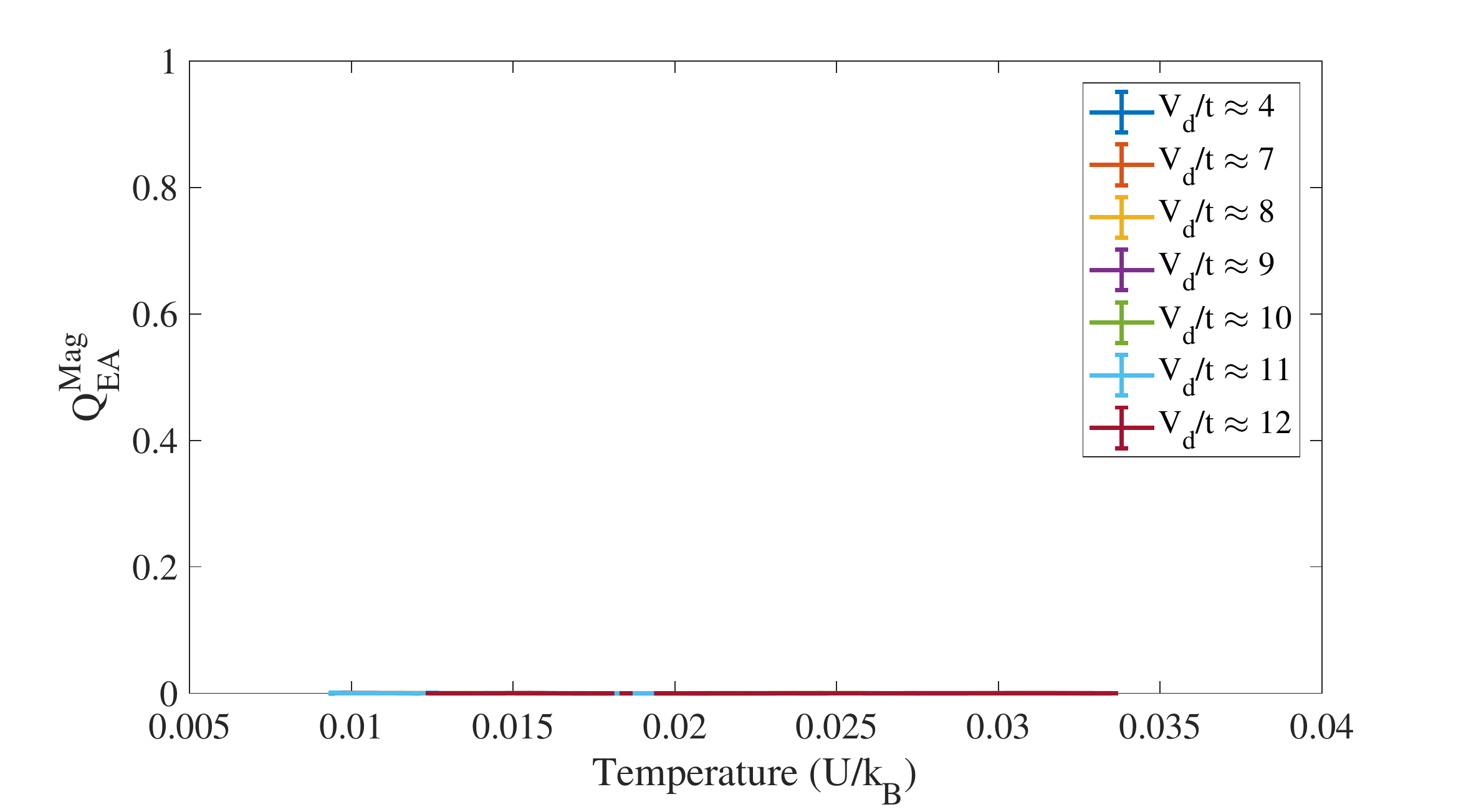}
        \caption{a) $Q^{\rm Pol}_{\rm EA}$; b) $Q^{\rm Mag}_{\rm EA}$ for fixed inter-dimer interaction strengths on a lattice of 10$^2$ = 100 dimers,
        averaged over 18 bond configurations.  The parameters used were $U/t=15.0113$, $V_{b}/t=1.0547$, $V_{p}/t =1.0623$, $V_{q}/t=1.0412$ for varying $V_{d}/t$.}
\label{fig:EA_Vd}
\end{figure}

\begin{figure}[thb]
   \includegraphics[width=1\linewidth]{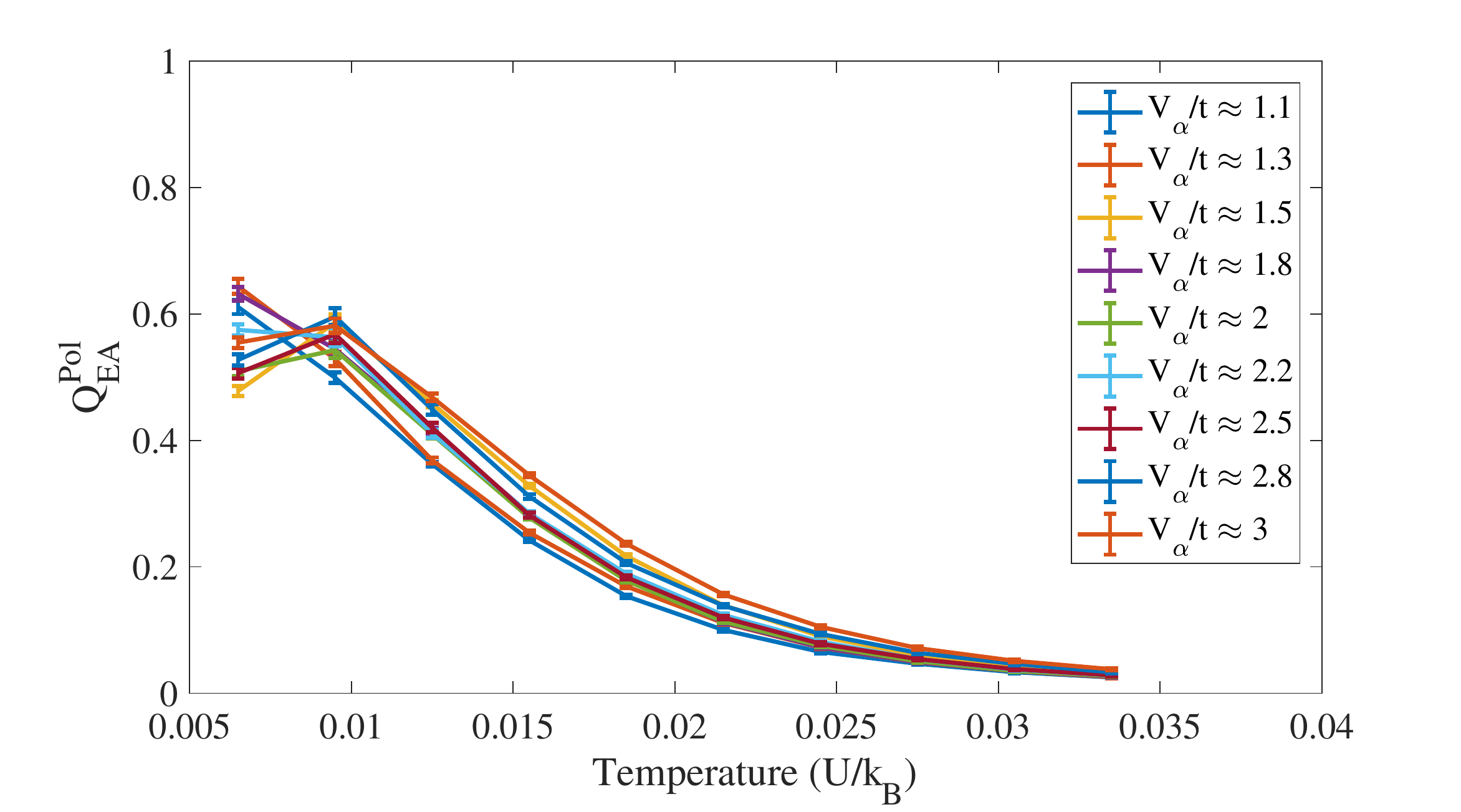}
   \includegraphics[width=1\linewidth]{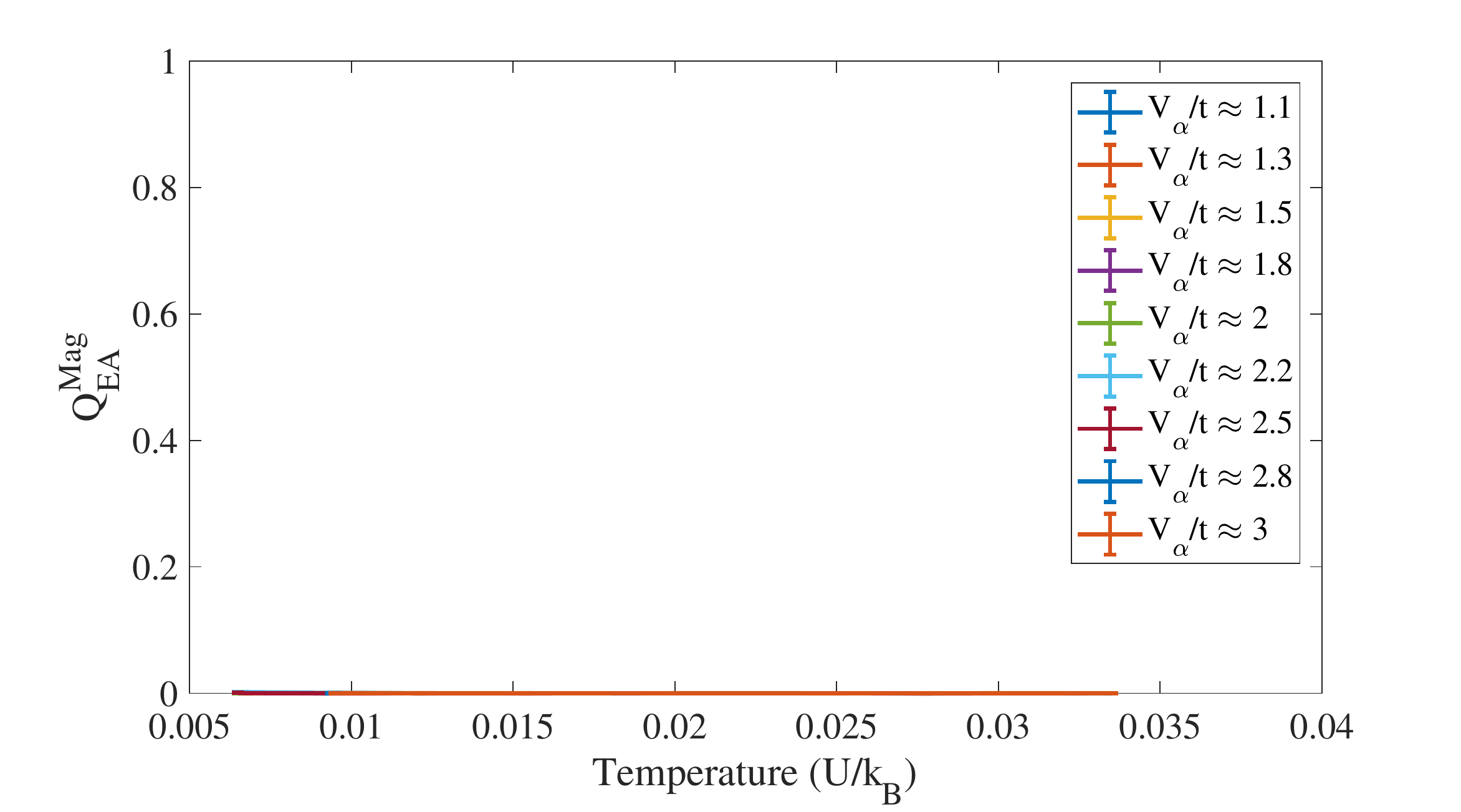}
\caption{a) $Q^{\rm Pol}_{\rm EA}$; b) $Q^{\rm Mag}_{\rm EA}$ for fixed intra-dimer interaction strengths on a lattice of 10$^2$ = 100 dimers,
        averaged over 18 bond configurations.  The parameters used were $U/t=15.0113$,  $V_{d}/t=10.0219$ for varying $V_{\alpha }$ with $\alpha = b$, $p$ and $q$.}
\label{fig:EA_Va}
\end{figure}

\begin{figure}[hbt]
  \includegraphics[width=8cm]{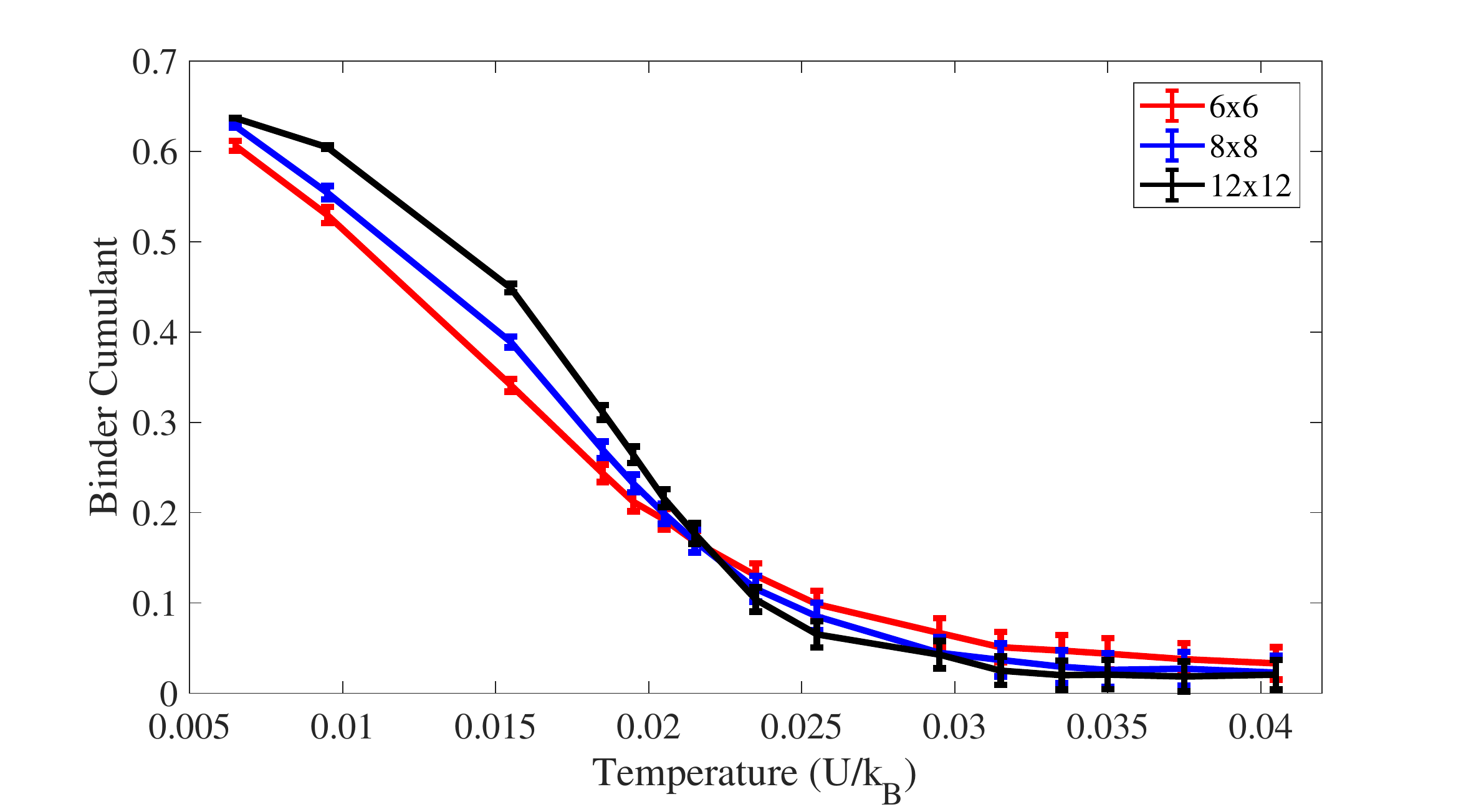}
  \caption{Binder cumulant plot of $V_{d}/t = 10.0219$ for lattice sizes $6\times 6$, $8\times 8$ and $12\times 12$. Other parameters used are $U/t=15.0113$, $V_{b}/t=1.0547$, 
  $V_{p}/t=1.0623$, $V_{q}/t=1.0412$.}
  \label{fig:Binder_10}
\end{figure}

Figure \ref{fig:Polarization} shows the polarization as a function of temperature for various intra- and inter-dimer interaction energies respectively. 
There is no evidence for ordering in the charge degrees of freedom for the parameters considered. We also see no ordering in the spin degrees of freedom shown in 
Fig.~\ref{fig:Magnetization}. 

Figures \ref{fig:El_Sus} and \ref{fig:Mag_Sus} show the electric and magnetic susceptibilities as a function of temperature 
respectively. We see similar behaviour for a wide range of $V_{d}$ and $V_{\alpha }$ for different lattice sizes of $N$ dimers.

In Fig.~\ref{fig:EA_Vd} we show $Q^{\rm Pol}_{\rm EA}$ and $Q^{\rm Mag}_{\rm EA}$ 
as a function of temperature. In the low temperature regime ($\lesssim 0.02\, U/k_{B}$) $Q^{\rm Pol}_{\rm EA}$ 
grows with decreasing temperature for $V_{d}/t \approx $ 10, 11, and 12. For intra-dimer interaction energies; $V_{d}/t \lesssim 10$, $Q^{\rm Pol}_{\rm EA}$ 
becomes appreciable for temperatures $\lesssim 0.03\, U/k_{B}$. 
Smaller intra-dimer interactions ($V_{d}/t \lesssim 10$) show stronger evidence 
of a non-zero Edwards-Anderson order parameter in the charge degrees of freedom at the chosen temperatures. 
Even though we see evidence of glassy ordering in charge degrees of freedom at low temperatures in Fig.~\ref{fig:EA_Vd}, we do not see any 
evidence of glassy ordering in the spin degrees of freedom.  $Q^{\rm Mag}_{\rm EA}$ takes on values indistinguishable 
from zero for all temperatures and $V_{d}/t$ values considered.

Similar results to those seen in Fig.~\ref{fig:EA_Vd} are shown in Fig.~\ref{fig:EA_Va} where we plot $Q^{\rm Pol}_{\rm EA}$ and $Q^{\rm Mag}_{\rm EA}$ 
as a function of temperature for different inter-dimer interaction energies. Here, we observe that small 
$V_{\alpha}/t$ values lead to a larger $Q^{\rm Pol}_{\rm EA}$ at low temperatures than large $V_{\alpha}/t$. 

A simple estimate of the phase transition is provided by the temperature at which the Edwards-Anderson order parameter becomes clearly distinct 
from zero, which can be determined from  Fig.~\ref{fig:EA_Vd} and \ref{fig:EA_Va} to be $k_{B}T \approx 0.025\, U/k_{B})$. 
In the following discussion we present a quantitative estimate of the phase transition in our model obtained from the Binder cumulant statistic.

Taking $V_d/t \simeq 10$, the intersection of the curves in Fig.~\ref{fig:Binder_10} determines an estimate for the critical temperature for the phase 
transition of $\approx 0.022\, U/k_{B}$, as compared to the naive estimate of $\approx 0.025\, U/k_{B}$ obtained from the $Q^{\rm Pol}_{EA}$ curve in Fig.~\ref{fig:EA_Vd}.
We observed similar behaviour for other parameter values $(e.g. V_{d}/t = 8$ and 9), but this is the clearest example of a transition.

\section{Discussion and Conclusions}
\label{sec:disc}

In this paper we have considered an extended Hubbard model of dimers on a triangular lattice, and obtained the low energy effective theory
in terms of separate spin and charge degrees of freedom in the one electron per dimer limit, taking into account the occupation of 
nearest neighbour sites, which extends a previous perturbative expansion of the model \cite{Hotta2010}.  In order to study the tendency
towards glassiness of charge degrees of freedom in the model, motivated by evidence of relaxor ferroelectric behaviour in 
the $\kappa -$(BEDT-TTF$)_{2}X$ family of salts, we made a classical approximation to the effective model, so that we could study 
it with Monte Carlo simulations.  The couplings in the effective model depend on the occupations of the neighbouring sites,
so in order to avoid recalculating the couplings at each step of the Monte Carlo simulation we calculated the full distribution of 
couplings and then drew couplings randomly from this distribution.  Under these conditions we have shown that a non-zero Edwards-Anderson order parameter
for charge degrees of freedom develops below a critical temperature $T_c$.

In order to make some connection to experiment, we take $U \sim 0.7$ eV \cite{Nakamura2012}, which would place $T_c = 0.022\, U/k_B \sim 180$ K, which is much 
higher than the temperatures at which glassy dynamics is seen in $\kappa -$(BEDT-TTF$)_{2}X$ salts, which is usually on the order of tens of kelvins.
However, several of the approximations we have made are likely to enhance ordering, so it is not surprising that our temperature estimate is considerably 
higher than experimental observations.  There are two main approximations we have made that likely enhance glassiness in the model.
First, we drop ``quantum'' terms in Eq.~(\ref{eq:Ham_C_ij}) to obtain our effective model, which corresponds to ignoring quantum fluctuations, that can be expected to depress $T_c$.  Second, we draw the couplings we use in the Monte Carlo simulation from a fixed distribution that is independent of time, rather than 
working with couplings that fluctuate with time.  This has the effect of generating quenched rather than time-dependent disorder, which will also enhance 
glassy tendencies in the model.  However, despite these approximations, glassiness appears quite readily in the charge degrees of freedom, so even if 
the two approximations we have made are relaxed, it seems likely that there will remain strong tendencies towards glassiness, that may well manifest 
themselves at lower temperature scales in the true system.  Previous work on models of classical models of spins coupled to charge degrees of freedom has
also demonstrated glassy dynamics, even in the absence of disorder \cite{Kennett2005}.

Another observation of note is that we find no ordering in the spin degrees of freedom and no development of charge polarization 
at the temperatures we can access, consistent with the experimental observations in $\kappa -$(BEDT-TTF$)_{2}X$ salts.  In addition, the observation of a 
non-zero Edwards-Anderson transition temperature in a two dimensional model might be surprising until one considers that it is on a triangular lattice, for
which each site has six neighbours, unlike the two dimensional Edwards-Anderson model on a square lattice with $\pm J$ couplings, for which there are
only four neighbours per site, that is a spin glass only at zero temperature \cite{Singh1986,Bhatt1988}.

We regard this work as a proof of principle that glassiness can arise in models that are relevant for $\kappa -$(BEDT-TTF$)_{2}X$ salts and see future 
avenues for exploration through i) studying a wider range of extended Hubbard model parameters, ii) allowing for time dependent couplings, and iii) investigating the effects of quantum terms in the effective model.  In addition to the equilibrium calculations considered here, the study of out of equilibrium dynamics in 
larger systems may help to determine the functional time dependence of aging dynamics that might also be accessible in experiments.

\section{Acknowledgements}
The authors acknowledge Compute Canada resources that were used to obtain the numerical 
results in this work. M. B. D. and M. P. K. were supported by NSERC.

\begin{appendix}
\begin{widetext}
\section{Commutators Used in the Strong Coupling Expansion} \label{App.Commutator}

In this Appendix we provide derivations of a number of commutators that prove useful in constructing the strong coupling expansion of the 
Hamiltonian. Many of the commutators derived here have corresponding simpler versions in Ref.~\cite{Farrell2013} 
in which the strong coupling expansion was performed on a  square lattice of sites rather than a triangular lattice of dimers.

\subsection{Useful Commutators}
Before deriving the commutator $\left[H_0, T_{\alpha}^{m, \{M_2\}, \{M_1\}}\right]$ it is helpful to first derive a few basic commutators that are used in the derivation of the commutator of $H_0$ with the hopping operator. We begin with 
\begin{align}\label{eq:numbercom}
	[n^{\vphantom{\dagger}}_{(x,y),i,\sigma},c^{\dagger}_{(x',y'), j, \sigma'} c^{\vphantom{\dagger}}_{(x'',y''), k, \sigma'}] 
=\, &\delta_{\sigma, \sigma'}\left[\delta_{(x,y),(x',y')} \delta_{i,j} - \delta_{(x,y),(x'',y'')} \delta_{i,k}\right]c^{\dagger}_{(x',y'), j, \sigma'}
	c^{\vphantom{\dagger}}_{(x'',y''), k, \sigma'}. 
\end{align}
Setting $(x'',y'') = (x',y'), k=j$ in this result it follows directly that 
\begin{align}
[n_{(x,y),i,\sigma},n_{(x',y'),j,\sigma'}] &=  0, 
\end{align}
which implies
\begin{align}
	[n_{(x,y),i,\sigma},h_{(x',y'),j,\sigma'}] = \left[n_{(x,y),i,\sigma}, O_{(x',y'),j}^\beta[\tilde{n}^\beta]\right] = \left[h_{(x,y),i,\sigma}, 
	O_{(x',y'),j}^\beta[\tilde{n}^\beta]\right]= 0.
\end{align}
The next set of commutators needed are those of the form $\big[n_{(x,y),i,\sigma'},(T_\alpha^m)_{(x',y';\, x'',y''),j,k,\sigma} \big]$ which must be calculated independently for each $m = -1,0,1$. 
In general they can be written as
\begin{align} \label{eq:n-hop-com}
\Big[n_{(x,y),i,\sigma'},&(T_\alpha^m)_{(x',y';\, x'',y''),j,k,\sigma} \Big]
	= (T_\alpha^m)_{(x',y';\, x'',y''),j,k,\sigma}\delta_{\sigma,\sigma'}\left[\delta_{(x,y),(x'',y'')}\delta_{i,k} - \delta_{(x,y),(x',y')}\delta_{i,j}\right] .
\end{align}

\subsection{On-site Interaction Commutator}
Using Eq.~(\ref{eq:n-hop-com}) one can show straightforwardly that
in general 
\begin{align}\label{eq:U_comm}
\left[H_U, T_{\alpha}^{m, \{M_2\}, \{M_1\}}\right] = m U T_{\alpha}^{m, \{M_2\}, \{M_1\}}.
\end{align}

\subsection{Nearest Neighbour Interaction Commutator}
Before deriving the next commutator it is useful to first rewrite the nearest neighbour interaction term as
\begin{align}
H_{V} = \frac{1}{2}\sum_{\gamma}V_\gamma\sum_{\mathclap{\substack{(x,y) \\ i}}}\hspace{4mm} \sum_{\mathclap{\substack{(\delta_{x_\gamma},\delta_{y_\gamma}) \\ \delta_\gamma}}}\hspace{3.5mm}\sum_{\sigma, \sigma'}n_{(x,y),i,\sigma}n_{(x+\delta_{x_\gamma},y+\delta_{y_\gamma}),i+\delta_\gamma,\sigma'},
\end{align}
where $\gamma = d, q, b, p$ and the factor of $\frac{1}{2}$ is to prevent double counting. It is possible to further rewrite this as
\begin{align} 
H_{V} = \frac{1}{2}\sum_{\gamma}V_\gamma \sum_{\mathclap{\substack{(x,y) \\ i}}}\sum_{\sigma}n_{(x,y),i,\sigma}\tilde{n}^\gamma_{(x,y),i},
\end{align}
where
\begin{align}
\tilde{n}^\gamma_{(x,y),i}=\hspace{3.5mm}\sum_{\mathclap{\substack{(\delta_{x_\gamma},\delta_{y_\gamma}) \\ \delta_\gamma}}}\hspace{3.5mm}\sum_{\tilde{\sigma}}n_{(x+\delta_{x_\gamma},y+\delta_{y_\gamma}),i+\delta_\gamma,\tilde{\sigma}}.
\end{align}
The desired commutator now takes the form
\begin{align} \label{eq:Hv-com-step1}
\left[H_{V}, T_{\alpha}^{m, \{M_2\}, \{M_1\}}\right] =& \frac{1}{2}\sum_{\gamma}V_\gamma\sum_{\mathclap{\substack{(x,y) \\\ i}}}\hspace{2mm}\sum_{\mathclap{\substack{(x',y') \\ j}}}\hspace{3.2mm}\sum_{\mathclap{\substack{(x'',y'') \\ k}}}{\vphantom{\sum}}'\left\{\prod_{\beta}\sum_{S[n_2^\beta] = M_2^\beta}\sum_{S[n_1^\beta]  = M_1^\beta}\right\}
	\left\{\prod_{\eta} O_{(x'',y''),k}^\eta[n_2^\eta]\right\} \nonumber  \\
 &\times  \sum_{\sigma,\sigma'} \left[n_{(x,y),i,\sigma'}\tilde{n}^\gamma_{(x,y),i},(T_\alpha^m)_{(x',y';\, x'',y''),j,k,\sigma}\right]
	\left\{\prod_{\xi} O_{(x',y'),j}^\xi[n_1^\xi]\right\} \nonumber \\
 =&  \frac{1}{2}\sum_{\gamma}V_\gamma\sum_{\mathclap{\substack{(x,y) \\\ i}}}\hspace{2mm}\sum_{\mathclap{\substack{(x',y') \\ j}}}\hspace{3.2mm}\sum_{\mathclap{\substack{(x'',y'') \\ k}}}{\vphantom{\sum}}'\left\{\prod_{\beta}\sum_{S[n_2^\beta] = M_2^\beta}\sum_{S[n_1^\beta]  = M_1^\beta}\right\}
	\left\{\prod_{\eta} O_{(x'',y''),k}^\eta[n_2^\eta]\right\} \nonumber \\
 &\times \sum_{\sigma,\sigma'} \Bigg\{ n_{(x,y),i,\sigma'}\left[\tilde{n}^\gamma_{(x,y),i},(T_\alpha^m)_{(x',y';\, x'',y''),j,k,\sigma}\right] \nonumber \\
&+ \left[n_{(x,y),i,\sigma'},(T_\alpha^m)_{(x',y';\, x'',y''),j,k,\sigma}\right]\tilde{n}^\gamma_{(x,y),i}\Bigg\}
	\left\{\prod_{\xi} O_{(x',y'),j}^\xi[n_1^\xi]\right\}. 
\end{align}
Equation (\ref{eq:n-hop-com}) can then be used to obtain
\begin{align}\label{eq:nt-hop-com}
&\left[\tilde{n}^\gamma_{(x,y),i},(T_\alpha^m)_{(x',y';\, x'',y''),j,k,\sigma}\right] \nonumber \\ &=\hspace{3.5mm} \sum_{\mathclap{\substack{(\delta_{x_\gamma},\delta_{y_\gamma}) \\\delta_\gamma}}}\hspace{3.5mm}\sum_{\tilde{\sigma}}(T_\alpha^m)_{(x',y';\, x'',y''),j,k,\sigma}\delta_{\sigma,\tilde{\sigma}}\left[\delta_{(x+\delta_{x_\gamma},y+\delta_{y_\gamma}),(x'',y'')}\delta_{i+\delta_\gamma,k}-\delta_{(x+\delta_{x_\gamma},y+\delta_{y_\gamma}),(x',y')}\delta_{i+\delta_\gamma,j}\right]. 
\end{align}
Applying Eq.~(\ref{eq:n-hop-com}) and relabeling some indicies leads to 
\begin{eqnarray} \label{eq:Hv-com-step2}
	\left[H_{V}, T_{\alpha}^{m, \{M_2\}, \{M_1\}}\right] &=& \frac{1}{2}\sum_{\gamma}V_\gamma \sum_{\mathclap{\substack{(x',y') \\ j}}}
	\hspace{3.2mm}\sum_{\mathclap{\substack{(x'',y'') \\ k}}}
            {\vphantom{\sum}}'\sum_{\sigma}\left\{\prod_{\beta}\sum_{S[n_2^\beta] = M_2^\beta}\sum_{S[n_1^\beta]  = M_1^\beta}\right\}
	\left\{\prod_{\eta} O_{(x'',y''),k}^\eta[n_2^\eta]\right\} \nonumber \\
	& & \times \Bigg\{ \left[\tilde{n}^\gamma_{(x'',y''),k}-\tilde{n}^\gamma_{(x',y'),j}\right] (T_\alpha^m)_{(x',y';\, x'',y''),j,k,\sigma} \nonumber \\
	&  & \hspace*{0.5cm}	+(T_\alpha^m)_{(x',y';\, x'',y''),j,k,\sigma}\left[\tilde{n}^\gamma_{(x'',y''),k}-\tilde{n}^\gamma_{(x',y'),j}\right]\Bigg\} 
  \left\{\prod_{\xi} O_{(x',y'),j}^\xi[n_1^\xi]\right\}. 
\end{eqnarray}
The central part of this equation can be rewritten as
\begin{align} \label{eq:Hv-com-step2.5}
\tilde{n}^\gamma_{(x'',y''),k}&(T_\alpha^m)_{(x',y';\, x'',y''),j,k,\sigma}-\tilde{n}^\gamma_{(x',y'),j}(T_\alpha^m)_{(x',y';\, x'',y''),j,k,\sigma} \nonumber \\ 
&+(T_\alpha^m)_{(x',y';\, x'',y''),j,k,\sigma}\tilde{n}^\gamma_{(x'',y''),k}-(T_\alpha^m)_{(x',y';\, x'',y''),j,k,\sigma}\tilde{n}^\gamma_{(x',y'),j} \nonumber \\
=& 2\left[\tilde{n}^\gamma_{(x'',y''),k}(T_\alpha^m)_{(x',y';\, x'',y''),j,k,\sigma}-(T_\alpha^m)_{(x',y';\, x'',y''),j,k,\sigma}\tilde{n}^\gamma_{(x',y'),j}\right] \nonumber \\
& + \left[(T_\alpha^m)_{(x',y';\, x'',y''),j,k,\sigma},\tilde{n}^\gamma_{(x'',y''),k}\right] - \left[\tilde{n}^\gamma_{(x',y'),j},(T_\alpha^m)_{(x',y';\, x'',y''),j,k,\sigma}\right] .
\end{align}
When Eq.~(\ref{eq:Hv-com-step2.5}) is inserted into Eq.~(\ref{eq:Hv-com-step2}) and  Eq.~(\ref{eq:nt-hop-com}) is also applied this leads to the result
\begin{align} \label{eq:Hv-com-step3}
\left[H_{V}, T_{\alpha}^{m, \{M_2\}, \{M_1\}}\right] =& \sum_{\gamma}V_\gamma\sum_{\mathclap{\substack{(x',y') \\ j}}}\hspace{3.2mm}\sum_{\mathclap{\substack{(x'',y'') \\ k}}}{\vphantom{\sum}}'\sum_{\sigma}\left\{\prod_{\beta}\sum_{S[n_2^\beta] = M_2^\beta}\sum_{S[n_1^\beta]  = M_1^\beta}\right\}
	\left\{\prod_{\eta} O_{(x'',y''),k}^\eta[n_2^\eta]\right\} \nonumber \\
 &\times  \left[\tilde{n}^\gamma_{(x'',y''),k}(T_\alpha^m)_{(x',y';\, x'',y''),j,k,\sigma} - (T_\alpha^m)_{(x',y';\, x'',y''),j,k,\sigma}\tilde{n}^\gamma_{(x',y'),j} \right] 
	\left\{\prod_{\xi} O_{(x',y'),j}^\xi[n_1^\xi]\right\}. 
\end{align}
and this becomes
\begin{align}
\left[H_{V}, T_{\alpha}^{m, \{M_2\}, \{M_1\}}\right] = \sum_{\gamma}V_\gamma(M_2^\gamma - M_1^\gamma)\, T_{\alpha}^{m, \{M_2\}, \{M_1\}},
\end{align}
which gives a commutator analogous to Eq.~(\ref{eq:U_comm}).

\section{Details of the Strong Coupling Expansion} \label{app:strong_coup}
In this Appendix we write the decomposed hopping operator as
\begin{align}
	T_{\alpha}^{m, \{M_{2}\}, \{M_1\}} = Y_{m,\alpha}^{\{M_{2}\},\{M_1\}},
\end{align}
which satisfies the commutator
\begin{align}
	\left[H_0,Y_{m,\alpha}^{ \{M_{2}\},\{M_1\}}\right] = \epsilon_{m}^{ \{M_{2}\},\{M_1\}}Y_{m,\alpha}^{\{M_{2}\},\{M_1\}},
\end{align}
with $\epsilon_{m}^{\{M_{2}\},\{M_1\}} = mU+\sum_{\gamma}V_\gamma(M_2^\gamma - M_1^\gamma)$ and $H_0 = H_U + H_V$. 
Recall that solutions for $S_n$ and $H_{T,n}'$ are required such that Eq.~(\ref{eq:EoMconstraint}) is satisfied to some desired order in $1/U$. For $S_1$ we must solve
\begin{align} \label{eq:s1_solve_1}
\big[H_0,[S_1,\tilde{H}_0]+H_T\big] = 0,
\end{align}
and remembering that $\tilde{H}_0 = H_0/U$ it is possible to rearrange this as
\begin{align} \label{eq:s1_solve_2}
\big[\tilde{H}_0,[\tilde{H}_0,S_1]]&= [\tilde{H}_0,H_T] 
	=\sum_{\mathclap{\substack{m,\alpha \\ \{M_{2}\},\{M_1\} }}}\left[\tilde{H}_0,Y_{m,\alpha}^{ \{M_{2}\},\{M_1\}}\right] 
	= \sum_{\mathclap{\substack{m,\alpha \\ \{M_{2}\},\{M_1\} }}}\frac{\epsilon_{m}^{\{M_{2}\},\{M_1\}}}{U} Y_{m,\alpha}^{ \{M_{2}\},\{M_1\}}. 
\end{align}
We can verify that $S_1$ takes the form:
\begin{align} \label{eq:s1_gen}
	S_1 = \tilde{\sum_{\mathclap{\substack{m,\alpha \\ \{M_{2}\},\{M_1\} }}}}\hspace{3mm} 
	\frac{U}{\epsilon_{m}^{\{M_{2}\},\{M_1\}}} Y_{m,\alpha}^{ \{M_{2}\},\{M_1\}},
\end{align}
where the tilde over the sum indicates that the all terms in the sum which have $\epsilon_{m}^{ \{M_{2}\},\{M_1\}} = 0$ are excluded. 
Substituting this into Eq.~(\ref{eq:s1_solve_2}) gives
\begin{align}
	\tilde{\sum_{\mathclap{\substack{m,\alpha \\ \{M_{2}\},\{M_1\} }}}}\hspace{3mm} \frac{U}{\epsilon_{m}^{ \{M_{2}\},\{M_1\}}}
	\left(\frac{\epsilon_{m}^{\{M_{2}\},\{M_1\}}}{U}\right)^2 Y_{m,\alpha}^{ \{M_{2}\},\{M_{1}\}} 
	= \sum_{\mathclap{\substack{m,\alpha \\ \{M_{2}\},\{M_1\} }}}\frac{\epsilon_{m}^{\{M_{2}\},\{M_1\}}}{U} Y_{m,\alpha}^{ \{M_{2}\},\{M_1\}},
\end{align}
and since it is possible to drop the $\epsilon_{m}^{\{M_{2}\},\{M_1\}} = 0$ terms from the sum on the right hand side with no consequence this becomes
\begin{align}
	\tilde{\sum_{\mathclap{\substack{m,\alpha \\ \{M_{2}\},\{M_1\} }}}}\frac{\epsilon_{m}^{ \{M_{2}\},\{M_1\}}}{U} Y_{m,\alpha}^{ \{M_{2}\},\{M_1\}}
	= \tilde{\sum_{\mathclap{\substack{m,\alpha \\ \{M_{2}\},\{M_1\} }}}}\frac{\epsilon_{m}^{\{M_{2}\},\{M_1\}}}{U} Y_{m,\alpha}^{\{M_{2}\},\{M_1\}},
\end{align}
verifying that the form of $S_1$ in Eq.~(\ref{eq:s1_gen}) solves Eq.~(\ref{eq:s1_solve_2}). Now that the form of $S_1$ is known it is possible to calculate the form of the first order correction 
\begin{eqnarray}
H_{T,1}' = H_T - [\tilde{H}_0,S_1] 
	&= &\sum_{\mathclap{\substack{m,\alpha \\ \{M_{2}\},\{M_1\} }}}Y_{m,\alpha}^{\{M_{2}\},\{M_1\}} -\tilde{\sum_{\mathclap{\substack{m,\alpha \\ \{M_{2}\},\{M_1\} }}}}\hspace{3mm} 
	\frac{U}{\epsilon_{m}^{\{M_{2}\},\{M_1\}}}\frac{\epsilon_{m}^{ \{M_{2}\},\{M_1\}}}{U} Y_{m,\alpha}^{ \{M_{2}\},\{M_1\}} \nonumber \\
	&= &\sum_{\mathclap{\substack{m,\alpha \\ \{M_{2}\},\{M_1\} }}}Y_{m,\alpha}^{\{M_{2}\},\{M_1\}} - \tilde{\sum_{\mathclap{\substack{m,\alpha \\ \{M_{2}\},\{M_1\} }}}}
	Y_{m,\alpha}^{ \{M_{2}\},\{M_1\}} \nonumber \\
	&= & \sum^*_{\mathclap{\substack{m,\alpha \\ \{M_{2}\},\{M_1\} }}}Y_{m,\alpha}^{ \{M_{2}\},\{M_1\}}. 
\end{eqnarray}
where the star above the sum denotes that the sum only includes terms in which $\epsilon_{m}^{ \{M_{2}\},\{M_1\}} = 0$. In the hopping operator notation this correction takes the form 
\begin{align}
H_{T,1}' &=\sum_{\alpha}\hspace{5mm}\sum^*_{\mathclap{\substack{m \\ \{M_{2}\},\{M_{1}\} }}}\hspace{2mm}T_{\alpha}^{m, \{M_{2}\}, \{M_1\}}.
\end{align}
Strictly, it is possible that the starred restriction is satisfied by having the $mU+\sum_{\gamma}V_\gamma(M_2^\gamma - M_1^\gamma)$ perfectly cancel for non-zero $m$, $\{M_1\}$
and $\{M_2\}$, however, if $U$ and the $V_\gamma$ are chosen so that this is never the case we will have $m=0, \{M_2\} = \{M_1\}$, allowing the first order correction to be written as
\begin{align}
H_{T,1}' &= \sum_{\alpha}\sum_{\{M\}}T_{\alpha}^{0, \{M\}, \{M\}},
\end{align}
with no additional conditions, completing the first order perturbation theory.

Before finding $S_2$ and the second order correction it is worthwhile to recall the the Jacobi identity
\begin{align}
\big[A,[B,C]\big] + \big[C,[A,B]\big] + \big[B,[C,A]\big] = 0,
\end{align}
which we can use to write
\begin{align}\label{eq:jacobi}
	\left[\tilde{H}_0,\left[Y_{m,\alpha}^{\{M_{2}\},\{M_1\}}, Y_{n,\nu}^{\{N_{2}\},\{N_1\}}\right]\right] %\nonumber \\
	= \frac{\epsilon_{m}^{ \{M_{2}\},\{M_1\}}+\epsilon_{n}^{ \{N_{2}\},\{N_1\}}}{U} 
	\left[Y_{m,\alpha}^{\{M_{2}\},\{M_1\}}, Y_{n,\nu}^{\{N_{2}\},\{N_1\}}\right].
\end{align}
For $S_2$ we must solve
\begin{align}\label{eq:s_2_int}
\left[\tilde{H}_0,[\tilde{H}_0,S_2]\right] =& \tfrac{1}{2}\left[\tilde{H}_0,\left[S_1,[S_1,\tilde{H}_0]\right]\right] + \left[\tilde{H}_0,\left[S_1,H_T\right]\right]  \nonumber  \\
	=& -\tilde{\sum_{\mathclap{\substack{m,\alpha \\ \{M_{2}\},\{M_1\} }}}}\hspace{10mm}\tilde{\sum_{\mathclap{\substack{n,\nu \\ \{N_{2}\},\{N_1\} }}}} 
	\frac{\epsilon_{m}^{ \{M_{2}\},\{M_1\}}+\epsilon_{n}^{ \{N_{2}\},\{N_1\}}}{2\epsilon_{m}^{\{M_{2}\},\{M_1\}}} \left[ Y_{m,\alpha}^{\{M_{2}\},\{M_1\}},
	Y_{n,\nu}^{\{N_{2}\},\{N_1\}}\right] \nonumber \\
	&+\tilde{\sum_{\mathclap{\substack{m,\alpha \\ \{M_{2}\},\{M_1\} }}}}\hspace{10mm}\sum_{\mathclap{\substack{n,\nu \\ \{N_{2}\},\{N_1\} }}}
	\frac{\epsilon_{m}^{ \{M_{2}\},\{M_1\}}
	+\epsilon_{n}^{ \{N_{2}\},\{N_1\}}}{\epsilon_{m}^{\{M_{2}\},\{M_1\}}} \left[ Y_{m,\alpha}^{\{M_{2}\},\{M_1\}}, 
	Y_{n,\nu}^{\{N_{2}\},\{N_1\}} \right] ,
\end{align}
where we used Eq.~(\ref{eq:jacobi}) in Eq.~(\ref{eq:s_2_int}). 
It is possible to rewrite 
\begin{equation}
	\sum_{\mathclap{\substack{n,\nu \\ \{N_{2}\},\{N_1\} }}} \hspace*{3mm}  = \hspace*{3mm} 
	\tilde{\sum_{\mathclap{\substack{n,\nu \\ \{N_{2}\},\{N_1\} }}}} \hspace*{3mm} + \hspace*{3mm} 
\sum^*_{\mathclap{\substack{n,\nu \\ \{N_{2}\},\{N_1\} }}} ,
	\label{eq:break_up}
\end{equation}
by breaking up the sum into parts which have $\epsilon_{n}^{ \{N_{2}\},\{N_1\}} \neq 0$ and parts which have $\epsilon_{n}^{ \{N_{2}\},\{N_1\}} = 0$. 
The commutator equation then becomes
\begin{align}
	\left[\tilde{H}_0,[\tilde{H}_0,S_2]\right] =& \tilde{\sum_{\mathclap{\substack{m,\alpha \\ \{M_{2}\},\{M_1\} }}}}\hspace{10mm}
	\tilde{\sum_{\mathclap{\substack{n,\nu \\ \{N_{2}\},\{N_1\} }}}} 
	\frac{\epsilon_{m}^{ \{M_{2}\},\{M_1\}}+\epsilon_{n}^{ \{N_{2}\},\{N_1\}}}{2\epsilon_{m}^{\{M_{2}\},\{M_1\}}} 
	\left[ Y_{m,\alpha}^{\{M_{2}\},\{M_1\}}, Y_{n,\nu}^{\{N_{2}\},\{N_1\}}\right] \nonumber \\
	&+\tilde{\sum_{\mathclap{\substack{m,\alpha \\ \{M_{2}\},\{M_1\} }}}}\hspace{10mm}\sum^*_{\mathclap{\substack{n,\nu \\ \{N_{2}\},\{N_1\} }}} 
	\frac{\epsilon_{m}^{ \{M_{2}\},\{M_1\}}+0}{\epsilon_{m}^{\{M_{2}\},\{M_1\}}} 
	\left[ Y_{m,\alpha}^{\{M_{2}\},\{M_1\}}, Y_{n,\nu}^{\{N_{2}\},\{N_1\}}\right]. 
\end{align}
Dropping the $\epsilon_{m}^{ \{M_{2}\},\{M_1\}}+\epsilon_{n}^{ \{N_{2}\},\{N_1\}} = 0$ terms from the first summation has no consequence allowing this to be written as
\begin{align} \label{eq:S2.LHS}
	\left[\tilde{H}_0,[\tilde{H}_0,S_2]\right] =& \overline{\sum_{\mathclap{\substack{m,\alpha  \\ \{M_{2}\},\{M_1\} }}}\hspace{10mm}
	\sum_{\mathclap{\substack{n,\nu \\ \{N_{2}\},\{N_1\} }}}}\hspace{3mm} 
	\frac{\epsilon_{m}^{ \{M_{2}\},\{M_1\}}+\epsilon_{n}^{ \{N_{2}\},\{N_1\}}}{2\epsilon_{m}^{\{M_{2}\},\{M_1\}}} 
	\left[ Y_{m,\alpha}^{\{M_{2}\},\{M_1\}}, Y_{n,\nu}^{\{N_{2}\},\{N_1\}}\right] \nonumber \\
	&+\tilde{\sum_{\mathclap{\substack{m,\alpha \\ \{M_{2}\},\{M_1\} }}}}\hspace{10mm}\sum^*_{\mathclap{\substack{n,\nu \\ \{N_{2}\},\{N_1\} }}} \,\, 
	\left[ Y_{m,\alpha}^{\{M_{2}\},\{M_1\}}, Y_{n,\nu}^{\{N_{2}\},\{N_1\}} \right] ,
\end{align}
where the over-bar indicates that the sums exclude all terms which have $\epsilon_{m}^{ \{M_{2}\},\{M_1\}} = 0$, $\epsilon_{n}^{ \{N_{2}\},\{N_1\}} = 0$, 
and $\epsilon_{m}^{ \{M_{2}\},\{M_1\}}+\epsilon_{n}^{ \{N_{2}\},\{N_1\}} = 0$. Again, rather than solving this equation explicitly for $S_2$ a solution of the form
\begin{align}
	S_2 =& \overline{\sum_{\mathclap{\substack{m,\alpha \\ \{M_{2}\},\{M_1\} }}}\hspace{10mm}
	\sum_{\mathclap{\substack{n,\nu \nonumber \\ \{N_{2}\},\{N_1\} }}}}\hspace{3mm} \frac{U^2}{2\epsilon_{m}^{\{M_{2}\},\{M_1\}}
	\left(\epsilon_{m}^{ \{M_{2}\},\{M_1\}}+\epsilon_{n}^{ \{N_{2}\},\{N_1\}}\right)} \left[ Y_{m,\alpha}^{\{M_{2}\},\{M_1\}},
	Y_{n,\nu}^{\{N_{2}\},\{N_1\}}\right] \\
	&+\tilde{ \sum_{\mathclap{\substack{m,\alpha \\ \{M_{2}\},\{M_1\} }}}}\hspace{10mm}\sum^*_{\mathclap{\substack{n,\nu \\ \{N_{2}\},\{N_1\} }}}\hspace{3mm} 
	\left(\frac{U}{\epsilon_{m}^{ \{M_{2}\},\{M_1\}}}\right)^2 
	\left[Y_{m,\alpha}^{\{M_{2}\},\{M_1\}}, Y_{n,\nu}^{\{N_{2}\},\{N_1\}}\right], 
\end{align}
can be guessed and easily verified by substitution into Eq.~(\ref{eq:S2.LHS}). Now that both $S_1$ and $S_2$ are known it is possible to calculate the second order correction
\begin{align}
H_{T,2}' =& \left[S_1,H_T\right]+\tfrac{1}{2}\left[S_1,[S_1,\tilde{H}_0]\right]+[S_2,\tilde{H}_0] \nonumber \\
	=&\tilde{\sum_{\mathclap{\substack{m,\alpha \\ \{M_{2}\},\{M_1\} }}}}\hspace{10mm}\sum_{\mathclap{\substack{n,\nu \\ \{N_{2}\},\{N_1\} }}}\hspace{3mm} 
	\frac{U}{\epsilon_{m}^{\{M_{2}\},\{M_1\}}}
	\left[Y_{m,\alpha}^{\{M_{2}\},\{M_1\}}, Y_{n,\nu}^{\{N_{2}\},\{N_1\}}\right] \nonumber \\
	&-\tilde{\sum_{\mathclap{\substack{m,\alpha \\ \{M_{2}\},\{M_1\} }}}}\hspace{10mm}\tilde{\sum_{\mathclap{\substack{n,\nu \\ \{N_{2}\},\{N_1\} }}}}\hspace{3mm} 
	\frac{U}{2\epsilon_{m}^{\{M_{2}\},\{M_1\}}}
	\left[Y_{m,\alpha}^{\{M_{2}\},\{M_1\}}, Y_{n,\nu}^{\{N_{2}\},\{N_1\}}\right] \nonumber \\
	&-\overline{\sum_{\mathclap{\substack{m,\alpha \\ \{M_{2}\},\{M_1\} }}}\hspace{10mm}\sum_{\mathclap{\substack{n,\nu \\ \{N_{2}\},\{N_1\} }}}}\hspace{3mm} 
	\frac{U}{2\epsilon_{m}^{\{M_{2}\},\{M_1\}}}\left[Y_{m,\alpha}^{\{M_{2}\},\{M_1\}}, Y_{n,\nu}^{\{N_{2}\},\{N_1\}}\right] \nonumber \\
	&- \tilde{\sum_{\mathclap{\substack{m,\alpha \\ \{M_{2}\},\{M_1\} }}}}\hspace{10mm}\sum^*_{\mathclap{\substack{n,\nu \\ \{N_{2}\},\{N_1}\} }}\hspace{3mm} 
	\frac{U}{\epsilon_{m}^{ \{M_{2}\},\{M_1\}}}\left[ Y_{m,\alpha}^{\{M_{2}\},\{M_1\}}, Y_{n,\nu}^{\{N_{2}\},\{N_1\}}\right] ,
\end{align}
and using Eq.~(\ref{eq:break_up}) to break up the sum leads directly to 
\begin{align}
	H_{T,2}' =& \tilde{\sum_{\mathclap{\substack{m,\alpha \\ \{M_{2}\},\{M_1\} }}}}\hspace{10mm}\tilde{\sum_{\mathclap{\substack{n,\nu \\ \{N_{2}\},\{N_1\} }}}}\hspace{3mm} 
	\frac{U}{2\epsilon_{m}^{\{M_{2}\},\{M_1\}}}\left[Y_{m,\alpha}^{\{M_{2}\},\{M_1\}}, Y_{n,\nu}^{\{N_{2}\},\{N_1\}}\right] \nonumber  \\
	&-\overline{\sum_{\mathclap{\substack{m,\alpha \\ \{M_{2}\},\{M_1\} }}}\hspace{10mm}\sum_{\mathclap{\substack{n,\nu \\ \{N_{2}\},\{N_1\} }}}}\hspace{3mm} 
	\frac{U}{2\epsilon_{m}^{\{M_{2}\},\{M_1\}}}\left[Y_{m,\alpha}^{\{M_{2}\},\{M_1\}}, Y_{n,\nu}^{\{N_{2}\},\{N_1\}}\right] ,
\end{align}
which can be simplified to 
\begin{align}
	H_{T,2}' &= \overline{\overline{\sum_{\mathclap{\substack{m,\alpha \\ \{M_{2}\},\{M_1\}, }}}\hspace{10mm}
	\sum_{\mathclap{\substack{n,\nu \\ \{N_{2}\},\{N_1\} }}}}}\hspace{3mm} 
	\frac{U}{\epsilon_{m}^{\{M_{2}\},\{M_1\}}}Y_{m,\alpha}^{\{M_{2}\},\{M_1\}}Y_{n,\nu}^{\{N_{2}\},\{N_1\}},
\end{align}
where the double over-bar indicates that the sums include only terms which satisfy $\epsilon_{m}^{ \{M_{2}\},\{M_1\}}+\epsilon_{n}^{ \{N_{2}\},\{N_1\}} = 0, 
\epsilon_{m}^{ \{M_{2}\},\{M_1\}} \neq 0, \epsilon_{n}^{ \{N_{2}\},\{N_1\}} \neq 0$. 
To simplify this further it is necessary to go back to the full hopping operator notation in which case
\begin{align}
H_{T,2}' =\, & U \sum_{\alpha,\nu}\hspace{5mm}\overline{\overline{\sum_{\mathclap{\substack{ \{M_{2}\}, \{M_{1}\} \\ \{N_{2}\}, \{N_{1}\}}}}
	\hspace{5.7mm}\sum_{\mathclap{\substack{ m, n }}}}} \hspace{1mm} 
	\frac{   T_{\alpha}^{m, \{M_{2}\}, \{M_1\}} T_{\nu}^{n, \{N_{2}\}, \{N_1\}}}{mU+
	\sum_{\gamma}V_\gamma(M_2^\gamma - M_1^\gamma)}. 
\end{align}
Now, for the double over-bar restriction to be satisfied for all $U$ and $V_\gamma$ the only 
terms that will be in the sum are those that have $n=-m, \{N_{1}\} = \{M_{2}\} - \{M_{1}\} + \{N_{2}\}$, 
which allows $H_{T,2}'$ to be reduced to
\begin{align}
	H_{T,2}' &= U \sum_{\alpha,\nu}\hspace{5mm}\tilde{\sum_{\mathclap{\substack{ \{M_{2}\}, \{M_{1}\} \\ \{N\},m}}}} 
	\hspace{3mm}\frac{   T_{\alpha}^{m, \{M_{2}\}, \{M_1\}}  T_{\nu}^{-m, \{N\}, \{M_{2}\} - \{M_{1}\} + \{N\}}}{mU+\sum_{\gamma}V_\gamma(M_2^\gamma - M_1^\gamma)},
\end{align}
with no additional conditions, which completes the second order perturbation theory.

\section{Second Order Processes and Contributions} 
\label{app:couplings}
There are eight different types of second order processes, each of which is associated with a particular coupling.
We list each of the eight couplings, $C_n$, with $n = 1$ to $8$, along with an illustrative example for each coupling in Table~\ref{tab:one},
where

\begin{eqnarray}
        C_1  = \frac{{t_b}^2}{\sum_\gamma V_\gamma (M_2^\gamma - M_1^\gamma) } , \quad  C_2  =  \frac{{t_b}^2}{-U + \sum_\gamma V_\gamma (M_2^\gamma - M_1^\gamma) } , \nonumber \\
        C_3  =   \frac{{t_p t_q}}{\sum_\gamma V_\gamma (M_2^\gamma - M_1^\gamma) } , \quad      C_4  =  \frac{{t_q t_p}}{-U + \sum_\gamma V_\gamma (M_2^\gamma - M_1^\gamma) }  , \nonumber \\
        C_5  =  \frac{{t_q}^2}{\sum_\gamma V_\gamma (M_2^\gamma - M_1^\gamma) }  ,  \quad  C_6  =    \frac{{t_p}^2}{\sum_\gamma V_\gamma (M_2^\gamma - M_1^\gamma) } , \nonumber \\
        C_7  =  \frac{{t_q}^2}{-U + \sum_\gamma V_\gamma (M_2^\gamma - M_1^\gamma) } , \quad    C_8  =  \frac{{t_p}^2}{-U + \sum_\gamma V_\gamma (M_2^\gamma - M_1^\gamma) } . \nonumber
\end{eqnarray}

\begin{center}
\begin{longtable}{|c|c|}
\caption{Examples of Second Order Processes for each of the couplings $C_1$ to $C_8$.}\\
\hline
	Process &  Contribution to Hamiltonian  \\ \hhline{|=|=|}
	\shortstack{\includegraphics[scale=0.2]{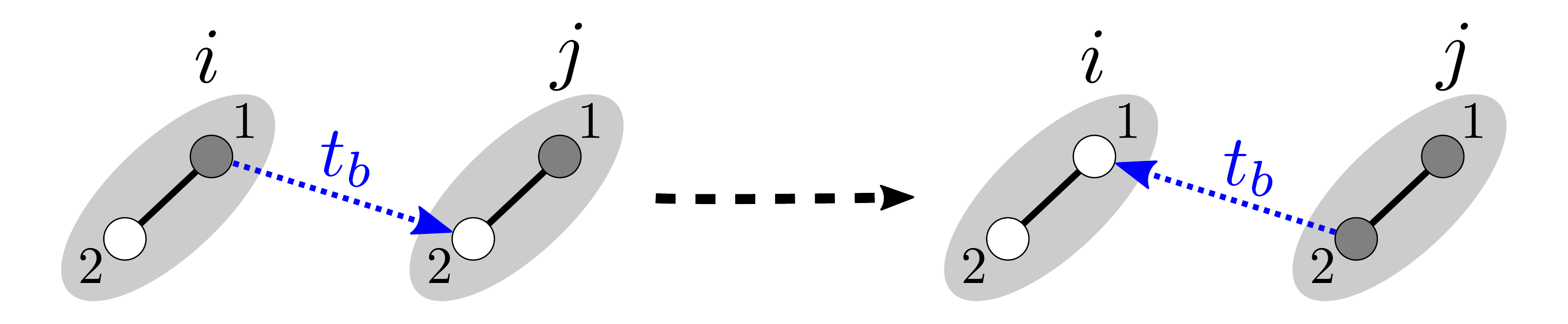} \\ $i=(x,y)$ and $j=(x+1,y)$ \\ \, } & 
	 \shortstack{ $C_1(\tfrac{1}{2}+P_i^z)(\tfrac{1}{2}+P_j^z)$ \\ \, \\ \, \\ \, \\ \, \\ \, \\ \, \\ \, }\\ \hline
	
	\shortstack{\includegraphics[scale=0.2]{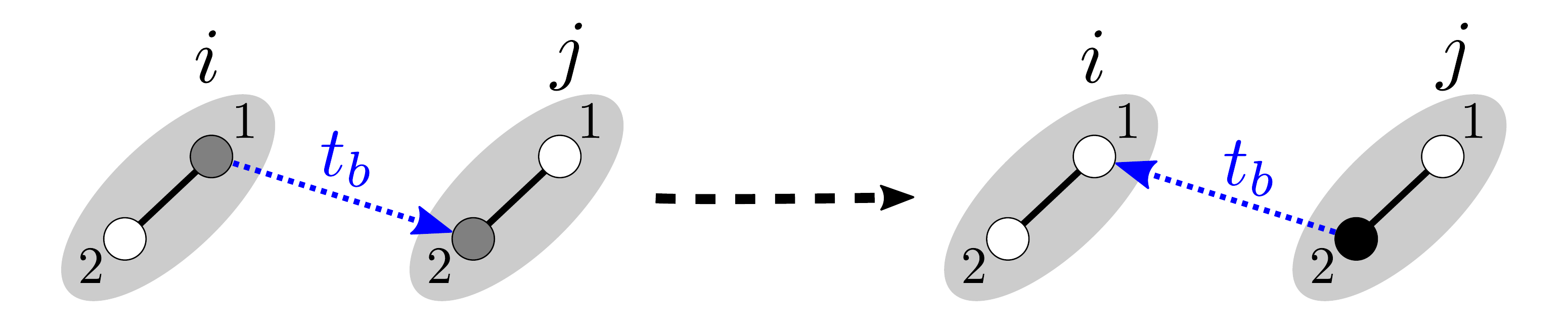}\\ $i=(x,y)$ and $j=(x+1,y)$ \\ \, } &
\shortstack{$C_2(\tfrac{1}{2}+P_i^z)(\tfrac{1}{2}-P_j^z)(\tfrac{1}{2}-2\vec{S_i}\cdot\vec{S_j})$
	 \\ \, \\ \, \\ \, \\ \, } \\ \hline

	\shortstack{\includegraphics[scale=0.2]{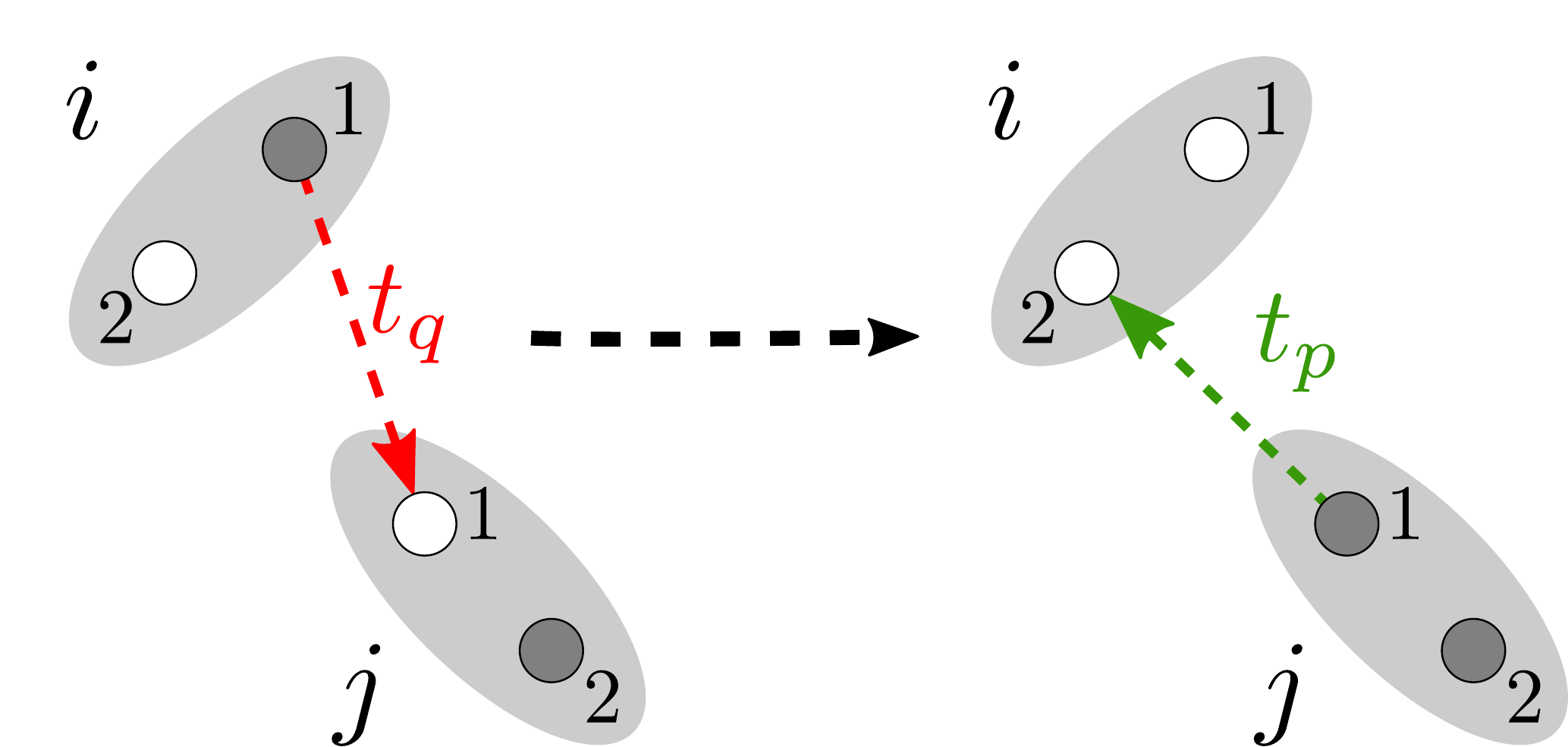} \\  $i=(x,y)$ and $j=(x+\tfrac{1}{2},y - \tfrac{1}{2})$ \\ \, } &
         \shortstack{$C_3(\tfrac{1}{2}-P_j^z)P_i^-$ \\ \, \\ \, \\ \,\\ \, \\ \, \\ \,}\\ \hline

	\shortstack{\includegraphics[scale=0.2]{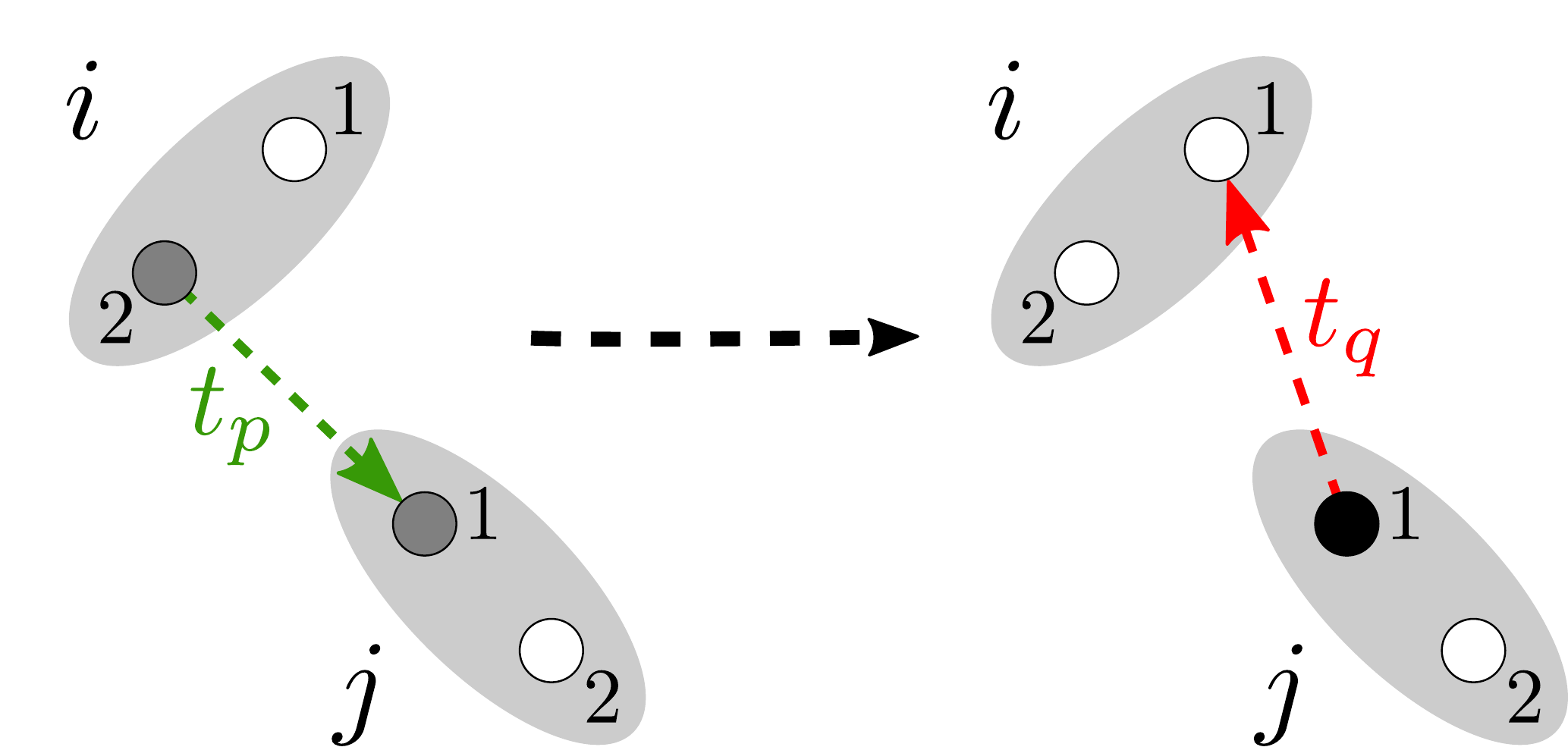} \\ $i=(x,y)$ and $j=(x+\tfrac{1}{2},y - \tfrac{1}{2})$ \\ \,} &
\shortstack{$C_4(\tfrac{1}{2}+P_j^z)P_i^-(\tfrac{1}{2}-2\vec{S_i}\cdot\vec{S_j})$\\ \, \\ \, \\ \, \\ \, } \\ \hline

	\shortstack{\includegraphics[scale=0.2]{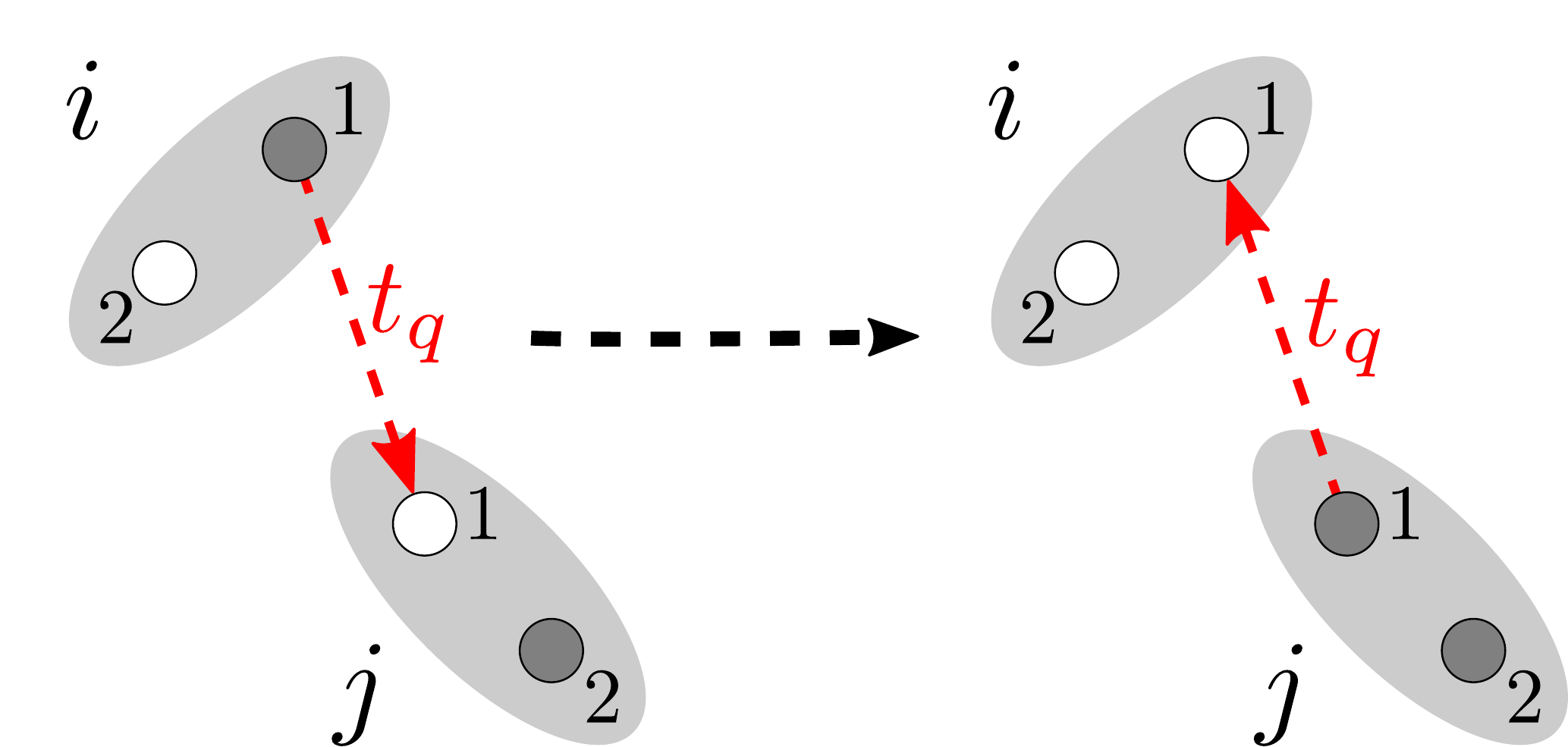} \\ $i=(x,y)$ and $j=(x+\tfrac{1}{2},y - \tfrac{1}{2})$ \\ \,} &
\shortstack{$C_{5}(\tfrac{1}{2}+P_i^z)(\tfrac{1}{2}-P_j^z)$ \\ \, \\ \, \\ \,\\ \, \\ \, \\ \,}\\ \hline

	\shortstack{\includegraphics[scale=0.2]{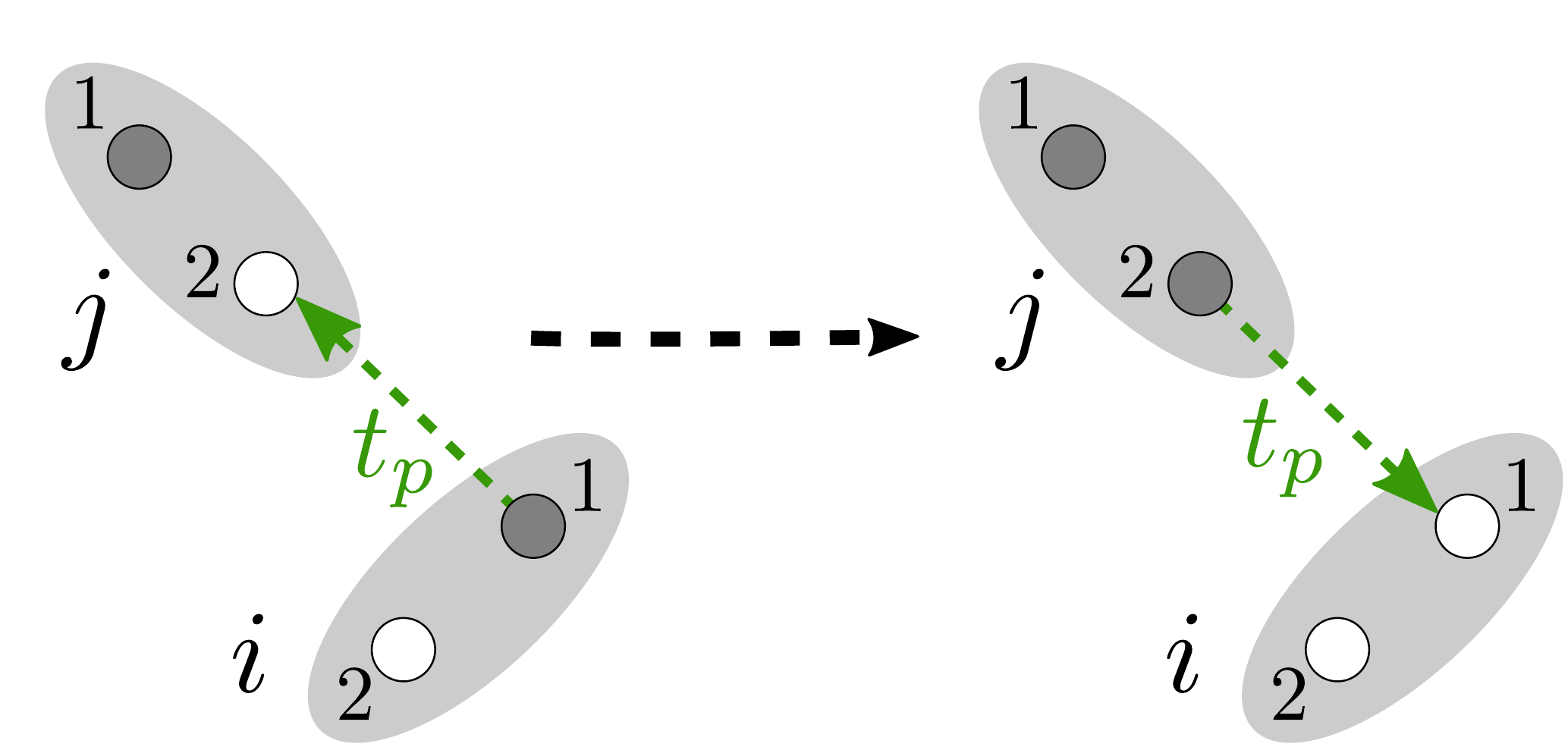} \\ $i=(x,y)$ and $j=(x-\tfrac{1}{2},y + \tfrac{1}{2})$ \\ \,} &
\shortstack{$C_6(\tfrac{1}{2}+P_i^z)(\tfrac{1}{2}+P_j^z)$ \\ \, \\ \, \\ \,\\ \, \\ \, \\ \,}\\ \hline

	\shortstack{\includegraphics[scale=0.2]{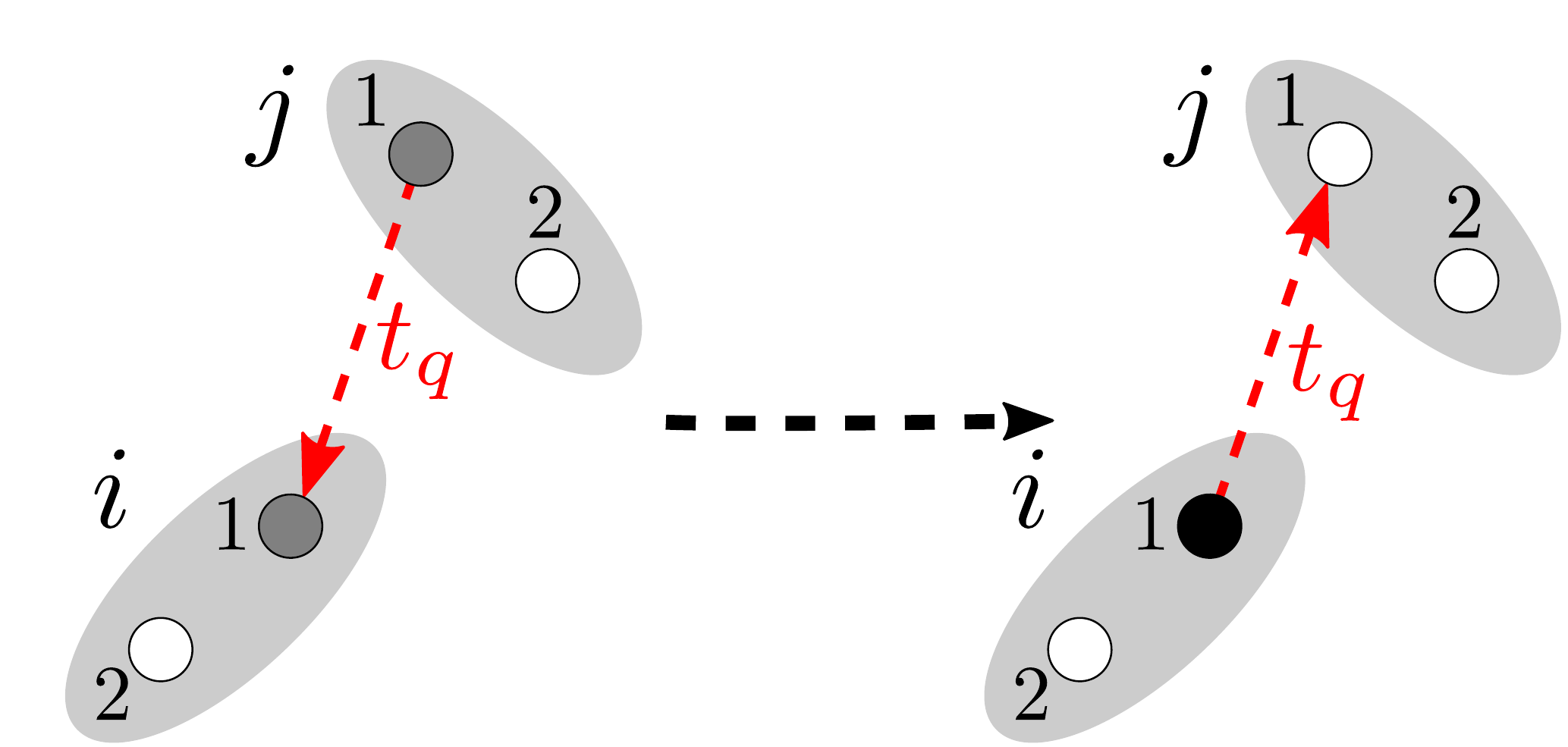} \\ $i=(x,y)$ and $j=(x+\tfrac{1}{2},y + \tfrac{1}{2})$ \\ \, } & 
\shortstack{$C_{7}(\tfrac{1}{2}+P_i^z)(\tfrac{1}{2}+P_j^z)(\tfrac{1}{2}-2\vec{S_i}\cdot\vec{S_j})$\\ \, \\ \,  \\ \, \\ \, \\ \,} \\ \hline

	\shortstack{\includegraphics[scale=0.2]{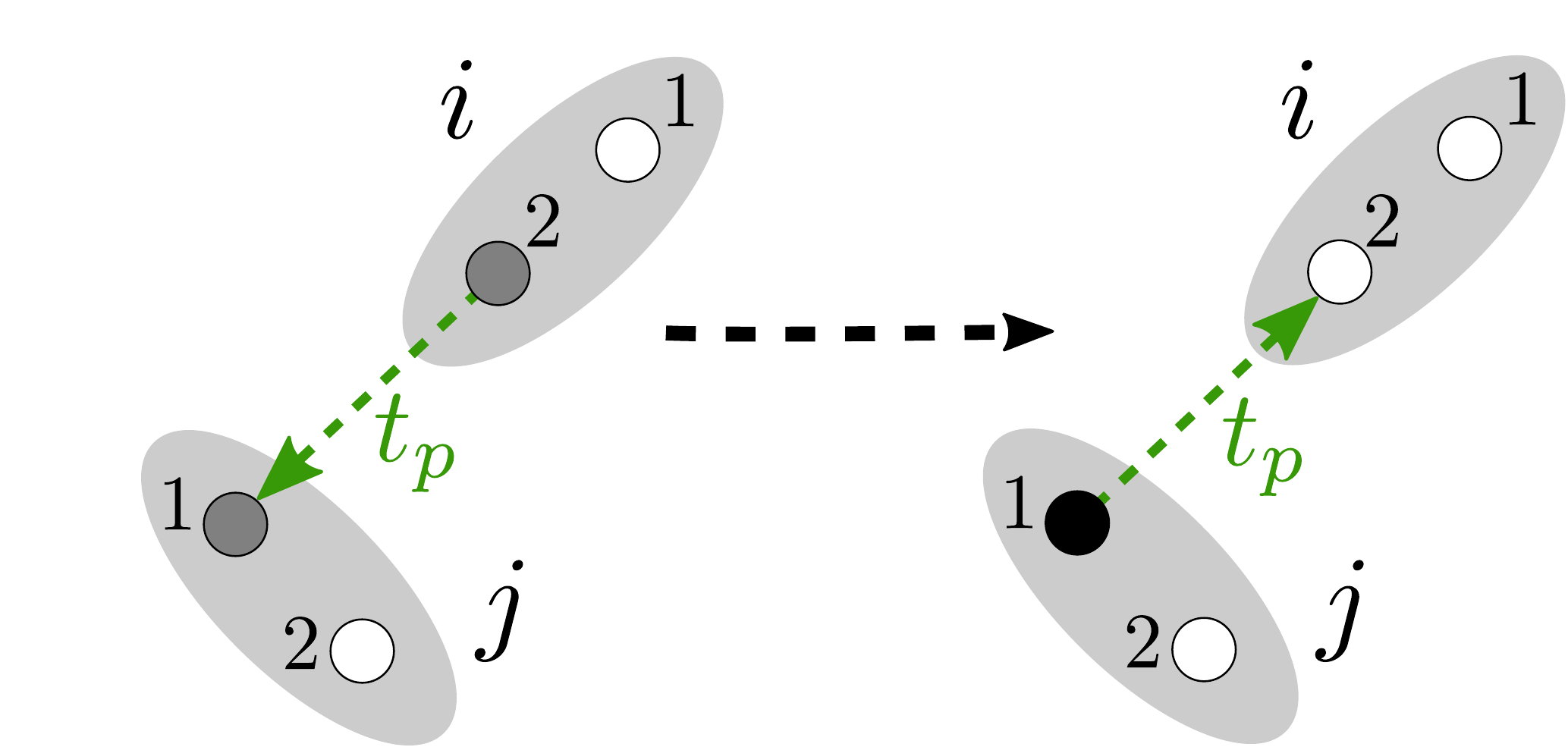} \\ $i=(x,y)$ and $j=(x-\tfrac{1}{2},y - \tfrac{1}{2})$ \\ \,} &
\shortstack{$C_8(\tfrac{1}{2}-P_i^z)(\tfrac{1}{2}+P_j^z)(\tfrac{1}{2}-2\vec{S_i}\cdot\vec{S_j})$\\ \, \\ \, \\ \, \\ \, \\ \, } \\ \hline
\label{tab:one}
\end{longtable}
\end{center}

\subsection{Tables of Coefficients} \label{App.CoefTable}
In tables \ref{tab:two} to \ref{tab:four} we set out the relationships between the coefficients $C_1$ to $C_8$ listed above and the coefficients that enter the
effective model Eq.~(\ref{eq:Ham_C_ij}).  The coefficients vary for each type of pair of dimers.  We only list non-zero 
coefficients

\renewcommand{\arraystretch}{1.3}
\begin{table}[ht]
\caption{Coefficients for $i = (x,y)$ and $j = (x\pm 1,y)$} \label{tab:coef_1}
\begin{center}
\begin{tabular}{|c|c|c|}
\hline
	Coupling & $i = (x,y)$ and $j = (x+ 1,y)$ & $i = (x,y)$ and $j = (x-1,y)$	\\
\hline
	$C_{i,j}^0$ & $\frac{C_2}{2}$ &  $-\frac{C_2}{2}$ \\
	$C_{i,j}^1$ & $-\frac{C_2}{2}$ &  $\frac{C_2}{2}$ \\
	$C_{i,j}^2$ & $2C_1 - C_2$ &  $2C_1 - C_2$ \\
	$C_{i,j}^3$ & $-C_2$ &  $-C_2$ \\
	$C_{i,j}^4$ & $-2C_2$ &  $2C_2$ \\
	$C_{i,j}^5$ & $2C_2$ &  $-2C_2$ \\
	$C_{i,j}^6$ & $4C_2$ &  $4C_2$ \\
\hline
\end{tabular}
	\label{tab:two}
\end{center}
\end{table}
\renewcommand{\arraystretch}{1}

\renewcommand{\arraystretch}{1.3}
\begin{table}[ht]
\caption{Coefficients for $i = (x,y)$ and $j = (x+\tfrac{1}{2},y-\tfrac{1}{2})$ or  $j = (x-\tfrac{1}{2},y+\tfrac{1}{2})$} \label{tab:coef_2}
\begin{center}
\begin{tabular}{|c|c|c|}
\hline
        Coupling & $i = (x,y)$ and $j = (x+\tfrac{1}{2},y - \tfrac{1}{2})$ & $i = (x,y)$ and $j = (x-\tfrac{1}{2},y+\tfrac{1}{2})$  \\
\hline
	$C_{i,j}^0$ & $\frac{1}{2}\left(C_7 - C_8\right)$ &  $-\frac{1}{2}\left(C_7 - C_8\right)$ \\
        $C_{i,j}^1$ & $\frac{1}{2}\left(C_7 + C_8\right)$ &  $-\frac{1}{2}\left(C_7 + C_8\right)$ \\
	$C_{i,j}^2$ & $2\left(C_6 - C_5\right) + \left(C_7 - C_8\right)$ &   $2\left(C_6 - C_5\right) - \left(C_7 - C_8\right)$ \\
	$C_{i,j}^3$ & $-\left(C_7 + C_8\right)$ &  $-\left(C_7 + C_8\right)$ \\
	$C_{i,j}^4$ & $2\left(C_8 - C_7\right)$ &  $2\left(C_7 - C_8\right)$ \\
        $C_{i,j}^5$ & $-2\left(C_7 + C_8\right)$ &  $2\left(C_7 + C_8\right)$ \\
	$C_{i,j}^6$ & $4\left(C_8 - C_7\right)$ &  $-4\left(C_7 - C_8\right)$ \\
	$C_{i,j}^7$ & $\frac{1}{4}\left(3C_3 + C_4\right)$ & $\frac{1}{4}\left(3C_3 + C_4\right)$ \\
	$C_{i,j}^8$ & $-\frac{1}{2}\left(C_3 - C_4\right)$ & $\frac{1}{2}\left(C_3 - C_4\right)$ \\
	$C_{i,j}^{9}$ & $C_3 - C_4$ & $C_3 - C_4$ \\
	$C_{i,j}^{10}$ & $2\left(C_3 - C_4\right)$ & $-2\left(C_3 - C_4\right)$ \\
\hline
\end{tabular}
	\label{tab:three}
\end{center}
\end{table}

\renewcommand{\arraystretch}{1.3}
\begin{table}[ht]
\caption{Coefficients for $i = (x,y)$ and $j = (x+\tfrac{1}{2},y+\tfrac{1}{2})$ or  $j = (x-\tfrac{1}{2},y-\tfrac{1}{2})$} \label{tab:coef_2}
\begin{center}
\begin{tabular}{|c|c|c|}
\hline
        Coupling & $i = (x,y)$ and $j = (x+\tfrac{1}{2},y + \tfrac{1}{2})$ & $i = (x,y)$ and $j = (x-\tfrac{1}{2},y-\tfrac{1}{2})$  \\
\hline
        $C_{i,j}^0$ & $\frac{1}{2}\left(C_7 + C_8\right)$ &  $-\frac{1}{2}\left(C_7 + C_8\right)$ \\
        $C_{i,j}^1$ & $\frac{1}{2}\left(C_7 - C_8\right)$ &  $-\frac{1}{2}\left(C_7 - C_8\right)$ \\
        $C_{i,j}^2$ & $2\left(C_6 - C_5\right) + \left(C_7 - C_8\right)$ &   $2\left(C_6 - C_5\right) + \left(C_7 - C_8\right)$ \\
        $C_{i,j}^3$ & $-\left(C_7 + C_8\right)$ &  $-\left(C_7 + C_8\right)$ \\
        $C_{i,j}^4$ & $-2\left(C_7 + C_8\right)$ &  $2\left(C_7 + C_8\right)$ \\
        $C_{i,j}^5$ & $-2\left(C_7 - C_8\right)$ &  $2\left(C_7 - C_8\right)$ \\
        $C_{i,j}^6$ & $-4\left(C_7 + C_8\right)$ &  $-4\left(C_7 - C_8\right)$ \\
	$C_{i,j}^{11}$ & $\frac{1}{4}\left(3C_3 + C_4\right)$ & $\frac{1}{4}\left(3C_3 + C_4\right)$ \\
	$C_{i,j}^{12}$ & $\frac{1}{2}\left(C_2 - C_3\right)$ & $\frac{1}{2}\left(C_3 - C_4\right)$ \\
        $C_{i,j}^{13}$ & $C_3 - C_2$ & $C_3 - C_4$ \\
        $C_{i,j}^{14}$ & $2\left(C_3 - C_2\right)$ & $-2\left(C_3 - C_4\right)$ \\
\hline
\end{tabular}
	\label{tab:four}
\end{center}
\end{table}

\end{widetext}

\end{appendix}

\bibliographystyle{apsrev4-1}
\bibliography{arxiv}

\end{document}